\documentclass[aps,pra,reprint,superscriptaddress,10pt,nofootinbib]{revtex4-2}

\usepackage[utf8]{inputenc}
\usepackage[english]{babel}
\usepackage{physics}
\usepackage{times}
\usepackage{graphicx,epsfig}
\graphicspath{{images/}}
\usepackage{color}
\usepackage[usenames,dvipsnames]{xcolor}
\usepackage{amsmath,bbm,amssymb, amsthm}
\usepackage{dsfont} 
\usepackage{stmaryrd}
\definecolor{linkcolor}{rgb}{0,0,0.6}		
\definecolor{bleu}{HTML}{1732a6}
\usepackage[colorlinks=true,pdfstartview=FitV,linkcolor=linkcolor,citecolor=linkcolor,urlcolor=linkcolor,hyperindex=true,hyperfigures=false]{hyperref}
\usepackage{numprint}
\usepackage[normalem]{ulem}

\addto\captionsenglish{}

\newcommand{\stirlingii}{\genfrac{\{}{\}}{0pt}{}}

\newcommand{\kket}[1]{{|#1 \rangle \rangle}}

\renewcommand{\ketbra}[2]{{|#1 \rangle \langle #2|}}
\newcommand{\kketbra}[2]{{|#1 \rangle \rangle \langle \langle #2|}}

\renewcommand{\braket}[2]{{\langle #1 | #2 \rangle}}
\newcommand{\id}{\mathbbm{1}}

\begin{document}

\title{Comparing the quantum switch and its simulations with energetically-constrained operations}

\author{Marco Fellous-Asiani}
\thanks{These authors contributed equally to this work.\\Corresponding authors:\\
m.fellous-asiani@cent.uw.edu.pl\\
raphael.mothe@neel.cnrs.fr}
\affiliation{Univ.\ Grenoble Alpes, CNRS, Grenoble INP\footnote{Institute of Engineering Univ. Grenoble Alpes}, Institut N\'eel, 38000 Grenoble, France}
\affiliation{Centre for Quantum Optical Technologies, Centre of New Technologies, University of Warsaw, Banacha 2c, 02-097 Warsaw, Poland}

\author{Raphaël Mothe}
\thanks{These authors contributed equally to this work.\\Corresponding authors:\\
m.fellous-asiani@cent.uw.edu.pl\\
raphael.mothe@neel.cnrs.fr}
\affiliation{Univ.\ Grenoble Alpes, CNRS, Grenoble INP\footnote{Institute of Engineering Univ. Grenoble Alpes}, Institut N\'eel, 38000 Grenoble, France}

\author{Léa Bresque}
\affiliation{Univ.\ Grenoble Alpes, CNRS, Grenoble INP\footnote{Institute of Engineering Univ. Grenoble Alpes}, Institut N\'eel, 38000 Grenoble, France}

\author{Hippolyte Dourdent}
\affiliation{Univ.\ Grenoble Alpes, CNRS, Grenoble INP\footnote{Institute of Engineering Univ. Grenoble Alpes}, Institut N\'eel, 38000 Grenoble, France}
\affiliation{ICFO-Institut de Ciencies Fotoniques, The Barcelona Institute of Science and Technology,\\ 08860 Castelldefels, Barcelona, Spain}

\author{Patrice A.\ Camati}
\affiliation{Univ.\ Grenoble Alpes, CNRS, Grenoble INP\footnote{Institute of Engineering Univ. Grenoble Alpes}, Institut N\'eel, 38000 Grenoble, France}

\author{Alastair A.\ Abbott}
\affiliation{Univ.\ Grenoble Alpes, Inria, 38000 Grenoble, France}

\author{Alexia Auffèves}
\affiliation{Univ.\ Grenoble Alpes, CNRS, Grenoble INP\footnote{Institute of Engineering Univ. Grenoble Alpes}, Institut N\'eel, 38000 Grenoble, France}

\author{Cyril Branciard}
\affiliation{Univ.\ Grenoble Alpes, CNRS, Grenoble INP\footnote{Institute of Engineering Univ. Grenoble Alpes}, Institut N\'eel, 38000 Grenoble, France}

\date{23 February 2023}

\begin{abstract}
Quantum mechanics allows processes to be superposed, leading to a genuinely quantum lack of causal structure. For example, the process known as the quantum switch applies two operations ${\cal A}$ and ${\cal B}$ in a superposition of the two possible orders, ${\cal A}$
before ${\cal B}$ and ${\cal B}$ before ${\cal A}$. Experimental implementations of the quantum switch have been challenged by some on the grounds that the operations ${\cal A}$ and ${\cal B}$ were implemented more than once, thereby simulating indefinite causal order rather than actually implementing it. Motivated by this debate, we consider a situation in which the quantum operations are physically described by a light-matter interaction model. While for our model the two processes are indistinguishable in the infinite energy regime, restricting the energy available for the implementation of the operations introduces imperfections, which allow one to distinguish processes using different number of operations.
We consider such an energetically-constrained scenario and compare the quantum switch to one of its natural simulations, where each operation is implemented twice. Considering a commuting-vs-anticommuting unitary discrimination task, we find that within our model the quantum switch performs better, for some fixed amount of energy, than its simulation. In addition to the known computational or communication advantages of causal superpositions, our work raises new questions about their potential energetic advantages. 
\end{abstract}

\maketitle

\section{Introduction}

In the standard view of both the classical and quantum worlds, processes normally occur with a fixed causal order: the order of successive operations is classically well defined. Nonetheless, in quantum theory it is possible to consider causally indefinite processes~\cite{oreshkov12,chiribella_quantum_2013}. For example, by using a quantum system in a superposition to coherently control the order in which operations are applied, one can obtain quantum processes in which the causal order is indefinite~\cite{chiribella_quantum_2013,wechs21}.
The question of whether or not there is an advantage (of any kind) in using superpositions of causal orders has been studied from different points of view. It has been shown, in particular, that the superposition of causal orders provides computational and communication advantages over any standard quantum circuit operating with a definite causal order~\cite{chiribella12,araujo_computational_2014,facchini15,feix15,guerin16,Ebler2018,salek2018quantum,taddei21,Chiribella_2021,Chiribella_2021b}.

The paradigmatic example of a quantum process with indefinite causal order is the ``quantum switch'' (QS)~\cite{chiribella_quantum_2013}. In this process, a two-level quantum control system $C$ is used to coherently control the order in which two quantum operations---any two completely positive (CP) maps---${\cal A}$ and ${\cal B}$ are applied to a target system $S$.
If $C$ is in the state $\ket{0_C}$ (resp.\ $\ket{1_C}$), then the order ${\cal A}$ before ${\cal B}$ (resp.\ ${\cal B}$ before ${\cal A}$) is realized.  When $C$ is in a superposition of these two control states, however, the causal order between ${\cal A}$ and ${\cal B}$ is itself superposed and hence indefinite.

Formally, the QS is defined as a quantum supermap~\cite{chiribella08a} that transforms the two operations ${\cal A}$ and ${\cal B}$ into a new one, which applies the latter in a coherently-controlled order. When ${\cal A}: \rho_S \mapsto U_A \rho_S U_A^\dagger$ and ${\cal B}: \rho_S \mapsto U_B \rho_S U_B^\dagger$ are unitary operations, then these are transformed into the new unitary operation ${\cal U}^\text{QS}: \rho_C\otimes\rho_S \mapsto U^\text{QS} (\rho_C\otimes\rho_S) U^{\text{QS}\dagger}$ acting on the control and the target systems, with
\begin{align}
    U^\text{QS}  &= U^\text{QS}(U_A,U_B) \notag \\
    & =\ketbra{0_C}{0_C}\otimes U_BU_A + \ketbra{1_C}{1_C}\otimes U_AU_B.
    \label{eq:def_supermap_switch}
\end{align}
A more general definition valid for any CP maps ${\cal A}$ and ${\cal B}$ (based on their Kraus decompositions) can be found in Ref.~\cite{chiribella_quantum_2013}.
In the case of a control system initially prepared in the state $\ket{+_C}=\frac{1}{\sqrt{2}}(\ket{0_C}+\ket{1_C})$---which, for concreteness, we shall henceforth restrict ourselves to---and for unitary operations ${\cal A}$ and ${\cal B}$ as above, the quantum switch then effectively applies the transformation
\begin{align}
    & \ket{+_C} \otimes \ket{\psi_S} \notag \\
    & \quad \mapsto \frac{1}{\sqrt{2}}\big(\ket{0_C} \otimes U_BU_A \ket{\psi_S}+\ket{1_C} \otimes U_AU_B \ket{\psi_S}\big) \label{eq:ideal_evol}
\end{align}
for any arbitrary initial target system state $\ket{\psi_S}$.
The fact that the evolution~\eqref{eq:ideal_evol} is obtained using, or implementing, each operation ${\cal A}$ and ${\cal B}$ only once is crucial for the causal indefiniteness of the QS.
Indeed, any quantum circuit with a well-defined causal structure simulating the evolution~\eqref{eq:ideal_evol} would necessarily require at least two uses (or ``implementations'') of either ${\cal A}$ or ${\cal B}$~\cite{chiribella_quantum_2013}.
This difference is behind many of the computational advantages offered by the QS~\cite{chiribella12,araujo_computational_2014,facchini15,feix15,taddei21}.

Over recent years, a number of experimental proposals~\cite{araujo_computational_2014,Friis2014,wechs21} and implementations~\cite{procopio_experimental_2015,rubino17,goswami18,Wei2019,Goswami2020,guo20,rubino_experimental_2021,Rubino2022experimental,cao22} of the QS have been presented.
Depending on the details of the implementations, and on which degrees of freedom were employed, a debate emerged as to whether these experiments truly realized the quantum switch, or whether they instead simply simulated the evolution of Eq.~\eqref{eq:ideal_evol}~\cite{maclean17,oreshkov19,paunkovic20,kristjansson20,vilasini22,ormrod22}.
For example, for many of the photonic demonstrations
it has been argued that the implementations of the operations ${\cal A}$ and ${\cal B}$ differed depending on which path the photon took (i.e., on the value of the control).
Hence, were ${\cal A}$ and ${\cal B}$ each really implemented once, rather than twice? 

Motivated by this debate, we investigate here a realistic scenario in which both the quantum switch and a natural simulation of it are applied to noisy operations.
Indeed, simulations exactly reproducing Eq.~\eqref{eq:ideal_evol} for perfect unitary operations may lead to different dynamics in the presence of imperfections.
To this end, we introduce an operational definition of a ``box''---a physical implementation of an abstract quantum operation---in which a system physically interacts once with an auxiliary quantum system in order to perform the desired operation. 
This allows us to concretely compare the QS, in which two boxes (realizing ${\cal A}$ and ${\cal B}$) are used, with simulations, such as the natural ``four box'' simulation (4B) in which four boxes (realizing ${\cal A}_0, {\cal A}_1$ and ${\cal B}_0, {\cal B}_1$) are used to superpose the causal orders ``${\cal A}_0$ before ${\cal B}_0$ and ${\cal B}_1$ before ${\cal A}_1$'' rather than directly ``${\cal A}$ before ${\cal B}$ and ${\cal B}$ before ${\cal A}$''.
We adopt a noise model---and thus a model for a box---that is motivated by the (approximate) implementation of unitary operations via the interaction with some auxiliary systems.
Specifically, we consider a cavity quantum electrodynamics (QED) setup, consisting of an atom that passes inside a cavity containing a single-mode quantum field. We employ the Jaynes-Cummings Hamiltonian as a light-matter model of interaction for the implementation of the unitary operations on the atomic qubit system. 
Due to the quantum nature of the field, it can generally become entangled with the system, effectively leading to a noisy operation on the target system. Only when the field contains an infinite amount of energy is the ideal unitary operation recovered. 

With this noise model, we show that the QS and the 4B lead to different dynamics and thus different final control-target states, allowing measurements to distinguish between these two setups. 
Furthermore, this allows us to assess the performance of this implementation of the QS from an energetic perspective. 
Considering a modified version of the commuting-vs-anticommuting discrimination task introduced in Ref.~\cite{chiribella12}---for which the QS is known, in the ideal case, to provide an advantage over all quantum circuits using two boxes in a fixed causal order---we find that in our model the QS also provides an energetic advantage over the 4B.
Beyond the computational and communication advantages brought about by the causal indefiniteness of the QS, our work paves the way to study some of its potential energetic advantages as well. Thus, our results complement recent theoretical~\cite{felce_quantum_2020,Guha2020,Simonov2022,Chen2021,Zhao2022,Nie2022} and experimental~\cite{Nie2020,cao21,Felce2021} interest in the potential utility of causal indefiniteness in quantum thermodynamics.

Our paper is organized as follows. 
In Sec.~\ref{sec:2} we introduce the key concepts we work with in this paper, introducing first our general definition of a box (Sec.~\ref{sec:box_def}) before presenting a specific box implementation based on the Jaynes-Cummings model (Sec.~\ref{sec:energetics}), which we will use to study the energetics of the QS and 4B protocols that we define in Sec.~\ref{sec:switch}. In Sec.~\ref{sec:comp_task} we define the discrimination task that we will use to benchmark the performance of the QS and the 4B. We then explain in Sec.~\ref{sec:finite_energy_comparison} how this allows us to compare the QS and the 4B under finite energy constraints on one of the operations, and we describe in Sec.~\ref{sec:part_comb} how we extend the comparison to circuits with fixed causal orders.
In Sec.~\ref{sec:results} we present and discuss our numerical and analytical results. We finally conclude in Sec.~\ref{sec:conclusions}.

\section{Preliminaries}\label{sec:2}

We begin this section by first abstracting the notion of an implementation of a quantum operation through the definition of a ``box''. Then, in the following subsection, we present the specific model of boxes as Jaynes-Cummings interactions that we will use throughout the paper. These models are independent of any particular experimental realization, which is beyond the scope of the present work. Nonetheless, in Appendix~\ref{app:experimental} (see also the caption of Fig.~\ref{fig:QS_4B}) we outline one possible realization for the QS and 4B within our framework.

\subsection{Implementing an operation: our definition of a ``box''}
\label{sec:box_def}

To clarify the differences between the QS and the 4B, let us properly define what we mean by a ``box''. We consider a target system $S$ on which one wants to implement a given operation. First, we need to distinguish the ideal operation one wishes to realize from its actual implementation. The ideal operation can be, for instance, a rotation in the Hilbert space while its implementation is how this is realized in practice. Does this implementation perfectly realise a unitary operation on the target system? Or does it effectively act as a trace-preserving (CPTP) map, which only approximates the desired dynamics?

\begin{figure}
    \centering
    \includegraphics[width=0.98\columnwidth]{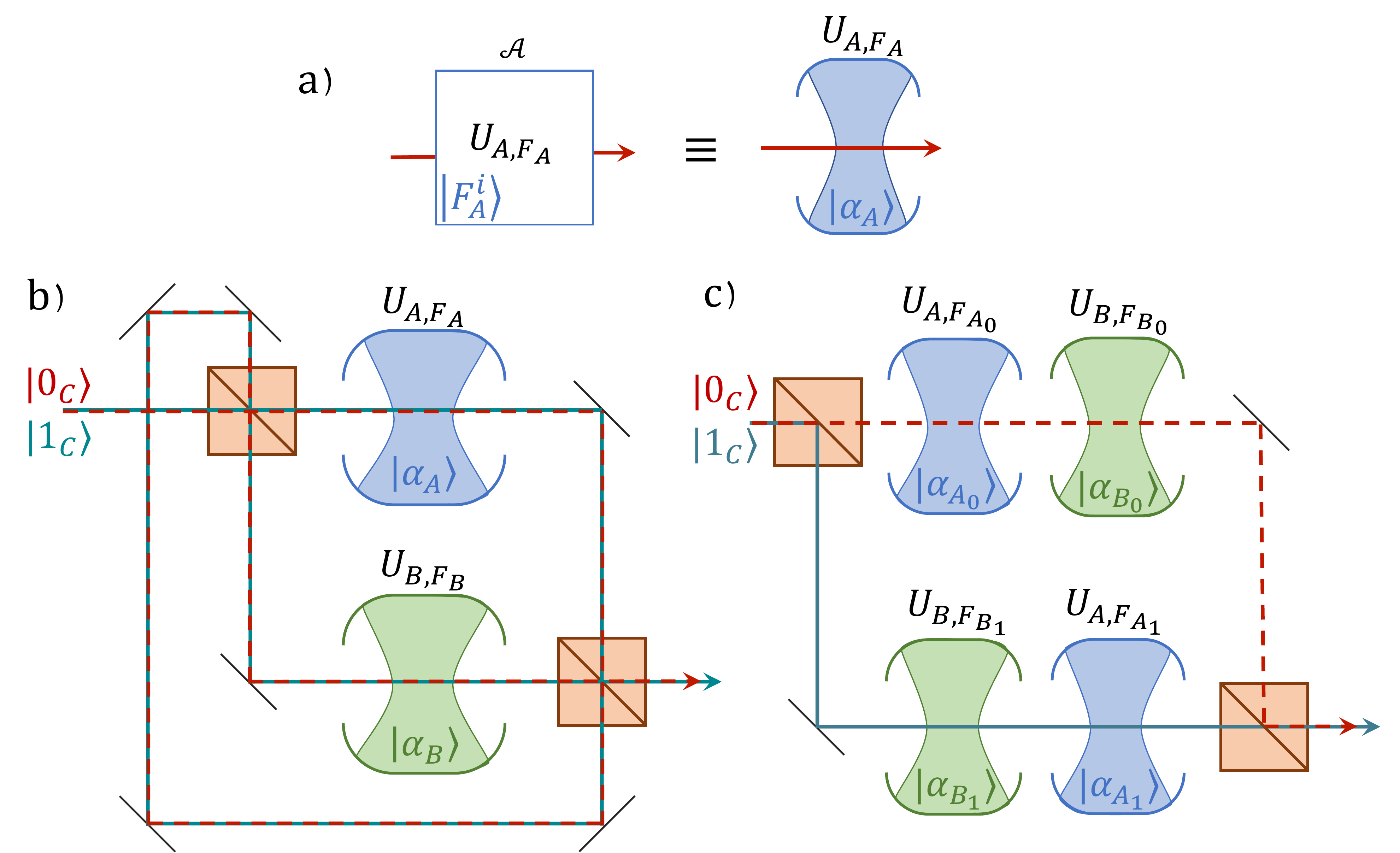}
    \caption{(a) We describe the implementation of an operation ${\cal A}$ by a ``box'' consisting of the pair $\{U_{A,F_A},\ket{F^\text{i}_A}\}$, where $U_{A,F_A}$ is a unitary operation acting on the target system and an auxiliary system initialized in the state $\ket{F^\text{i}_A}$. Such an implementation is meant to approximate an ideal operation $U_A$ (in the unitary case) on the target system; see Subsec.~\ref{sec:box_def}. In the more specific model of Subsec.~\ref{sec:energetics} that we use in our study, the auxiliary system is taken to be a field in a cavity, initially prepared in a coherent state $\ket{F_A^\text{i}}=\ket{\alpha_A}$. General depiction of (b) the quantum switch (QS) versus (c) its four-box (4B) simulation, employing interferometric-like setups. In both cases, the control qubit starts in a superposition of the states $\ket{0_C}$ and $\ket{1_C}$. The path taken by the target system (solid or
    dashed lines), and thus the order of the operations, is coherently controlled by the state of the control system. The coherent splitting of the paths is generically depicted by orange squares akin to polarizing beam-splitters. The role of the depicted paths and orange squares is to flip on which branch of the superposition the boxes act. As these sketches illustrate, the basic conceptual difference between the QS and 4B setups is the number of physical boxes needed to obtain the final target state. A possible way to physically realise both experiments (and in particular, the splitting by the orange squares) is proposed in Appendix~\ref{app:experimental}.}
    \label{fig:QS_4B}
\end{figure}

In order to implement a given, but otherwise arbitrary, operation
$\mathcal{A}$ in a controlled way on a quantum system $S$, one would typically couple $S$ with some auxiliary system $F_{A}$. The global
Hamiltonian describing the evolution of $S$ and $F_{A}$ will generally
have an expression of the form $H=H_{S}+H_{F_{A}}+H_{SF_{A}}^{\text{int}}$,
where $H_{S}$ and $H_{F_{A}}$ are the free Hamiltonians of $S$
and $F_{A}$, respectively, and $H_{SF_{A}}^{\text{int}}$ is the
interaction Hamiltonian. Here we will assume that the free Hamiltonians are given and always present, while the controllable quantities 
are $H_{SF_{A}}^{\text{int}}$ (choosing $F_{A}$ appropriately),
the time of interaction and the initial state of $F_{A}$. We will consider that the operation
$\mathcal{A}$ corresponds only to the controllable part of the dynamics,
i.e.\ without the free evolution, by considering the interaction picture~\cite{haroche2006exploring}.

More formally, for the implementation of operation $\mathcal{A}$
through the Hamiltonian $H$, the time evolution operator in the
interaction picture with respect to $H_{S}+H_{F_{A}}$ is given by
$U_{A,F_{A}}=\mathcal{T}\exp\left\{ -\frac{i}{\hbar}\int_{0}^{t}ds\,H_{I}\left(s\right)\right\} $,
where $\mathcal{T}$ is the time-ordering operator, $t$ is the
interaction time, and $H_{I}\left(t\right)=e^{+\frac{i}{\hbar}t\left(H_{S}+H_{F_{A}}\right)}H_{SF_{A}}^{\text{int}}e^{-\frac{i}{\hbar}t\left(H_{S}+H_{F_{A}}\right)}$
is the interaction Hamiltonian in the interaction picture. The pair $\left\{ U_{A,F_{A}},\left|F_{A}^{\text{i}}\right\rangle \right\}$, where $\left|F_{A}^{\text{i}}\right\rangle$ is the initial state of the auxiliary system, thus describes the controllable quantities; this is what we take to define the ``box'' implementing operation $\mathcal{A}$, see Fig.~\ref{fig:QS_4B}(a).

Throughout this article we will take the ideal operations that one wishes to perform on the target systems to be unitary, described by unitary operators $U_A$ (or $U_B$).
The goal of the interaction considered above is, ideally, to induce the evolution
\begin{align}
&U_{A,F_A}\big(\ket{\psi_S} \otimes \ket*{F_A^\text{i}}\big) = \big(U_A \ket{\psi_S} \big) \otimes \ket*{F_A^\text{f}} \quad\forall\left|\psi_{S}\right\rangle,
\label{eq:1}
\end{align}
where $\ket*{F_A^\text{f}}$ is some final state of $F_A$ which does not depend on the target system's initial state $\ket{\psi_S}$. However, in general, one may only be able to approximate such an evolution, the final state typically being entangled (at least for some initial state $\ket{\psi_S}$). The operation effectively performed on the target system will then be obtained by tracing out the auxiliary system from the output state, and will be found to be a noisy version of $U_A$ rather than its ideal implementation~\cite{Barnes99,Gea-Banacloche02,Gea-Banacloche02_2,Ozawa02}.

\subsection{Energetic model}
\label{sec:energetics}

The model of a box described above is deliberately rather general.
In this paper, we will focus on a specific model of such an interaction, in which the auxiliary systems are electromagnetic fields with finite energy. 
The noise in the operations will thus originate from the fact that these fields become entangled with the system $S$; the less energy there is in the fields, the noisier the operations will be.

We thereby consider an atomic system, and take $S$ to a be a two-level (i.e., qubit) system defined by two energy levels of the atom with the free Hamiltonian $H_S=-\frac{\hbar \omega_0}{2} \sigma_z$, where $\omega_0$ is the frequency of the system $S$ (and $\sigma_z$ is the usual Pauli matrix). The atom is coupled to a resonant electromagnetic field through a Jaynes-Cummings interaction. Specifically, the free Hamiltonian of the field and the interaction Hamiltonian can be expressed using the bosonic annihilation (resp.\ creation) operator $a$ (resp.\ $a^\dagger$) and the lowering (resp.\ raising) operator of the atom $\sigma_-$ (resp.\ $\sigma_+=\sigma_-^\dagger$) as
\begin{align}
&H_{F_A}=\hbar \omega_0 a^{\dagger} a\\
&H^{\text{int}}_{SF_A}= 
\frac{\hbar \Omega_0}{2} \left( e^{i \phi} \, \sigma_+ \otimes a + e^{-i\phi} \, \sigma_- \otimes a^{\dagger} \right)
\label{HJC_arbitraire}
\end{align}
respectively, where $\Omega_0$ is the vacuum Rabi frequency and $\phi$ is a phase that will determine the axis of rotation of the qubit (see below).
Since the electromagnetic field is resonant with the system qubit, it shares the same frequency $\omega_0$.
In the interaction picture with respect to $H_S+H_{F_A}$, after having performed the rotating wave approximation~\cite{haroche2006exploring}, the effective Hamiltonian is simply that of the interaction, i.e., $H=H^{\text{int}}_{SF_A}$, with $H^{\text{int}}_{SF_A}$ as defined in Eq.~\eqref{HJC_arbitraire}.

We take the auxiliary system of the box in this model to be an electromagnetic field initialized in a coherent state $\ket{F_A^\text{i}} = \ket{\alpha}= e^{-\frac{|\alpha|^2}{2}} \sum_{n=0}^{+\infty} \frac{\alpha^n}{\sqrt{n!}} \ket{n}$, where $\ket{n}$ are the Fock states (and where we take $\alpha>0$ without loss of generality). 
The average number of photons inside the coherent field is $\bar{n}=|\alpha|^2$, the energy in the field simply being $E=\hbar \omega_0 \bar{n}$. The energy contained in the field defines the energy of the box, which we take to quantify the energy invested to realise the corresponding operation; and the energy invested in a process using more than one box will simply be the sum of the energies of each box.

As we show explicitly in Appendix~\ref{app:jaynes_cumming}, for a time of interaction $t=\theta/(\Omega_0 \sqrt{\bar{n}})$ and in the infinite-energy limit $\bar{n} \to +\infty$, such a Jaynes-Cummings interaction with such an initial coherent state for the field allow one to implement a perfect rotation of angle $\theta$ around the axis $\vec{\mathbf{u}}=(\cos\phi,\sin\phi,0)_{x,y,z}$ in the Bloch sphere. In the finite-energy regime, this same choice of interaction time only gives an approximation---i.e., a noisy implementation---of the same rotation; see also Appendix~\ref{app:Kraus_ops_asympt}.
We note that  in such a model not all rotations can thus be approximated by means of this interaction, but only rotations around an axis in the equatorial plane of the Bloch sphere. This restriction is due \emph{(i)} to the model of interaction we choose and \emph{(ii)} to the initial field state, and this scheme is standard in light-matter experiments in cavity~\cite{haroche2006exploring,AspectandFabrebook}. 
This restriction will also motivate our choice of operations $U_A$ that we will consider when benchmarking the QS and its 4B simulation in Sec.~\ref{sec:benchmarking} below.

\subsection{The quantum switch and its simulated implementation}
\label{sec:switch}

As described in the introduction, the QS induces, in the ideal unitary case, the dynamics of Eq.~\eqref{eq:def_supermap_switch} on a qubit ``control'' system $C$ and a ``target'' system $S$.
The situation with noisy operations ${\cal A}$ and ${\cal B}$, implemented as boxes as per Sec.~\ref{sec:box_def}, is however more subtle, and different implementations or simulations of the QS will lead to different output states.
To be able to distinguish the different cases, we shall define the evolution induced on the Hilbert spaces containing the control, target and auxiliary systems. The analysis here applies to the general box model of Sec.~\ref{sec:box_def}, and we will comment on the specific energetic model described in Sec.~\ref{sec:energetics} at the end of this section.

A first natural possibility is to consider using a unique box each for ${\cal A}$ and ${\cal B}$, and hence having one auxiliary system for the box ${\cal A}$ (living in the space $F_A$) and one auxiliary system for the box ${\cal B}$ (living in the space $F_B$), as sketched in Fig.~\ref{fig:QS_4B}(b).
In such an implementation the (unitary) evolution induced by the QS on $C, S, F_A$ and $F_B$ becomes, following Eq.~\eqref{eq:def_supermap_switch},
\begin{align}
    U^\text{QS}(U_{A,F_A},U_{B,F_B})=&\ketbra{0_C}{0_C} \otimes U_{B,F_B} U_{A,F_A} \notag\\
    &+ \ketbra{1_C}{1_C} \otimes U_{A,F_A}U_{B,F_B}  ,
    \label{eq:W_QS}
\end{align}
where $U_{A,F_A}$ (resp.\ $U_{B,F_B}$) makes $S$ and $F_A$ (resp.\ $F_B$) interact through the model of interaction considered above (and acts with the identity on any other system, which for simplicity we do not write explicitly).
The final state of $C$, $S$ and the auxiliary systems, recalling that for simplicity we take $C$ to start in the state $\ket{+_C}$, is then
\begin{equation}
    \ket{\psi^{\text{f,QS}}_{CSF_AF_B}}= \frac{1}{\sqrt{2}} \left( \ket{0_C} \ket*{\Psi_0^{\text{QS}}} + \ket{1_C}\ket*{\Psi_1^{\text{QS}}} \right),
    \label{eq:QS_fieldlevel_ABimperfect}
\end{equation}
with
\begin{align}
    &\ket*{\Psi_0^{\text{QS}}}=U_{B,F_B} U_{A,F_A} \ket*{\psi_S}\ket*{F^\text{i}_{A}}\ket*{F^\text{i}_{B}}, \label{psi_QS_0}\\
    &\ket*{\Psi_1^{\text{QS}}}=U_{A,F_A} U_{B,F_B} \ket*{\psi_S}\ket*{F^\text{i}_{A}}\ket*{F^\text{i}_{B}} \label{psi_QS_1}
\end{align}
(and with implicit tensor products).
By tracing over the auxiliary degrees of freedom, one notices that the final density matrix obtained on $CS$ corresponds to the one obtained from the general definition of the QS process for arbitrary CPTP operations ${\cal A}$ and ${\cal B}$~\cite{chiribella_quantum_2013}. 
Hence, our vision of using one box for ${\cal A}$ and one box for ${\cal B}$ is consistent with the standard definition of the quantum switch for noisy operations.

A second possibility is to try and implement the evolution of Eq.~\eqref{eq:def_supermap_switch} with what we call the ``four box'' (4B) setup, which would require two boxes for ${\cal A}$ and two boxes for ${\cal B}$, and hence four auxiliary systems living in the spaces $F_{A_0},F_{A_1},F_{B_0},F_{B_1}$: see Fig.~\ref{fig:QS_4B}(c). In this case, the induced (unitary) evolution of $C,S$ and all $F_X$'s is
\begin{align}
    U^\text{4B}(U_{A,F_A},U_{B,F_B})=&\ketbra{0_C}{0_C}\otimes U_{B,F_{B_0}} U_{A,F_{A_0}} \notag\\
    &+ \ketbra{1_C}{1_C}\otimes U_{A,F_{A_1}}U_{B,F_{B_1}} .
    \label{eq:W_4B}
\end{align}
Here again, the $U_{A,F_{A_k}}$ ($k \in \{0,1\}$) only make $S$ and $F_{A_k}$ interact, leaving any other degree of freedom intact, and we assume that $U_{A,F_{A_1}}$ and $U_{A,F_{A_2}}$ have identical actions on their respective spaces (with analogous behaviour for the $U_{B,F_{B_k}}$).
For a control initially in $\ket{+_C}$, the final state of $C$, $S$ and the auxiliary systems is then
\begin{equation}
    \ket{\psi^{\text{f,4B}}_{CSF_{A_0}F_{B_0}F_{A_1}F_{B_1}}}= \frac{1}{\sqrt{2}} \left( \ket{0_C} \ket{\Psi_0^{\text{4B}}} + \ket{1_C}\ket{\Psi_1^{\text{4B}}} \right)
    \label{eq:4B_fieldlevel_ABimperfect}
\end{equation}
with
\begin{align}
    &\ket{\Psi_0^{\text{4B}}}=U_{B,F_{B_0}} U_{A,F_{A_0}} \ket{\psi_S}\ket*{F^\text{i}_{A_0}}\ket*{F^\text{i}_{B_0}}\ket*{F^\text{i}_{A_1}}\ket*{F^\text{i}_{B_1}}, \label{psi_4B_0}\\
    &\ket{\Psi_1^{\text{4B}}}=U_{A,F_{A_1}} U_{B,F_{B_1}} \ket{\psi_S}\ket*{F^\text{i}_{A_0}}\ket*{F^\text{i}_{B_0}}\ket*{F^\text{i}_{A_1}}\ket*{F^\text{i}_{B_1}}.\label{psi_4B_1}
\end{align}

We thus observe a mathematical difference between the QS and the 4B, even when the implementations of the operations are perfect as one can see from Eqs.~\eqref{eq:W_QS} and~\eqref{eq:W_4B}. Formally, when including the description of the auxiliary systems, the mathematical structure of the evolution induced by the QS consists in taking the ideal evolution (when we can ignore the auxiliary systems, as in Eq.~\eqref{eq:def_supermap_switch}), and performing for this case the mapping $U_A \to U_{A,F_A}$, $U_B \to U_{B,F_B}$, $\ket{\psi_S} \to \ket{\psi_S}\ket*{F^\text{i}_{A}}\ket*{F^\text{i}_{B}}$. For the 4B, as two different boxes are used for the two implementations of both ${\cal A}$ and ${\cal B}$, there is no such mapping. For instance, $U_A$ could be mapped to either $U_{A,F_{A_0}}$ or $U_{A,F_{A_1}}$. This remark highlights the fact that it is important to consider the way the operations are implemented to distinguish the QS from circuits that are simulating it, as our practical definition of a box allows us to do.
Note finally that, in a noiseless scenario (i.e., in the infinite energy regime, when the auxiliary systems never become entangled with $S$), tracing out the auxiliary systems would give the same final state, and thus effective evolution, for both the 4B and the QS, thereby rendering the two setups indistinguishable.

In order to study the energetics of the QS and the 4B, we will naturally adopt the specific box model described in Sec.~\ref{sec:energetics}. There, the target is hence a qubit system corresponding to an atom flying through the setups of Fig.~\ref{fig:QS_4B}, while the control system $C$ is encoded in the spatial degree of freedom of the atom.
An explicit possible implementation for both setups is outlined in more detail in Appendix~\ref{app:experimental}.

\section{Benchmarking via a discrimination task}
\label{sec:benchmarking}

As we saw above, the QS and 4B do not implement the same evolution in the noisy case.
This motivates the main question we address in this work: 
is there an advantage in using one setup rather than the other when the energy we can invest is limited? For a fixed amount of energy to implement a given operation, should one concentrate all of it in one box and exploit causal indefiniteness (as in the QS), or do we get the same by distributing it in more boxes and only simulating the quantum switch (as in the 4B)?

Here, we use the concrete energetic model we introduced to study this question.
To this end, we will benchmark the performance of the QS and the 4B in a concrete scenario---namely, at performing a commuting-vs-anticommuting discrimination task along the lines of Ref.~\cite{chiribella12}---when the available energy in the fields is constrained.
As a baseline performance indicator, we will also compare these processes to the standard model of quantum circuits with fixed causal order (FCO), where one looks for the optimal quantum circuit to solve the same task~\cite{Chiribella08}.

\subsection{The commuting-vs-anticommuting discrimination task}
\label{sec:comp_task}

Reference~\cite{chiribella12} introduced the following unitary discrimination task, for which an advantage of the QS was found over all FCO quantum circuits.
Assume one is given two black boxes implementing some unitary operations $U_A$ and $U_B$, with the promise that they either commute or anticommute.
The goal is to determine which of these two possibilities is the case.
The probability that a given strategy or process does so defines its \textit{success probability}.

Note that one only has ``black-box'' access to $U_A$ and $U_B$, meaning that one has no classical description of them.
Here we will adapt the task as follows: instead of having access to a black box implementing $U_A$ perfectly, we assume one is given access to a noisy implementation of $U_A$, implemented as boxes in the finite-energy regime following the model described in Sec.~\ref{sec:energetics}. For simplicity we still assume, however, that $U_B$ can be implemented perfectly (not through a Jaynes-Cummings interaction). One is again asked to determine whether $U_A$ (the ideal unitary that the black-box is meant to approximate) and $U_B$ commute or anticommute.

For concreteness, it is necessary to fix the sets $\mathcal{S}_{[,]}$ and $\mathcal{S}_{\{,\}}$ of commuting and anticommuting unitaries that the pairs $(U_A,U_B)$ are drawn from, and how they are sampled: different sets, with different distributions of the unitaries, will indeed give different success probabilities for our task.
Since we consider a two-level target system, the unitaries $U_A$ and $U_B$ are rotations on the Bloch sphere;
we shall generically denote by $R_{\vec{\mathbf{u}}}(\theta)$ a rotation by an angle $\phi$ around an axis specified by the unit vector $\vec{\mathbf{u}}$ (irrespectively of the global phase that it may introduce, which is irrelevant in this paper). One can verify that two such rotations with nontrivial rotation angles (i.e., with $\theta \neq 0 \mod 2\pi$) commute if and only if their rotation axes are colinear, and they anticommute if and only if their rotation axes are orthogonal and the rotation angles are $\pi$.
To comply with the physical model we introduced for the implementation of $U_A$, we will take its rotation axis in the equatorial plane $(Oxy)$ of the Bloch sphere (see Sec.~\ref{sec:energetics}), but we again do not impose any restriction on $U_B$. This leads us to define the sets $\mathcal{S}_{[,]}$ and $\mathcal{S}_{\{,\}}$ of pairs $(U_A,U_B)$ as follows:
\begin{align}
    \mathcal{S}_{[,]} & =\{ ( R_{\vec{\mathbf{u}}}(\theta_A), R_{\vec{\mathbf{u}}}(\theta_B) ), \vec{\mathbf{u}} \in (Oxy)\},
    \label{eq:set_com} \\
    \mathcal{S}_{\{ ,\}} & =\{(R_{\vec{\mathbf{u}}_A}(\pi), R_{\vec{\mathbf{u}}_B}(\pi)), \vec{\mathbf{u}}_A  \in (Oxy), \vec{\mathbf{u}}_B \perp \vec{\mathbf{u}}_A \}.
    \label{eq:set_anticom}
\end{align}
In our task we will take the pairs $(U_A,U_B)$ to be drawn from either $\mathcal{S}_{[,]}$ or $\mathcal{S}_{\{ ,\}}$ with equal probability. Furthermore, we take the axes $\vec{\mathbf{u}}$ and $\vec{\mathbf{u}}_A$ above to be uniformly distributed in $(Oxy)$, $\vec{\mathbf{u}}_B$ to be uniformly distributed in the plane orthogonal to $\vec{\mathbf{u}}_A$, and the rotation angles $\theta_A,\theta_B$ to be uniformly distributed in $[-\pi,\pi]$. (Note that these sets of commuting or anticommuting unitaries differ from those considered in Ref.~\cite{araujo_witnessing_2015}, which did not single out any particular orientation in the Bloch sphere.)

With this in place, we can now benchmark how the QS and 4B perform on this commuting-vs-anticommuting discrimination task by looking at how the probability of successfully guessing from which set, $\mathcal{S}_{[,]}$ or $\mathcal{S}_{\{ ,\}}$, $U_A$ and $U_B$ are drawn from depends on the energy available to approximate $U_A$ within the box model considered.

\subsection{Formal description of the benchmark protocol}
\label{sec:finite_energy_comparison}

\subsubsection{The QS and 4B strategies to perform the task}

In the case of ideally implemented unitaries $U_A, U_B$, the QS and 4B can perform the task perfectly~\cite{chiribella12}. To see this, note that the state of Eq.~\eqref{eq:ideal_evol}
can be rewritten in the $\{\ket{+_C},\ket{-_C}\}$ basis for the control as
\begin{align}
    \ket*{\psi^\text{f}_{CS}}= \frac{1}{2} \left( \ket{+_C} \{U_A,U_B\} \ket{\psi_S} - \ket{-_C}[U_A,U_B] \ket{\psi_S}\right),
    \label{eq:QS_perfect_game}
\end{align}
where $\{\cdot,\cdot\}$ and $[\cdot,\cdot]$ are the anticommutator and commutator, respectively. Recall that one has the promise that one of the two is null; it then suffices to measure the control qubit in the $\{\ket{\pm_C}\}$ basis to see which term survives in Eq.~\eqref{eq:QS_perfect_game}, and thereby determine perfectly (i.e., with probability 1) whether the unitaries commute or anticommute. Note that this strategy works for any distribution of the unitaries, and any initial state of the target system $\ket{\psi_S}$.

In the finite-energy regime, we shall consider the same strategy for the QS and the 4B: simply measure the control qubit at the output in the $\{\ket{\pm_C}\}$ basis. If the result is $\ket{+_C}$, resp.\ $\ket{-_C}$, we shall make the guess that $U_A$ and $U_B$ commute, resp.\ anticommute. Here, the noise induced by the entanglement between the target qubit and the quantum field will make the success probability lower than 1. In the situation we consider, where the operation $U_B$ is ideal but the implementation of $U_A$ is not, the QS uses one quantum field as an auxiliary system to approximate $U_A$, while the 4B uses two, so one may expect their probabilities of correctly guessing the commuting-vs-anticommuting property to also be different. The question of the energetic costs associated to the QS and the 4B to solve this task is then legitimate to ask, so as to see what the most efficient way to invest the energy is. We will thus compare the two processes under the constraint that they use the same total amount of energy in the boxes implementing $U_A$: an average of $\bar{n}$ photons in the single box of the QS, and $\bar{n}/2$ in each of the corresponding two boxes of the 4B (for simplicity we take the natural choice of sharing the energy evenly between the boxes). 

\subsubsection{Success probabilities of the QS and 4B}
\label{sec:proba_success_QS_4B}

For $(U_A,U_B) \in \mathcal{S}_{[,]}$, the task is thus successfully completed if the outcome of the measurement on the control qubit is $\ket{+_C}$; for $(U_A,U_B) \in \mathcal{S}_{\{,\}}$ on the other hand, it is successfully completed if one gets the result $\ket{-_C}$. From Eqs.~\eqref{eq:QS_fieldlevel_ABimperfect} and~\eqref{eq:4B_fieldlevel_ABimperfect}, one easily finds that the probabilities for these events, for both the QS and 4B---denoted below by $\epsilon \in \{\text{QS},\text{4B}\}$---are
\begin{align}
&p_{\text{success}}^{\left[ ,\right],\epsilon }(U_A,U_B)=\frac{1+\text{Re}\left[ \braket{\Phi^{\epsilon}_0 (U_A,U_B)}{\Phi^{\epsilon}_1 (U_A,U_B)}  \right]}{2}, \label{eq:success_com}\\
&p_{\text{success}}^{\left\{ ,\right\},\epsilon }(U_A,U_B)=\frac{1-\text{Re}\left[ \braket{\Phi^{\epsilon}_0(U_A,U_B)}{\Phi^{\epsilon}_1(U_A,U_B)} \right]}{2}\label{eq:success_anticom},
\end{align}
where we made the dependency on $(U_A,U_B)$ explicit, and with
\begin{align}
    \ket*{\Phi^{\text{QS}}_0(U_A,U_B)}&=U_{B} U_{A,F_{A}} \ket{\psi_S}\ket{\alpha_{A}}, \label{eq:Phi0QS}\\
    \ket*{\Phi^{\text{QS}}_1(U_A,U_B)}&=U_{A,F_{A}} U_{B} \ket{\psi_S}\ket{\alpha_{A}}, \label{eq:Phi1QS}\\
    \ket*{\Phi^{\text{4B}}_0(U_A,U_B)}&=U_{B}U_{A,F_{A_0}} \ket{\psi_S}\ket{\alpha_{A_0}}\ket{\alpha_{A_1}}, \label{eq:Phi04B}\\
    \ket*{\Phi^{\text{4B}}_1(U_A,U_B)}&=U_{A,F_{A_1}}U_{B} \ket{\psi_S}\ket{\alpha_{A_0}}\ket{\alpha_{A_1}}, \label{eq:Phi14B}
\end{align}
in accordance with Eqs.~\eqref{psi_QS_0}--\eqref{psi_QS_1} and~\eqref{psi_4B_0}--\eqref{psi_4B_1}, recalling that $U_A$ is approximated through the application of $U_{A,F_{A_{(k)}}}$ involving some auxiliary systems initialized in the states $\ket*{F^\text{i}_{A_{(k)}}} = \ket{\alpha_{A_{(k)}}}$ (cf.\ Fig.~\ref{fig:QS_4B}) and that $U_B$ is assumed to be implemented perfectly (without involving any auxiliary system). As mentioned before, we shall ensure that the QS and 4B are provided with the same total amount of energy $\hbar \omega_0 \bar{n}$, so that we will take $\alpha_A = \sqrt{\bar{n}}$ and $\alpha_{A_0} = \alpha_{A_1} = \sqrt{\bar{n}/2}$.

Averaging over the choice of unitaries $(U_A,U_B)$, either from $\mathcal{S}_{[,]}$ or from $\mathcal{S}_{\{,\}}$ with probability $\frac12$, we thus find that the average success probabilities in the commuting-vs-anticommuting discrimation task are
\begin{align}
    \langle p_{\text{success}}^{\epsilon}\rangle=&\frac{1}{2}\int_{\mathcal{S}_{[,]}} \textup{d}\mu_{\left[,\right]}({U_A,U_B}) \, p_{\text{success}}^{\epsilon,\left[,\right]}({U_A,U_B}) \notag \\
    & +\frac{1}{2}\int_{\mathcal{S}_{\{,\}}} \!\textup{d}\mu_{\left\{ ,\right\} }({U_A,U_B}) \, p_{\text{success}}^{\epsilon,\left\{ ,\right\} }({U_A,U_B}),
    \label{eq:p_succ_avg}
\end{align}
where $\textup{d}\mu_{[,]}$ and $\textup{d}\mu_{\{,\}}$ are the measures on the sets $\mathcal{S}_{[,]}$ and $\mathcal{S}_{\{,\}}$ that correspond to the way we chose to sample $(U_A,U_B)$ as described following Eqs.~\eqref{eq:set_com} and \eqref{eq:set_anticom}.

Note that the success probabilities in Eqs.~\eqref{eq:success_com}--\eqref{eq:success_anticom} and~\eqref{eq:p_succ_avg} depend on the initial state $\ket{\psi_S}$ of the target system. In fact, expanding these equations (using Eqs.~\eqref{eq:Phi0QS}--\eqref{eq:Phi14B}) it is easily seen that these are linear in $\ketbra{\psi_S}{\psi_S}$. Hence, from these calculations we can also directly obtain the success probabilities for any mixed initial state $\rho_S$, after replacing $\ketbra{\psi_S}{\psi_S}$ by $\rho_S$.
For concreteness, and in order not to favour any specific orientation of the Bloch sphere, in the following we will take $\rho_S = \id/2$; see however Appendix~\ref{app:success_probability_derivation} for calculations that apply to any $\rho_S$.

From here we can now compute and compare the success probabilities for the QS and 4B, for any fixed value of $\bar{n}$. This will allow us to see which of the two setups performs the best at fixed energy---or reciprocally, for a fixed success probability that one targets, which process uses the least amount of energy, i.e., is the most energy efficient. To gain more insights about the performances of the two setups and relate them to what is more commonly done in the literature, we shall also compare these to general circuits with fixed causal order that we now introduce.

\subsection{Circuits with fixed causal order}
\label{sec:part_comb}

The QS is a circuit with no definite causal order, using each of the operations ${\cal A}$ and ${\cal B}$ once and only once. It is indeed generally contrasted in the literature with quantum circuits that apply ${\cal A}$ and ${\cal B}$ (once each) in a fixed causal order (FCO): either with ${\cal A}$ before ${\cal B}$, or with ${\cal B}$ before ${\cal A}$.

\begin{figure}
\centering   
\includegraphics[width=80mm]{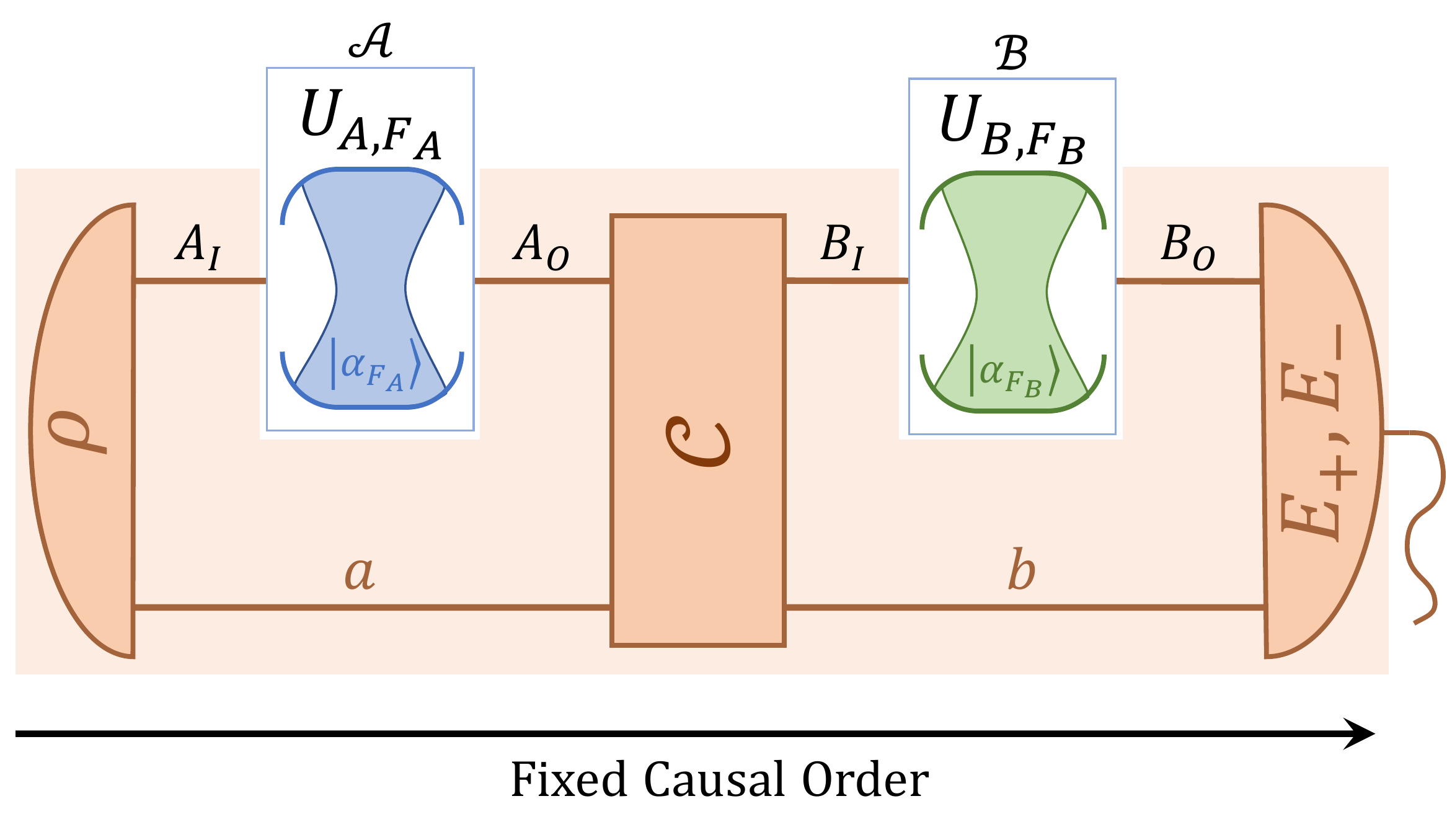}
\caption{A general circuit with FCO where ${\cal A}$ is applied before ${\cal B}$. The target qubit and (possibly) some entangled ``memory'' system $a$ are initially prepared in some state $\rho$. The target system then undergoes operation ${\cal A}$, after which the target and memory systems evolve through some channel $\mathcal{C}$. Operation ${\cal B}$ is then applied to the target system, and a measurement is finally performed on the target and new memory system $b$, described by a POVM $(E_+,E_-)$. In our unitary discrimination task, the result of this POVM is taken as a guess for whether $U_A$ and $U_B$ (the unitaries that the operations ${\cal A}$ and ${\cal B}$ implement or approximate) commute or anticommute.}
\label{comb}
\end{figure}

Such circuits take the general form of Fig.~\ref{comb}, with the operation ${\cal A}$ and ${\cal B}$ being inserted between an initial state preparation, a transformation and a final measurement.
Using the framework of quantum circuits~\cite{Chiribella08,Chiribella09}, it is however not necessary to look into the internal details of these circuit operations. One can instead encode all the circuit elements other than ${\cal A}$ and ${\cal B}$ in a single mathematical object, a so-called ``quantum tester'': namely, a pair of positive semidefinite operators $(W_+,W_-)$ acting on the input and output spaces of the operations ${\cal A}$ and ${\cal B}$ (denoted $A_I, A_O, B_I, B_O$ in Fig.~\ref{comb}).
Each operator $W_\pm$ corresponds to a possible outcome (a measurement result  `$+$' or `$-$') of the circuit; its probability is obtained via the generalized Born rule:
\begin{align}
    p(\pm|{\cal A},{\cal B}) = \Tr\big[W_\pm^T(\underline{A}\otimes\underline{B})\big], \label{eq:probas_FCO}
\end{align}
where $\underline{A}$ and $\underline{B}$ are the Choi matrices of the operations ${\cal A}$ and ${\cal B}$ (as defined in Eq.~\eqref{eq:def_Choi} in Appendix~\ref{app:asymptotic_behaviours}), and $T$ denotes the transpose.
Hence, $(W_+,W_-)$ can be seen as a generalization for quantum processes of positive-operator valued measures (POVMs, i.e., general quantum measurements on quantum states).
The fact that ${\cal A}$ and ${\cal B}$ are applied in a fixed order in the circuit imposes some specific constraints on $(W_+,W_-)$; the details are provided in Appendix~\ref{app:proba_circuit}. 

In our task, associating the measurement outcome `$+$', resp.~`$-$', with the guess that $U_A$ and $U_B$---the unitaries that ${\cal A}$ and ${\cal B}$ are meant to implement---commute, resp.~anticommute, one can obtain the success probability of a circuit with FCO at performing the task by integrating over the commuting and anticommuting sets $\mathcal{S}_{[,]}$ and $\mathcal{S}_{\{,\}}$, similarly to Eq.~\eqref{eq:p_succ_avg}:
\begin{align}
    \langle p_{\text{success}}^{\text{FCO}} \rangle=&\frac{1}{2}\int_{\mathcal{S}_{[,]}} \textup{d}\mu_{\left[,\right]}({U_A,U_B}) \, p(+|{\cal A},{\cal B}) \notag \\
    & +\frac{1}{2}\int_{\mathcal{S}_{\{,\}}} \textup{d}\mu_{\left\{ ,\right\} }({U_A,U_B}) \, p(-|{\cal A},{\cal B}) \notag \\
    = & \frac{1}{2} \Tr[W_+^T G_+] + \frac{1}{2} \Tr[W_-^T G_-] \label{eq:Psucc_FCO}
\end{align}
with
\begin{align}
    G_+ = & \int_{\mathcal{S}_{[,]}} \textup{d}\mu_{\left[,\right]}(U_A,U_B) \, (\underline{A}\otimes\underline{B}), \label{eq:def_Gp} \\
    G_- = & \int_{\mathcal{S}_{[,]}} \textup{d}\mu_{\{,\}}(U_A,U_B) \, (\underline{A}\otimes\underline{B}). \label{eq:def_Gm}
\end{align}
To find the circuit with a given FCO that performs the best at the task under consideration, one can directly maximize the probability of success in Eq.~\eqref{eq:Psucc_FCO} over all pairs $(W_+,W_-)$ satisfying the required constraints; see Appendix~\ref{app:subsec:psucc_FCO}. 

As it turns out, by solving this optimization problem we find that one can obtain a success probability of 1 when using a FCO circuit with ${\cal B}$ before ${\cal A}$ in the ideal, unitary case.
Indeed there exists a quantum circuit with FCO that allows one to perfectly discriminate between the sets of commuting or anticommuting operations considered here; see Appendix~\ref{combBA}.
It is known, however, that for some other sets $\mathcal{S}_{[,]},\mathcal{S}_{\{,\}}$ the probability of success of the best circuit with FCO is strictly lower than 1 (in contrast with the QS)~\cite{chiribella12,araujo_witnessing_2015}. The reason we could obtain a probability 1 for FCO circuits here is that the sets $\mathcal{S}_{[,]}$ and $\mathcal{S}_{\{,\}}$ we consider are too restrictive---restricted in particular to unitaries $U_A$ with a rotation axis in the equatorial plane defined by the orientation of the $z$-axis. The FCO strategy giving a success probability of 1 for the sets we use is optimized for this particular orientation: it would generally give a strictly lower success probability for unitaries with different rotation axes, i.e., for other more general sets of commuting or anticommuting operations.

The strategies of the QS or 4B, on the other hand, are oblivious to the specific choice of operations $(U_A,U_B)$, and therefore of the orientation used to define the sets $\mathcal{S}_{[,]}$ and $\mathcal{S}_{\{,\}}$. 
For a perhaps fairer comparison, and to obtain some more insight, we will look below at how FCO circuits perform when we moreover require that they should not single out any particular orientation.
Instead, we require that (just as for the QS and the 4B) they act in the same way---i.e., give the same statistics---for all orientations of the axis used to define the equatorial plane and the sets $\mathcal{S}_{[,]}$ and $\mathcal{S}_{\{,\}}$, making them ``isotropic". The restricted subset of such isotropic FCO circuits is formally defined in Appendix~\ref{app:isoFCO}. 

\section{Results}
\label{sec:results}

\subsection{Comparing success probabilities across strategies}

In this part, we study quantitatively the energetic differences between using the QS or the 4B. We find that, in the commuting-vs-anticommuting discrimination task we considered and within our specific energetic model, the QS is more efficient than the 4B for a fixed energy limitation. We then compare the QS and the 4B to the ensemble of all isotropic circuits with FCO. The energetic constraint is quantified by the average total number of photons $\bar{n}$ that each setup is given. We present numerical results for finite $\bar{n}$ and analytical results in the limit where $\bar{n}$ is large enough.

\begin{figure}
\centering   
\includegraphics[width=80mm]{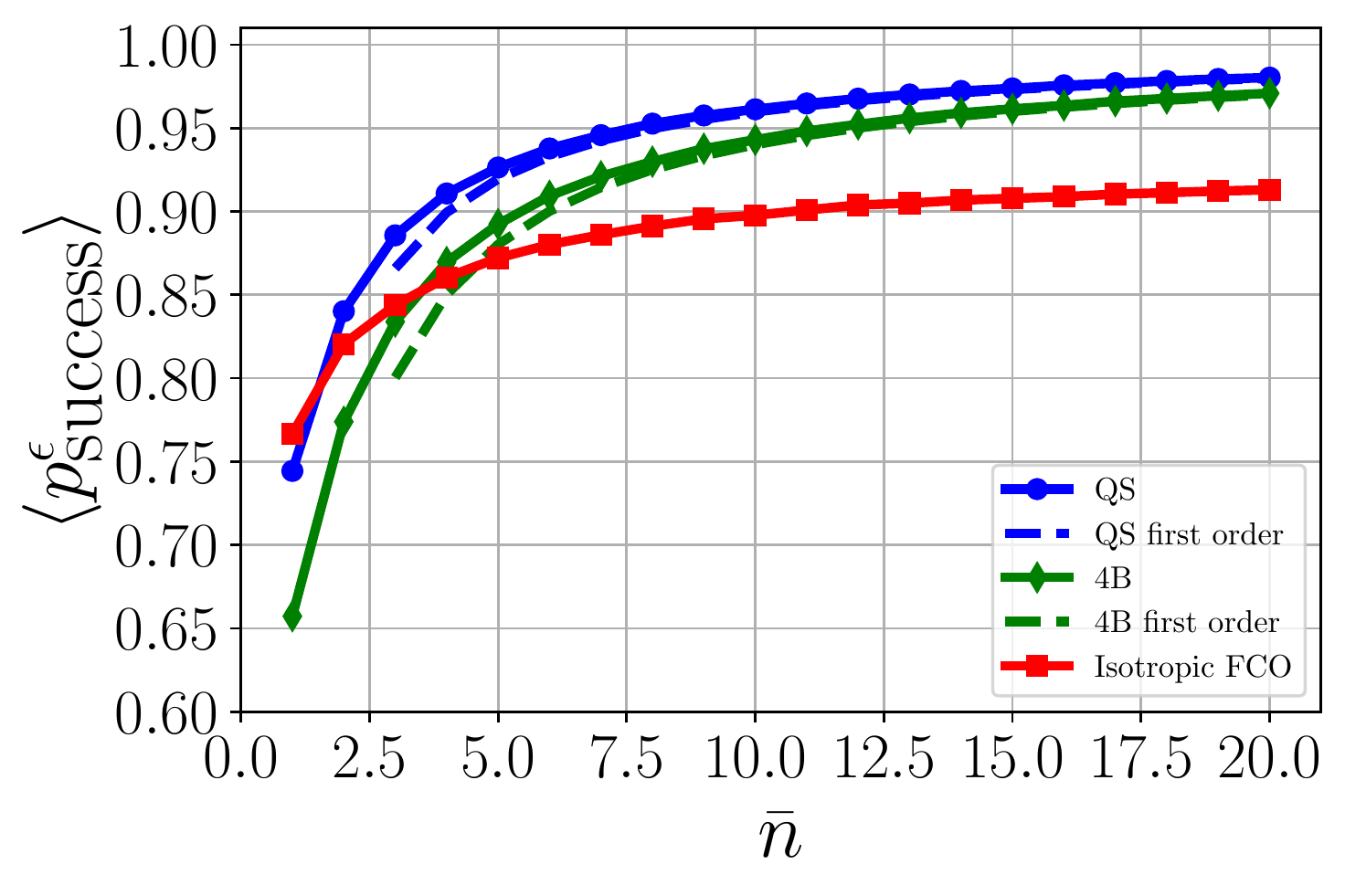}
\caption{Average success probabilities for our commuting-vs-anticommuting discrimination task of the QS, the 4B and the best isotropic FCO circuit, for $1 \le \bar{n}\le 20$. The connected points are obtained numerically for different values of $\bar{n}$, while the dashed lines (for the QS and 4B) are the analytical first-order expansions in $1/\bar{n}$, see Eqs.~\eqref{eq:psucc_QS}--\eqref{eq:psucc_4B}.
}
\label{avg}
\end{figure}

In Fig.~\ref{avg}, the average success probabilities for the QS and the 4B are shown for $1 \le \bar{n}\le 20$. As expected, the success probabilities of the QS and 4B tend to 1 for large $\bar{n}$. All success probabilities decrease when the amount of energy (i.e., $\bar{n}$) decreases, as the implementation of the operation ${\cal A}$ becomes noisier.
For a finite number of photons, we observe a clear advantage of the QS over the 4B, with its average success probability being above that of the 4B for any given value of $\bar{n}$. We also observe that for $\bar{n}$ large enough, the average success probabilities of the QS and the 4B are above those of the best isotropic FCO circuits, which are optimized (and in general different) for each value of $\bar{n}$, and whose limit for $\bar{n}\rightarrow\infty$ is only $\simeq 0.93$ (see Appendix~\ref{app:isoFCO}). In the lower $\bar{n}$ regime, we observe two crossings such that both the QS and the 4B are outperformed by the optimized isotropic FCO circuits. 
This can be understood by recalling that these FCO circuits are optimized for each $\bar{n}$, whereas the QS and the 4B use a fixed strategy that becomes very poor for low $\bar{n}$. (For $\bar{n}$ lower than the values shown in Fig.~\ref{avg}, the approximation of $U_A$ then becomes too bad for the QS and 4B strategies to be judicious, and for their analyses to be relevant.)

In Appendix~\ref{app:success_probability_derivation}, we also perform perturbative expansions of the behaviour of the success probability for the QS and the 4B in the large $\bar{n}$ limit. For an initial target system in the state $\rho_S = \id/2$, we found:
\begin{align}
   &\langle p_{\text{success}}^{\text{QS}} \rangle  = 1-\frac{3+\pi^2}{32\bar{n}}+O\left( \frac{1}{\bar{n}^2} \right), \label{eq:psucc_QS} \\
   &\langle p_{\text{success}}^{\text{4B}} \rangle  = 1-\frac{6+4\pi^2/3}{32 \bar{n}}+O\left( \frac{1}{\bar{n}^2} \right),\label{eq:psucc_4B} \\
   &\langle p_{\text{success}}^{\text{QS}} \rangle - \langle p_{\text{success}}^{\text{4B}} \rangle  = \frac{3+\pi^2/3}{32 \bar{n}}+O\left( \frac{1}{\bar{n}^2} \right),
\end{align}
which indeed shows formally that in this limit, the QS outperforms the 4B. The success probabilities reach 1 at order 0 in $1/\bar{n}$, in agreement with Fig.~\ref{avg}, and with the fact that the QS and 4B always succeed in the task when $U_A$ and $U_B$ are commuting or anticommuting unitaries \cite{chiribella_quantum_2013,araujo_witnessing_2015}. Graphically, the first-order approximation agrees well for $\bar{n}\gtrsim 8$ (see Fig.~\ref{avg}).

\subsection{Connecting the success probability with entanglement}

A natural question raised by these results is why the QS outperforms the 4B in the presence of noise in our benchmark task.
We can gain some insight into this question by looking more closely at how and why the success probabilities decrease in both setups.
In particular, we will see here that the success probabilities are directly related to a more fundamental quantity, the entanglement entropy between the control and the other systems.

The entanglement entropy between the control $C$ and the other systems ($SF_{A}F_{B}$ for the QS and $SF_{A_{0}}F_{A_{1}}F_{B_{0}}F_{B_{1}}$ in the case of the 4B), for any fixed $(U_A,U_B)$, is defined as the von Neumann entropy $S(\rho^\epsilon_C(U_A,U_B))$ (with $\epsilon\in\{\text{QS},\text{4B}\}$)  of the reduced state of the control.
As we saw (see Eqs.~\eqref{eq:success_com} and~\eqref{eq:success_anticom}), for an initial pure state $\ket{\psi_S}$, the success probabilities for the QS and 4B are directly related to $\Re[ \braket{\Phi^{\epsilon}_0 (U_A,U_B)}{\Phi^{\epsilon}_1 (U_A,U_B)}]$.
One can easily show that, in this same case, $S(\rho^\epsilon_C(U_A,U_B))$  is directly related to $| \braket{\Phi^{\epsilon}_0 (U_A,U_B)}{\Phi^{\epsilon}_1 (U_A,U_B)} |$ as
\begin{equation}
    S(\rho^\epsilon_C(U_A,U_B)) = H\left(\tfrac{1+| \braket{\Phi^{\epsilon}_0 (U_A,U_B)}{\Phi^{\epsilon}_1 (U_A,U_B)} |}{2}\right),
\end{equation}
where $H(x)=-x\log(x)-(1-x)\log(1-x)$ is the binary entropy.

For the specific discrimination task we considered, the series expansions given in Appendix~\ref{app:asymptotic_behaviours} readily allow us to see that $\Re[ \braket{\Phi^{\epsilon}_0 (U_A,U_B)}{\Phi^{\epsilon}_1 (U_A,U_B)}  ] = \pm(1 - \frac{c^\epsilon}{\bar{n}} + O\big(\frac{1}{\bar{n}^2})\big)$ for some constants $c^\epsilon$ depending in general on $(U_A,U_B)$. A similar calculation shows that $ \Im[ \braket{\Phi^{\epsilon}_0 (U_A,U_B)}{\Phi^{\epsilon}_1 (U_A,U_B)}  ] = \frac{d^\epsilon}{\bar{n}} + O(\tfrac{1}{\bar{n}^2})$ for some other constants $d^\epsilon$ that again depend in general on $(U_A,U_B)$,
from which we see
\begin{equation}
    | \braket{\Phi^{\epsilon}_0 (U_A,U_B)}{\Phi^{\epsilon}_1 (U_A,U_B)} | = 1 - \frac{c^\epsilon}{\bar{n}} + O\left(\frac{1}{\bar{n}^2}\right),
\end{equation}
for the same constant $c^\epsilon$, independently of $d^\epsilon$.
We thus observe that, at first order in $\frac{1}{\bar{n}}$, there is a direct monotonous connection between the success probability and the entanglement entropy of the control: one has $S(\rho^\epsilon_C(U_A,U_B)) \simeq H(p^{\cdot,\epsilon}_\text{success}(U_A,U_B))$, with $p^{\cdot,\epsilon}_\text{success} = p^{[],\epsilon}_\text{success}$ for each pair $(U_A,U_B)$ in $\mathcal{S}_{[,]}$ and $p^{\cdot,\epsilon}_\text{success} = p^{\{,\},\epsilon}_\text{success}$ for $(U_A,U_B)$ in $\mathcal{S}_{\{,\}}$.

In the large $\bar{n}$ limit (in which the QS and 4B induce the same effective transformation), the control does not become entangled with the other systems. The reduction in success probability for finite $\bar{n}$, i.e., when the operations become noisy, can thus be explained by the loss of information due to the control becoming entangled with the other systems (and, notably, the inaccessible fields).
Moreover, the difference in performance for a given $\bar{n}$ between the QS and the 4B can hence primarily be attributed to the fact that less entanglement is created by the QS (which effectively reduces the length of the Bloch vector of $\rho^\epsilon_C(U_A,U_B)$), rather than, for example, being due to an effective rotation of the Bloch vector meaning a measurement in the $\{\ket{\pm_C}\}$ basis may no longer be optimal. Understanding why less entanglement is created by the QS in this task, and whether this is a general feature beyond what we consider here, is an intriguing open question that may help to further understand the differences between the QS and the 4B.

\section{Conclusions}
\label{sec:conclusions}

The quantum switch (QS), and causal indefiniteness more generally, has attracted significant recent interest as a potential computational resource. This has led to some debate around different experiments striving to implement the quantum switch as to whether they are faithful implementations or just simulations~\cite{maclean17,oreshkov19,paunkovic20,kristjansson20,vilasini22,ormrod22}.
Motivated by these questions, in this paper we introduced a practical definition of an operation as a ``box'' relating it to its physical implementation, and based on which we investigated  some of the physical differences between the quantum switch (the QS) and a natural ``four box'' simulation of it (the 4B).
We employed a novel energetic approach for this comparison, modelling the implemented operations as an atom interacting with a coherent state of light through the Jaynes-Cummings model, and where the noise is a consequence of the limited energy budget in the coherent state. We used this model to study which of these processes is the most energy efficient in performing a specific benchmark task, involving determining whether two operations commute or anti-commute. 

More precisely, by considering a specific set of single-qubit rotations around the equatorial plane as in Eqs.~\eqref{eq:set_com}--\eqref{eq:set_anticom}, and assuming an ideal implementation of the unitary $U_B$, we showed numerically (and analytically in the high-energy limit) that the QS performs better than the 4B for a fixed amount of energy, or equivalently, that it requires less energy to reach a desired success probability for the benchmark task. 
We also showed that the QS and the 4B outperform (except in the very low-energy regime) a natural class of quantum circuits with fixed causal order (FCO) that, like both the QS and the 4B, are ``isotropic'' and thus operate independent to the reference frame used to define the operations of our benchmark task.
We note, incidentally, that this class of isotropic circuits with FCO that we introduced here may be of independent interest beyond the context of this work.

In addition to shedding light on the differences between the QS and its simulations, these results highlight the potential of superpositions of causal orders as energy-efficient quantum processes. 
We provided some initial insight into why this might be the case, showing that the advantage of the QS in our benchmarking task over the 4B is closely related to the amount of entanglement the processes generate for a given energy budget. 
Nonetheless, much work remains to understand the generality of our results and the potential energetic advantages that can be obtained: 
to what extent can they be generalized beyond the specific physical model and task we considered here? Do they still hold if both $U_A$ and $U_B$ are taken to be noisy or if the energy is not required to be shared evenly (i.e., $\bar{n}/2$ photons per cavity) between the two ``copies'' of the operation?
Our work thus motivates a more systematic study of the energetics of the QS, causally indefinite processes, and their simulations.

\begin{acknowledgments}
This work is supported by the Agence Nationale de la Recherche under the programme ``Investissements d'avenir'' (ANR-15-IDEX-02), the “Laboratoire d’Excellence” (Labex) “Laboratoire d’Alliances Nanosciences—Energies du Futur” (LANEF), the Templeton World Charity Foundation, Inc. (grant number TWCF0338), the EU NextGen Funds, the Government of Spain (FIS2020-TRANQI and Severo Ochoa CEX2019-000910-S), Fundació Cellex, Fundació Mir-Puig, Generalitat de Catalunya (CERCA program) and the “Quantum Optical Technologies” project,
carried out within the International Research Agendas programme of the
Foundation for Polish Science co-financed by the European Union under
the European Regional Development Fund. We thank Igor Dotsenko for discussions on the experimental implementation. 

\end{acknowledgments}

\appendix

\section{Experimental implementation under consideration}
\label{app:experimental}

\begin{figure*}
\centering   
\includegraphics[width=180mm]{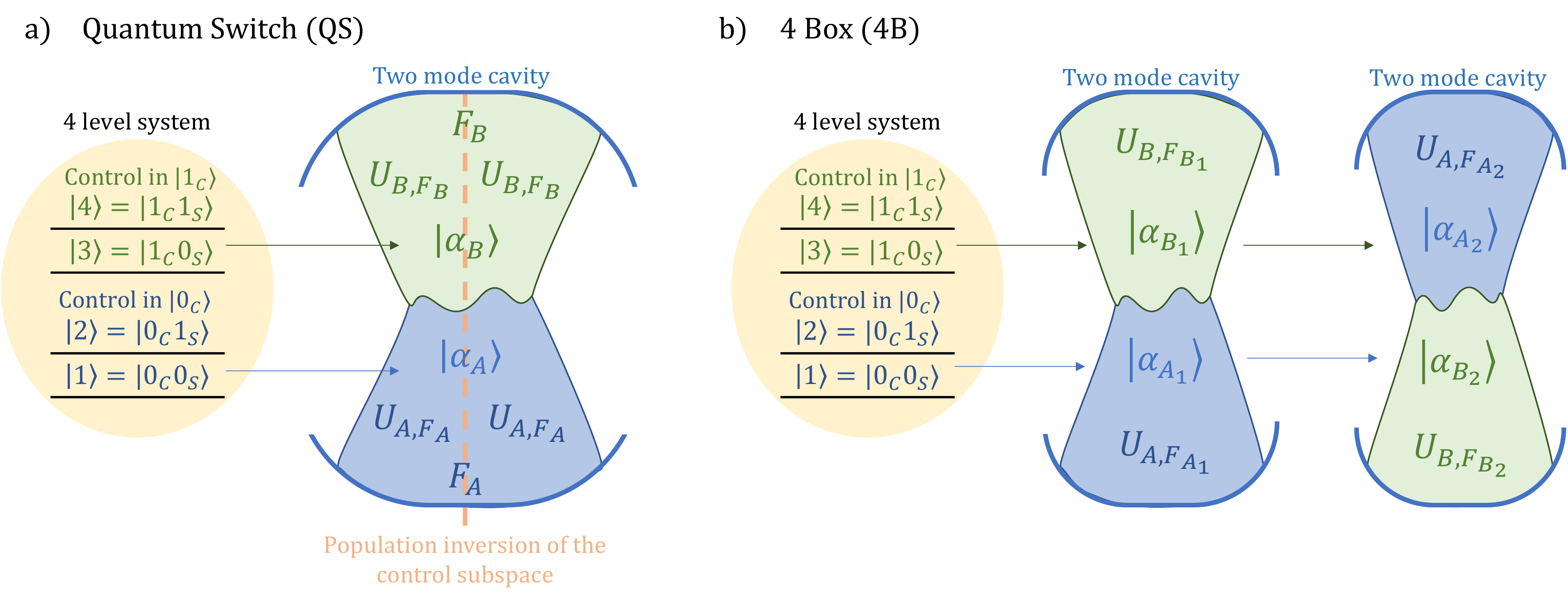}
\caption{Proposed implementation for (a) the QS and (b) the 4B setups with two-mode cavities and a four-level atom. The four-level system
encodes both the target and the control qubits. Each mode of the two-mode
cavity field extends over the whole interior of the cavity; for a
pictorial representation, however, we represent these modes in the
upper and lower parts of the cavity. For each cavity, the field mode
$F_{A}$ couples to the lower atomic subspace $\left\{ \left|1\right\rangle ,\left|2\right\rangle \right\}$ (corresponding to the control state $\ket{0_C}$)
while the field mode $F_{B}$ couples to the upper atomic subspace
$\left\{ \left|3\right\rangle ,\left|4\right\rangle \right\} $ (control state $\ket{1_C}$). The
QS is implemented by the atom traversing a single cavity. The population inversion in the middle enables the two operations $\mathcal{A}$ and $\mathcal{B}$ (or the ``boxes'' $\{U_{A,F_A},\ket{\alpha_A}\}$ and $\{U_{B,F_B},\ket{\alpha_B}\}$, see Sec.~\ref{sec:box_def}) to be implemented sequentially in the same cavity, with their order depending on the control state. For the 4B two such cavities are employed and the atom traverses
sequentially the two cavities.} \label{Exp_impl}
\end{figure*}

In the last decade, there has been a number of experimental investigations
of the quantum switch, and more generally the coherent control of unknown quantum operations, mainly using photonic setups~\cite{procopio_experimental_2015,rubino17,goswami18,Wei2019,Goswami2020,guo20,rubino_experimental_2021,Rubino2022experimental,cao22}, but also spins in nuclear magnetic resonance~\cite{Nie2020} or superconducting circuits~\cite{Felce2021,cuomo}. A proposal with trapped ions was also presented~\cite{Friis2014}. 
Contrarily to the model we considered, however, in these implementations the operations $\mathcal{A}$ and $\mathcal{B}$ are not implemented via the Jaynes-Cummings interaction with a single bosonic mode whose average number of excitations can be easily tuned. A natural setup to implement an energy constrained realization of the QS and 4B, as we considered in this work, would be a cavity quantum electrodynamics
platform. Indeed, such coupling is naturally realized by the electric
dipole Hamiltonian between atoms and field in the rotating wave approximation~\cite{haroche2006exploring}. We therefore propose a new experimental
implementation of the QS and 4B in this platform.

More specifically, we consider a four-level atom on which the control and target states are encoded. This atom interacts via a Jaynes-Cummings Hamiltonian with a single two-mode cavity in the case of the QS, and with two such two-modes cavities in the case of the 4B. These interactions mediate the implementation of the operations $\mathcal{A}$ and $\mathcal{B}$.

The four-level atom can be mapped to two effective qubits, encoding the target and control, respectively. By properly addressing the different levels, it is possible to change the state of one of them without affecting the other. Let us denote by $\left\{ \left|1\right\rangle ,\left|2\right\rangle ,\left|3\right\rangle ,\left|4\right\rangle \right\} $
these four atomic states. We consider the
subspace spanned by the two lower levels to encode the target state
when the control state is $\left|0_C\right\rangle $, and the subspace
spanned by the two upper levels to encode the target state when the
control state is $\left|1_C\right\rangle $. This defines the mapping $\big\{ \left|1\right\rangle =\left|0_{C}0_{S}\right\rangle$, $\left|2\right\rangle =\left|0_{C}1_{S}\right\rangle$, $\left|3\right\rangle =\left|1_{C}0_{S}\right\rangle $, $\left|4\right\rangle =\left|1_{C}1_{S}\right\rangle \big\} $
between the Hilbert space of the atom and the Hilbert space of the
two qubits. 
The initial state of the atom should be such that it corresponds to a product state between the system and the control.

For the QS, as illustrated in Fig.~\ref{Exp_impl}(a), the passage of the atom through the first half of the cavity will coherently implement the operations ${\cal A}$ and ${\cal B}$ depending on the state of the control, $\ket{0_C}$ or $\ket{1_C}$ respectively. When the atom reaches the middle of the cavity, a fast electric pulse is applied, inverting the populations between the upper and lower atomic subspaces and thus effectively implementing a Pauli $\sigma_{x}$ gate on the control qubit.
In this way, the target state component that underwent operation ${\cal A}$ (i.e., that was encoded in the lower subspace) now becomes coupled to the field mode that implements
operation ${\cal B}$ (i.e., that is encoded in the upper subspace). Similarly, the target state component previously encoded in the upper subspace is now encoded in the lower subspace. When the atom passes through
the second half of the cavity, the operations ${\cal A}$ and ${\cal B}$ are again implemented,
but to the target state component corresponding to the other control state; i.e., ${\cal B}$ is implemented to the component that underwent ${\cal A}$ in the first half, and vice versa. Just like the orange squares in Fig.~\ref{fig:QS_4B}, the role of $\sigma_x$ is to flip on which branch of the superposition the boxes act. When the
atom exits the cavity, its state corresponds to the bipartite control
and target qubit state after the operation of the quantum switch, as in Eq.~\eqref{eq:QS_fieldlevel_ABimperfect} (up to a flip of the control qubit). 

In order to implement the 4B, two such cavities can be used. The atom passes through the first cavity, coherently implementing ${\cal A}$ on the upper subspace and ${\cal B}$
on the lower subspace. In the passage through the second cavity, the
gates ${\cal B}$ and ${\cal A}$ are coherently implemented in the upper and lower
subspaces, respectively, thus implementing the 4B, see Fig.~\ref{Exp_impl}(b). In contrast to
the quantum switch, which employs two quantum fields, one for each mode, the 4B employs
four quantum fields, in agreement with the description of Sec.~\ref{sec:switch}.

The implementation described above provides a potential means to experimentally observe an energy advantage of using superpositions of causal orders. We thereby challenge experimental groups to realise this novel implementation.

\section{Approximating unitary evolutions in the Jaynes-Cummings model}
\label{app:jaynes_cumming}

Recall the Hamiltonian in the Jaynes-Cummings model:
\begin{align}
    H= \frac{\hbar \Omega_0}{2} \left( e^{i \phi} \, \sigma_+ \otimes a + e^{-i\phi} \, \sigma_- \otimes a^{\dagger} \right)
\end{align}
with $\sigma_- = \ketbra{g}{e}$ and $\sigma_+=\sigma_-^\dagger = \ketbra{e}{g}$ (and with $\ket{g}$ and $\ket{e}$ denoting the ground and excited states of the atom).
Applying this Hamiltonian for a given time $t$, we get the unitary $U(t) = e^{-i \frac{H t}{\hbar}} = e^{-i \frac{\Omega_0 t}{2} \left( e^{i \phi} \, \sigma_+ \otimes a + e^{-i\phi} \, \sigma_- \otimes a^{\dagger} \right)}$. Expanding the exponential, using the (easily verified) facts that
$\left( e^{i \phi} \, \sigma_+ \otimes a + e^{-i\phi} \, \sigma_- \otimes a^{\dagger} \right)^{2k} = \ketbra{g}{g} \otimes (a^\dagger a)^k + \ketbra{e}{e} \otimes (a a^\dagger)^k$ and $\left( e^{i \phi} \, \sigma_+ \otimes a + e^{-i\phi} \, \sigma_- \otimes a^{\dagger} \right)^{2k+1} = e^{i \phi} \, \ketbra{e}{g} \otimes a (a^\dagger a)^k + e^{-i\phi} \, \ketbra{g}{e} \otimes a^{\dagger} (a a^\dagger)^k$,
and introducing the photon number operator $N = a^\dagger a (= a a^\dagger - 1)$, we can write $U(t)$ in the $\{\ket{g}, \ket{e}\}$ basis for the atom, as a block operator
\begin{align}
    U(t) = 
    \begin{pmatrix}
    \cos\left(\frac{\Omega_{0}t}{2}\sqrt{N}\right) &
    -ie^{-i\phi}a^{\dagger}\frac{\sin\left(\frac{\Omega_{0}t}{2}\sqrt{N+1}\right)}{\sqrt{N+1}}\\
    -ie^{+i\phi}a\frac{\sin\left(\frac{\Omega_{0}t}{2}\sqrt{N}\right)}{\sqrt{N}} &
    \cos\left(\frac{\Omega_{0}t}{2}\sqrt{N+1}\right)
    \end{pmatrix}. \label{eq:Ut}
\end{align}

Consider applying this unitary to a product state $\ket{\psi}\otimes\ket{\alpha}$ of the atom-field system, with $\ket{\psi} = c_g \ket{g} + c_e \ket{e}$ and $\ket{\alpha}$ a coherent state with amplitude%
\footnote{Taking any complex value for $\alpha$ would simply shift the angle $\phi$ that defines the rotation axis: there is no loss of generality in considering $\alpha > 0$ here.}
$\alpha > 0$ and mean photon number $\bar{n} = |\alpha|^2$---i.e., written in the Foch basis, $\ket{\alpha} = \sum_{n\ge 0} c_n^{(\bar{n})} \ket{n}$ with $c_n^{(\bar{n})} = e^{-\bar{n}/2} \, \sqrt{\bar{n}}^n / \sqrt{n!}$. Using the above expression, we get
\begin{widetext}
\begin{align}
    U(t) \ket{\psi}\otimes\ket{\alpha} & = \sum_{n\ge 0} c_n^{(\bar{n})} \Big[ \big( \cos\left(\frac{\Omega_{0}t}{2}\sqrt{n}\right) c_g \ket{g} + \cos\left(\frac{\Omega_{0}t}{2}\sqrt{n+1}\right) c_e \ket{e} \big) \otimes \ket{n} \notag \\[-1mm]
    & \hspace{15mm} -ie^{-i\phi}\sin\left(\frac{\Omega_{0}t}{2}\sqrt{n+1}\right) c_e \ket{g} \otimes \ket{n+1} -ie^{+i\phi}\sin\left(\frac{\Omega_{0}t}{2}\sqrt{n}\right) c_g \ket{e} \otimes \ket{n-1} \Big] \notag \\[1mm]
    & = \sum_{n\ge 0} \Big[ c_n^{(\bar{n})} \big( \cos\left(\frac{\Omega_{0}t}{2}\sqrt{n}\right) c_g \ket{g} + \cos\left(\frac{\Omega_{0}t}{2}\sqrt{n+1}\right) c_e \ket{e} \big) \notag \\[-1mm]
    & \hspace{15mm} -c^{(\bar{n})}_{n-1} ie^{-i\phi}\sin\left(\frac{\Omega_{0}t}{2}\sqrt{n}\right) c_e \ket{g} -c^{(\bar{n})}_{n+1} ie^{+i\phi}\sin\left(\frac{\Omega_{0}t}{2}\sqrt{n+1}\right) c_g \ket{e} \Big] \otimes \ket{n}. \label{eq:Ut_psi_alpha}
\end{align}
\end{widetext}

Assume now that $\ket{\alpha}$ has a large mean photon number: $\bar{n} \gg 1$. In that case the (Poissonian) distribution of the weights $|c_n^{(\bar{n})}|^2$ is peaked around $\bar{n}$, with a width of the order of $\sqrt{\bar{n}}$; beyond this peak, the weights $c_n^{(\bar{n})}$ are negligible.
Expanding $\sqrt{n}$ around its value for $n = \bar{n}$, we have $\sqrt{n} \approx \sqrt{\bar{n}} + \frac{n-\bar{n}}{2\sqrt{\bar{n}}}$, and we can thus write $\cos\left(\frac{\Omega_{0}t}{2}\sqrt{n}\right) \approx \cos\left(\frac{\Omega_{0}t}{2}\sqrt{\bar{n}} + \frac{n-\bar{n}}{4\sqrt{\bar{n}}}\Omega_{0}t\right)$.
Assuming now that the interaction time $t$ is small enough so that $\Omega_0 t \ll 1$, then for all $n$ whose corresponding weight $c_n^{(\bar{n})}$ is non-negligible (for which $\frac{n-\bar{n}}{4\sqrt{\bar{n}}}$ is typically smaller than 1), the second term in the cosine above is negligible (and so would be all higher-order terms in ($n-\bar{n}$)). We thus get, in the relevant range of $n$, $\cos\left(\frac{\Omega_{0}t}{2}\sqrt{n}\right) \approx \cos\left(\frac{\Omega_{0}t}{2}\sqrt{\bar{n}}\right)$, and similarly, $\cos\left(\frac{\Omega_{0}t}{2}\sqrt{n+1}\right) \approx \cos\left(\frac{\Omega_{0}t}{2}\sqrt{\bar{n}}\right)$ and $\sin\left(\frac{\Omega_{0}t}{2}\sqrt{n}\right) \approx \sin\left(\frac{\Omega_{0}t}{2}\sqrt{n+1}\right) \approx \sin\left(\frac{\Omega_{0}t}{2}\sqrt{\bar{n}}\right)$.

Notice now that $c^{(\bar{n})}_{n-1} = \frac{\sqrt{n}}{\sqrt{\bar{n}}} c_n^{(\bar{n})}$ and $c^{(\bar{n})}_{n+1} = \frac{\sqrt{\bar{n}}}{\sqrt{n+1}} c_n^{(\bar{n})}$. Still in the relevant range of $n$ around $\bar{n}$, we have $\frac{\sqrt{n}}{\sqrt{\bar{n}}} \approx \frac{\sqrt{\bar{n}}}{\sqrt{n+1}} \approx 1$, so that $c^{(\bar{n})}_{n-1} \approx c^{(\bar{n})}_{n+1} \approx c_n^{(\bar{n})}$. All in all, we can then approximate Eq.~\eqref{eq:Ut_psi_alpha} as~\cite{AspectandFabrebook}
\begin{widetext}
\begin{align}
    U(t) \ket{\psi}\otimes\ket{\alpha} & \approx \sum_{n\ge 0} \Big[  \cos\left(\frac{\Omega_{0}t}{2}\sqrt{\bar{n}}\right) \big( c_g \ket{g} + c_e \ket{e} \big) - i \sin\left(\frac{\Omega_{0}t}{2}\sqrt{\bar{n}}\right) \big( e^{-i\phi} c_e \ket{g} + e^{+i\phi} c_g \ket{e} \big) \Big] \otimes c_n^{(\bar{n})} \ket{n} \notag \\[1mm]
    & \approx \left(R_\phi\big(\Omega_0 t \sqrt{\bar{n}}\big)\ket{\psi}\right)\otimes\ket{\alpha}, \label{eq:Ut_psi_alpha_approx}
\end{align}
\end{widetext}
where $R_\phi(\theta) = e^{-i\frac{\theta}{2}(\cos\phi \, \sigma_x + \sin\phi \, \sigma_y)}$ denotes a rotation on the Bloch sphere by an angle $\theta$, around an equatorial axis defined by its azimuthal angle $\phi$ (thus corresponding to the notation $R_{\vec{\mathbf{u}}}(\theta)$ introduced in the main text, with $\vec{\mathbf{u}}=(\cos\phi,\sin\phi,0)_{x,y,z}$; $\sigma_x$ and $\sigma_y$ above are Pauli matrices).

One thus recognises that in the large photon number limit, and under the above assumptions, the joint unitary $U(t)$ applied to both the atom and the cavity initialized in a coherent state effectively approximates a rotation $R_\phi\big(\Omega_0 t \sqrt{\bar{n}}\big)$ of the state of the atom alone, while leaving the state of the cavity essentially unchanged. If one aims at approximating a rotation by a given angle $\theta$, for a given (large enough) average number of photons $\bar{n}$ in the cavity, one should thus choose the time of interaction $t$ such that $\Omega_0 t \approx \theta / \sqrt{\bar{n}}$ (which indeed is such that $\Omega_0 t \ll 1$ when $\bar{n} \gg 1$, as assumed above).
Note that the somewhat hand-waving calculations and approximations presented here can be made rigorous: we will see in Appendix~\ref{app:Kraus_ops_asympt} how precisely the above choice approximates the desired rotations in the asymptotic limit of large $\bar{n}$.

\begin{figure}
\centering   
\includegraphics[width=\columnwidth]{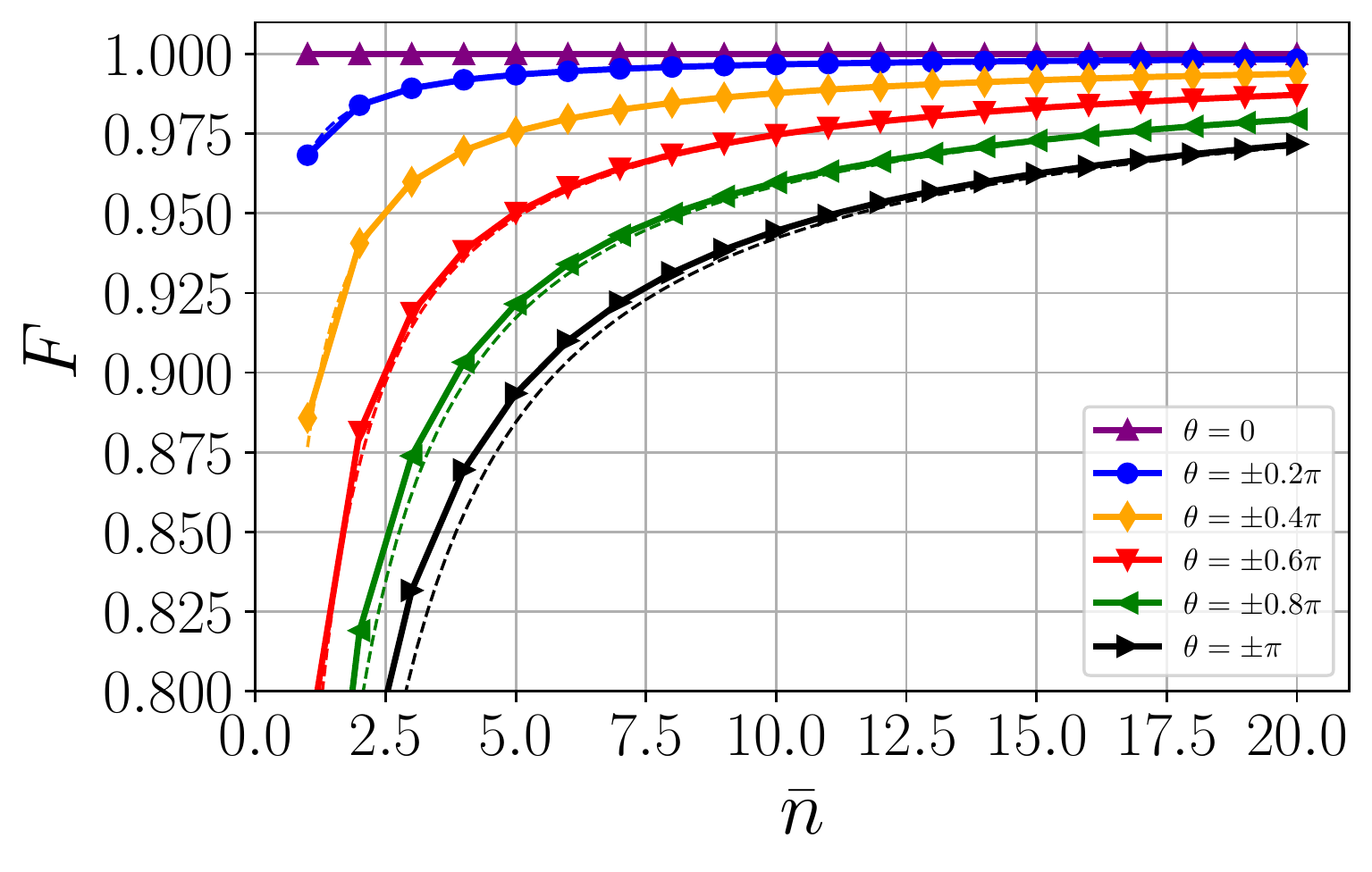}
\caption{Fidelity of the approximate rotations in the finite-energy regime. Here the fidelity of an operation ${\cal F}$ with respect to a unitary $U$ is defined as the fidelity of the output state with respect to the ideal one, averaged over all input states; for 2-dimensional systems it can be calculated as $F = \frac12+\frac{1}{12}\sum_{j=x,y,z}\Tr[U\sigma_jU^\dagger{\mathcal F}(\sigma_j)]$~\cite{Bowdrey2002}. The connected points are obtained numerically for different values of $\bar{n}$, while the dashed lines are the analytical first-order expansions in $1/\bar{n}$ (namely, $F = 1 - \frac{1-\cos\theta+\theta^2/2}{12 \bar{n}}+O\left( \frac{1}{\bar{n}^2} \right)$).}
\label{fig:F_vs_nbar}
\end{figure}

In the finite $\bar{n}$ regime, $U(t)$ does of course not induce a perfect rotation of the state of the atom only. The time of interaction $t$ could in principle be optimized, for each finite value of $\bar{n}$, so that the Jaynes-Cummings considered here gives the best possible approximation of a given desired rotation. However, for simplicity and as a rule of thumb we will simply take this time of interaction to be $\Omega_0 t = \theta / \sqrt{\bar{n}}$, as dictated by the approximation in the large $\bar{n}$ limit.%
\footnote{We do not claim that this choice is necessarily optimal. E.g., choosing $\Omega_0 t = \theta / \sqrt{\bar{n}+1}$, instead, or $\Omega_0 t = \theta / \sqrt{\bar{n}+\delta}$ for some other value of $\delta$, could give slightly larger fidelities in Fig.~\ref{fig:F_vs_nbar}---and correspondingly, slightly larger success probabilities for the task considered in the main text, shown on Fig.~\ref{avg}. However, this would not qualitatively change our comparison between the QS and 4B setups or any of our results (e.g., our first order expansions in $\frac{1}{\bar{n}}$ of the fidelity above, and of the success probabilities in Appendix~\ref{app:asymptotic_game} would not depend on the (fixed) value of $\delta$), so for simplicity we will stick to the choice $\Omega_0 t = \theta / \sqrt{\bar{n}}$.}
In Fig.~\ref{fig:F_vs_nbar} we show the fidelity of the approximate rotation induced on the atom as a function of $\bar{n}$, for various values of $\theta$, to see how well this approaches the ideal unitary case. Note that the smaller (in absolute value) the angle $\theta$, the better the approximation (for a fixed value of $\bar{n}$; in the large-$\bar{n}$ limit, based on our analysis in Appendix~\ref{app:Kraus_ops_asympt} below, we find $F = 1 - \frac{1-\cos\theta+\theta^2/2}{12 \bar{n}}+O\left( \frac{1}{\bar{n}^2} \right)$). In particular, one loses the $2\pi$-periodicity for the approximations; in this paper we therefore consider all rotation angles to be in the interval $[-\pi,\pi]$.

\section{Kraus operators and asymptotic behaviours of the induced linear maps}
\label{app:Kraus_ops_asympt}
\subsection{Kraus operators}

From Eq.~\eqref{eq:Ut}, we can obtain the reduced dynamics of the two-level system induced by the Jaynes-Cummings Hamiltonian. For that, we consider as above that the field is initialized in a coherent state $\ket{\alpha}$ where $\alpha=\sqrt{\bar{n}}$, $\bar{n}$ being the average number of photons in this field. We also consider a time of interaction $t=\theta/(\Omega_0 \sqrt{\bar{n}})$ ($\theta$ being the angle of rotation we wish to perform), as prescribed in Appendix~\ref{app:jaynes_cumming}, and denote by $U^{(\bar{n})}$ the corresponding unitary operator: $U^{(\bar{n})} = U\big(\frac{\theta}{\Omega_0 \sqrt{\bar{n}}}\big)$. The reduced dynamics is then given by the map $\mathcal{F}^{(\bar{n})}$ such that for any initial density matrix $\rho_S$ for the two level system, 
\begin{align}
    \mathcal{F}^{(\bar{n})}(\rho_S)=\sum_{n=0}^{+\infty}A_n^{(\bar{n})} \rho_S A_n^{(\bar{n})\dagger}, \label{eq:def_Ga}
\end{align}
where the Kraus operators $A_n^{(\bar{n})}$ are defined as
\begin{align}
    A_n^{(\bar{n})} & = (\id\otimes\bra{n}) \, U^{(\bar{n})} \, (\id\otimes\ket{\alpha}) \notag \\
    & = \sum_{m\ge 0} c_m^{(\bar{n})} (\id\otimes\bra{n}) \, U^{(\bar{n})} \, (\id\otimes\ket{m}) \notag \\
    & = \!\begin{pmatrix}
    c_n^{(\bar{n})} \cos\left(\frac{\theta}{2} \frac{\sqrt{n}}{\sqrt{\bar{n}}}\right) &
    \!\!\!-ie^{-i\phi} c_{n-1}^{(\bar{n})} \sin\!\left(\frac{\theta}{2} \frac{\sqrt{n}}{\sqrt{\bar{n}}}\right)\!\! \\[2mm]
    \!-ie^{+i\phi} c_{n+1}^{(\bar{n})} \sin\!\left(\frac{\theta}{2} \frac{\sqrt{n+1}}{\sqrt{\bar{n}}}\right)\!\!\! &
    c_n^{(\bar{n})} \cos\left(\frac{\theta}{2} \frac{\sqrt{n+1}}{\sqrt{\bar{n}}}\right)
    \end{pmatrix}\!. \label{eq:Kraus_An}
\end{align}

\subsection{Asymptotic behaviours}
\label{app:asymptotic_behaviours}

We aim here at describing the dynamics in the regime of large $\bar{n}$. For this purpose, in order to make the calculations more compact, we will describe the maps under consideration by making use of their Choi representation~\cite{choi75}.
The Choi matrix of a linear map ${\cal M}: {\cal L}({\cal H}^I) \to {\cal L}({\cal H}^O)$, from some ``input'' Hilbert space ${\cal H}^I$ to some ``output'' Hilbert space ${\cal H}^O$ and where $\mathcal{L}(\mathcal{H})$ represents the set of linear operators acting on the Hilbert space $\mathcal{H}$, is defined as
\begin{align}
    \underline{M} = \sum_{i,k} \ketbra{i}{k}^I \otimes {\cal M}(\ketbra{i}{k}^I) \ \ \in {\cal L}({\cal H}^I)\otimes {\cal L}({\cal H}^O), \label{eq:def_Choi}
\end{align}
where $\{\ket{i}^I\}$ is a fixed (``computational'') basis of ${\cal H}^I$. The Choi matrix elements are thus $\bra{i,j}\underline{M}\ket{k,l} = \bra{j}{\mathcal{M}}\big(\ketbra{i}{k}\big)\ket{l}$.

We will look below at the asymptotic behaviors for two different linear maps, in the large $\bar{n}$ regime. For that, using the facts that $c_{n-1}^{(\bar{n})} = \frac{\sqrt{n}}{\sqrt{\bar{n}}}c_{n}^{(\bar{n})}$ and $c_{n+1}^{(\bar{n})} = \frac{\sqrt{\bar{n}}}{\sqrt{n+1}}c_{n}^{(\bar{n})}$, it will be convenient to write the Kraus operators of Eq.~\eqref{eq:Kraus_An} above as
\begin{widetext}
\begin{align}
    A_n^{(\bar{n})} & = c_n^{(\bar{n})} \begin{pmatrix}
    \cos\left(\frac{\theta}{2}\frac{\sqrt{n}}{\sqrt{\bar{n}}}\right) &
    -ie^{-i\phi} \frac{\sqrt{n}}{\sqrt{\bar{n}}} \sin\left(\frac{\theta}{2}\frac{\sqrt{n}}{\sqrt{\bar{n}}}\right) \\[2mm]
    -ie^{+i\phi} \frac{\sqrt{\bar{n}}}{\sqrt{n+1}} \sin\left(\frac{\theta}{2}\frac{\sqrt{n+1}}{\sqrt{\bar{n}}}\right) &
    \cos\left(\frac{\theta}{2}\frac{\sqrt{n+1}}{\sqrt{\bar{n}}}\right)
    \end{pmatrix} \notag \\[3mm]
    & = c_n^{(\bar{n})} \begin{pmatrix}
    \sum_{p=0}^\infty\frac{(-1)^p}{(2p)!}\left(\frac{\theta}{2}\right)^{2p}\frac{n^p}{\bar{n}^p} &
    -ie^{-i\phi} \sum_{p=0}^\infty\frac{(-1)^p}{(2p+1)!}\left(\frac{\theta}{2}\right)^{2p+1}\frac{n^{p+1}}{\bar{n}^{p+1}} \\[2mm]
    -ie^{+i\phi} \sum_{p=0}^\infty\frac{(-1)^p}{(2p+1)!}\left(\frac{\theta}{2}\right)^{2p+1}\frac{(n+1)^p}{\bar{n}^p} &
    \sum_{p=0}^\infty\frac{(-1)^p}{(2p)!}\left(\frac{\theta}{2}\right)^{2p}\frac{(n+1)^p}{\bar{n}^p}
    \end{pmatrix}\!. \label{eq:Kraus_An_v2}
\end{align}
\end{widetext}

\subsubsection{Effective map applied to the atom, $\mathcal{F}^{(\bar{n})}: \rho \mapsto \sum_n A_n^{(\bar{n})} \rho A_n^{(\bar{n})\dagger}$}
\label{app:def_G}

We first calculate the Choi matrix of the map $\mathcal{F}^{(\bar{n})}$ defined through Eq.~\eqref{eq:def_Ga}. In order to keep this section short, we will detail the calculations for a single coefficient of the Choi matrix only. All other coefficients are obtained in a similar way. This Choi matrix will then be used in Appendix~\ref{app:asymptotic_game} to estimate the success probability of the QS at the commuting-vs-anticommuting task, in the large-$\bar{n}$ limit.

Using Eq.~\eqref{eq:Kraus_An_v2}, we have:
\begin{align}
    \bra{00}\underline{F}^{(\bar{n})}\!\ket{00} & = \bra{0}\mathcal{F}^{(\bar{n})}\big(\ketbra{0}{0}\big)\ket{0} \notag \\[1mm]
    & \hspace{-15mm} = \sum_{n\ge 0} \bra{0}\!A_n^{(\bar{n})} \ketbra{0}{0} A_n^{(\bar{n})\dagger}\!\ket{0} \notag \\[-2mm]
%    & \hspace{-15mm} = \sum_{n\ge 0} (c_n^{(\bar{n})})^2 \cos^2 \left(\frac{\theta}{2} \sqrt{\frac{n}{\bar{n}}} \right)  \notag \\[-1mm]
     & \hspace{-15mm} = \sum_{n \ge 0} (c_n^{(\bar{n})})^2 \sum_{p \geq 0} \frac{(-1)^{p}}{(2p)!}\!\left(\frac{\theta}{2}\right)^{\!\!2p}\frac{n^{p}}{\bar{n}^{p}}\sum_{q \geq 0} \frac{(-1)^{q}}{(2q)!}\!\left(\frac{\theta}{2}\right)^{\!\!2q}\frac{n^{q}}{\bar{n}^{q}} \notag \\
    & \hspace{-15mm} = \sum_{p\ge 0} \sum_{q\ge 0}\frac{(-1)^{p+q}}{(2p)!(2q)!}\!\left(\frac{\theta}{2}\right)^{\!2p+2q} \!\frac{1}{\bar{n}^{p+q}}\sum_{n\ge 0} (c_n^{(\bar{n})})^2 n^{p+q}, \label{eq:Galpha00}
\end{align}
where we used the fact that the triple series is absolutely convergent%
\footnote{Indeed, the summands $s_{n,p,q} = (c_n^{(\bar{n})})^2 \frac{(-1)^{p}}{(2p)!}\!\left(\frac{\theta}{2}\right)^{\!\!2p}\frac{n^{p}}{\bar{n}^{p}} \frac{(-1)^{q}}{(2q)!}\!\left(\frac{\theta}{2}\right)^{\!\!2q}\frac{n^{q}}{\bar{n}^{q}}$ satisfy $|s_{n,p,q}| \leq (c_n^{(\bar{n})})^2 \frac{1}{(2p)!}\!\left(\frac{\theta}{2}\right)^{\!\!2p}\frac{n^{p}}{\bar{n}^{p}} \frac{1}{(2q)!}\!\left(\frac{\theta}{2}\right)^{\!\!2q}\frac{n^{q}}{\bar{n}^{q}}$, so that $\sum_{p,q} |s_{n,p,q}| \leq (c_n^{(\bar{n})})^2 \cosh^2\left(\frac{\theta}{2}\frac{\sqrt{n}}{\sqrt{\bar{n}}}\right)$ for each $n$. Using for instance Stirling's formula, it is then straightforward to see that $\lim_{n\to+\infty} n^2 [(c_n^{(\bar{n})})^2 \cosh^2\left(\frac{\theta}{2}\frac{\sqrt{n}}{\sqrt{\bar{n}}}\right)] = 0$, which ensures that $\sum_{n,p,q} |s_{n,p,q}| < +\infty$.}
to swap the sums in the last line.

In the last sum above we recognize the $(p+q)^\text{th}$ moment of the Poisson distribution with mean value $\bar{n}$, which can be written as~\cite{Haight1967handbook}
\begin{align}
\sum_{n\ge 0} (c_n^{(\bar{n})})^2 n^{p+q} = \sum_{j = 0}^{p+q} \stirlingii{p+q}{p+q-j} \bar{n}^{p+q-j}, \label{eq:higher_moment}
\end{align}
where $\stirlingii{p+q}{p+q-j}$ denotes a Stirling number of the second kind.
Inserting this into Eq.~\eqref{eq:Galpha00} and taking the convention that $\stirlingii{p+q}{p+q-j} = 0$ for $j > p+q$, we can then swap the sums again so as to obtain
\begin{align}
    & \bra{00}\underline{F}^{(\bar{n})}\!\ket{00} \notag \\
    & = \sum_{j\ge 0} \left( \sum_{p\ge 0} \sum_{q\ge 0}\frac{(-1)^{p+q}}{(2p)!(2q)!}\!\left(\frac{\theta}{2}\right)^{\!2p+2q} \stirlingii{p+q}{p+q-j} \right) \frac{1}{\bar{n}^j}.
\end{align}
Evaluating the first two terms of the sum over $j$ (using $\stirlingii{p+q}{p+q} = 1$ and $\stirlingii{p+q}{p+q-1} = \frac{(p+q)(p+q-1)}{2}$) and truncating the higher-order terms in $\frac{1}{\bar{n}}$, we obtain, after some algebraic manipulations,
\begin{align}
    & \bra{00}\underline{F}^{(\bar{n})}\!\ket{00} \notag \\
    & = \frac12(1+\cos\theta) + \frac{1}{16\bar{n}} (\theta\sin\theta - \theta^2\cos\theta) + O\left(\frac{1}{\bar{n}^2}\right).
\end{align}
Beyond this first term, the other elements of the Choi matrix are $\bra{i,j}\underline{F}^{(\bar{n})}\ket{k,l} = \bra{j}\mathcal{F}^{(\bar{n})}\big(\ketbra{i}{k}\big)\ket{l} = \sum_n \bra{j}A_n^{(\bar{n})} \ketbra{i}{k} A_n^{(\bar{n})\dagger}\ket{l}$, i.e., are obtained as sums over $n$ of products of 2 elements (one being conjugated) of the Kraus operators of Eq.~\eqref{eq:Kraus_An_v2} (from which we get 2 more sums $\sum_p \sum_q$, as above). After exchanging the sums $\sum_n$ with the sums $\sum_p \sum_q$ (as allowed by the fact that the series are still absolutely convergent), we get terms of the form $\sum_{n\ge 0} (c_{n}^{(\bar{n})})^2 n^r(n+1)^s$, which can now be written (after expanding $(n+1)^s$, using Eq.~\eqref{eq:higher_moment} and rearranging the sums) as $\sum_{j=0}^{r+s} (\sum_{k=0}^{\min(s,j)}\binom{s}{k}\stirlingii{r+s-k}{r+s-j})\bar{n}^{r+s-j} = \bar{n}^{r+s}(1 + \frac{(r+s)(r+s-1)+2s}{2\bar{n}} + O\!\left(\frac{1}{\bar{n}^2}\right))$. After swapping the sums again, we can evaluate the leading terms (for $j=0,1$), as we did above.

For the case where $\phi=0$, these calculations lead to
\begin{widetext}
{\small\begin{align}
&\underline{F}^{(\bar{n})} = \underline{R}_0(\theta) + \frac{1}{16 \bar{n}}\! \left(\!\!\!
\begin{array}{cccc}
 \theta  s_{\theta }{-}\theta ^2 c_{\theta } & \!-i \left(s_{\theta
   }{-}\theta(2{-}c_{\theta }){+}\theta ^2 s_{\theta }\right)\! & -i \left(s_{\theta }{-}\theta  c_{\theta }{+}\theta ^2 s_{\theta }\right) & -\theta  s_{\theta }{-}\theta ^2 c_{\theta } \\
 i \left(s_{\theta
   }{-}\theta(2{-}c_{\theta }){+}\theta ^2 s_{\theta }\right) & -\theta  s_{\theta }{+}\theta ^2c_{\theta } & -4(1{-}c_{\theta }){+}\theta  s_{\theta }{+}\theta ^2c_{\theta } & i \left(s_{\theta }{-}\theta  c_{\theta }{+}\theta ^2 s_{\theta
   }\right) \\
 i \left(s_{\theta }{-}\theta  c_{\theta }{+}\theta ^2 s_{\theta }\right) & -4(1{-}c_{\theta }){+}\theta  s_{\theta }{+}\theta ^2c_{\theta } & 3 \theta  s_{\theta }{+}\theta ^2c_{\theta } & i
   \left(s_{\theta }{+}\theta(2{-}3 c_{\theta }){+}\theta ^2 s_{\theta }\right) \\
 -\theta  s_{\theta }{-}\theta ^2 c_{\theta } & -i \left(s_{\theta }{-}\theta  c_{\theta }{+}\theta ^2 s_{\theta
   }\right) & \!-i \left(s_{\theta }{+}\theta(2{-}3 c_{\theta }){+}\theta ^2 s_{\theta }\right)\! & 
   -3 \theta  s_{\theta }{-}\theta ^2c_{\theta } \\
\end{array}
\!\!\!\right) \!+ O\!\left(\frac{1}{\bar{n}^2}\right) \label{eq:choi_Fnbar}
\end{align}}
\end{widetext}
with $c_\theta = \cos\theta$, $s_\theta = \sin\theta$, and where
\begin{align}
&\underline{R}_0(\theta) = \frac12 \left(\!
\begin{array}{cccc}
 1{+}c_{\theta } & i s_{\theta } & i s_{\theta } & 1{+}c_{\theta } \\
 -i s_{\theta } & 1{-}c_{\theta } & 1{-}c_{\theta } &
   -i s_{\theta } \\
 -i s_{\theta } & 1{-}c_{\theta } & 1{-}c_{\theta }
   & -i s_{\theta } \\
 1{+}c_{\theta } & i s_{\theta } & i s_{\theta } & 1{+}c_{\theta } \\
\end{array}
\!\right) \label{eq:ChoiRO_theta}
\end{align}
is the Choi matrix of the map $\rho \mapsto R_0(\theta) \rho R_0(\theta)^\dagger$ that applies a perfect rotation by an angle $\theta$ around the $x$ axis of the Bloch sphere.
For some azimuthal angle $\phi\neq 0$, one just has to multiply the 2nd row and 3rd columns of the Choi matrices above by $e^{i \phi }$, and the 3rd row and 2nd columns by $e^{-i \phi }$ (in accordance with how $\phi$ appears in the Kraus operators $A_n^{(\bar{n})}$, see Eq.~\eqref{eq:Kraus_An}%
\footnote{Note that $\underline{F}^{(\bar{n})}$ thus remains Hermitian, and even positive semidefinite, as required for the Choi matrix of completely positive (CP) map.}).

Thus, we clearly see from Eq.~\eqref{eq:choi_Fnbar}, that in the large-$\bar{n}$ limit, the operation $\mathcal{F}^{(\bar{n})}$ (obtained as described above from the Jaynes-Cummings interaction, with the choice $\Omega_0 t = \theta / \sqrt{\bar{n}}$) indeed tends to a perfect rotation by an angle $\theta$, which confirms the approximate calculations of Appendix~\ref{app:jaynes_cumming}. One can also see that the elements of the first-order correction matrix in Eq.~\eqref{eq:choi_Fnbar} are null for $\theta=0$, and increase (up to a certain point) with $|\theta|$; in particular, as also noticed before, we lose the $2\pi$-periodicity.

\subsubsection{Induced map $\mathcal{G}^{(\bar{n})}: \rho \mapsto G^{(\bar{n})} \rho G^{(\bar{n})\dagger}$ with $G^{(\bar{n})} = (\id\otimes\bra{\alpha/\sqrt{2}}) \, U^{(\bar{n}/2)} \, (\id\otimes\ket{\alpha/\sqrt{2})}$}
\label{app:def_F}

In our calculations of the success probability of the 4B at the commuting-vs-anticommuting task we consider, we are also led to consider the linear map $\mathcal{G}^{(\bar{n})}$ defined in the title of this subsection. As we did for $\mathcal{F}^{(\bar{n})}$ above, we will derive here a perturbative expansion in the large-$\bar{n}$ limit.

Contrary to $\mathcal{F}^{(\bar{n})}$ that involved infinitely many Kraus operators $A_n^{(\bar{n})}$, the map $\mathcal{G}^{(\bar{n})}$ is defined in terms of a single Kraus operator $G^{(\bar{n})}$. We can thus first consider the expansion of $G^{(\bar{n})}$ in the large-$\bar{n}$ limit. We do this in a similar way to what we did for $\mathcal{F}^{(\bar{n})}$ above, noting that $G^{(\bar{n})} = \sum_n \big(c_n^{(\bar{n}/2)}\big)^* A_n^{(\bar{n}/2)}$ and starting from the form of Eq.~\eqref{eq:Kraus_An_v2} for the operators $A_n^{(\bar{n}/2)}$; the difference with the previous calculations being that here we only have single sums $\sum_p$ (as opposed to double sums $\sum_p\sum_q$), to be swapped with the sums $\sum_n$ and then $\sum_j$.

For the case where $\phi=0$ we thus get
\begin{widetext}
{\small\begin{align}
    G^{(\bar{n})} & = \begin{pmatrix}
    \cos\!\left(\frac{\theta}{2}\right) &
    -i\sin\!\left(\frac{\theta}{2}\right) \\[1mm]
    -i\sin\!\left(\frac{\theta}{2}\right) &
    \cos\!\left(\frac{\theta}{2}\right)
    \end{pmatrix} + \frac{1}{16 \bar{n}} \begin{pmatrix}
    2\theta\sin\!\left(\frac{\theta}{2}\right)-\theta^2\cos\!\left(\frac{\theta}{2}\right) &
    i\left(4\sin\!\left(\frac{\theta}{2}\right)-2\theta\cos\!\left(\frac{\theta}{2}\right)+\theta^2\sin\!\left(\frac{\theta}{2}\right)\right) \\[1mm]
    i\left(4\sin\!\left(\frac{\theta}{2}\right)-2\theta\cos\!\left(\frac{\theta}{2}\right)+\theta^2\sin\!\left(\frac{\theta}{2}\right)\right) &
    -6\theta\sin\!\left(\frac{\theta}{2}\right)-\theta^2\cos\!\left(\frac{\theta}{2}\right)
    \end{pmatrix} + O\!\left(\frac{1}{\bar{n}^2}\right) , \label{eq:Gnbar}
\end{align}}
from which we then easily obtain the Choi matrix of the linear map $\mathcal{G}^{(\bar{n})}$:
{\small\begin{align}
&\underline{G}^{(\bar{n})} = \underline{R}_0(\theta){+}\frac{1}{8 \bar{n}}\! \left(\!\!\!
\begin{array}{cccc}
 \theta  s_{\theta }{-}\frac{\theta ^2}{2} (1{+}c_{\theta }) & -i \left(s_{\theta
   }{-}\theta{+}\frac{\theta ^2}{2} s_{\theta }\right) & -i \left(s_{\theta
   }{-}\theta{+}\frac{\theta ^2}{2} s_{\theta }\right) & -\theta  s_{\theta }{-}\frac{\theta ^2}{2} (1{+}c_{\theta }) \\
 i \left(s_{\theta
   }{-}\theta{+}\frac{\theta ^2}{2} s_{\theta }\right) & -2(1{-}c_{\theta }){+}\theta  s_{\theta }{-}\frac{\theta ^2}{2} (1{-}c_{\theta }) & -2(1{-}c_{\theta }){+}\theta  s_{\theta }{-}\frac{\theta ^2}{2} (1{-}c_{\theta }) & i \left(s_{\theta }{+}\theta  (1{-}2c_{\theta }){+}\frac{\theta ^2}{2} s_{\theta
   }\right) \\
 i \left(s_{\theta
   }{-}\theta{+}\frac{\theta ^2}{2} s_{\theta }\right) & -2(1{-}c_{\theta }){+}\theta  s_{\theta }{-}\frac{\theta ^2}{2} (1{-}c_{\theta }) & -2(1{-}c_{\theta }){+}\theta  s_{\theta }{-}\frac{\theta ^2}{2} (1{-}c_{\theta }) & i \left(s_{\theta }{+}\theta  (1{-}2c_{\theta }){+}\frac{\theta ^2}{2} s_{\theta
   }\right) \\
 -\theta  s_{\theta }{-}\frac{\theta ^2}{2} (1{+}c_{\theta }) & -i \left(s_{\theta }{+}\theta  (1{-}2c_{\theta }){+}\frac{\theta ^2}{2} s_{\theta
   }\right) & -i \left(s_{\theta }{+}\theta  (1{-}2c_{\theta }){+}\frac{\theta ^2}{2} s_{\theta
   }\right) & 
   -3 \theta  s_{\theta }{-}\frac{\theta ^2}{2}(1{+}c_{\theta }) \\
\end{array}
\!\!\!\right) \!+ O\!\left(\frac{1}{\bar{n}^2}\right) \label{eq:choi_Gnbar}
\end{align}}
\end{widetext}
(again with $c_\theta = \cos\theta$ and $s_\theta = \sin\theta$). As before, for $\phi\neq 0$ one just needs to multiply the appropriate rows or columns of the Choi matrices above by either $e^{i\phi}$ or $e^{-i\phi}$.

\section{Success probabilities for the QS and 4B at the commuting-vs-anticommuting discrimination task}
\label{app:success_probability_derivation}

\subsection{Exact expressions}

Here we provide the exact analytic expressions for the success probabilities of Eqs.~\eqref{eq:success_com}--\eqref{eq:success_anticom}, when averaged following Eq.~\eqref{eq:p_succ_avg}. For this purpose, we start by recognizing, using Eqs.~\eqref{eq:Phi0QS}--\eqref{eq:Phi14B}, that
\begin{align}
    &\braket{\Phi^{\text{QS}}_0 (U_A,U_B)}{\Phi^{\text{QS}}_1(U_A,U_B)} =\Tr\left( \mathcal{F}^{(\bar{n})}\big(U_B \rho_S\big) U_B^{\dagger} \right),
    \label{eq:app:overlapfuncG}\\
    &\braket{\Phi^{\text{4B}}_0 (U_A,U_B)}{\Phi^{\text{4B}}_1(U_A,U_B)} =\Tr\left(\mathcal{G}^{(\bar{n})}\big(U_B \rho_S\big) U_B^{\dagger} \right),
    \label{eq:app:overlapfuncF}
\end{align}
where $\mathcal{F}^{(\bar{n})}$ and $\mathcal{G}^{(\bar{n})}$ are the linear maps previously defined in Secs.~\ref{app:def_G} and~\ref{app:def_F} (which appear in the calculation after tracing over the cavity fields), and where $\rho_S = \ketbra{\psi_S}{\psi_S}$. Note that these expressions, as well as the success probabilities below, are all linear in $\rho_S$, so that these remain valid for any mixed input states of the target system.
The average success probabilities for the QS and 4B in the commuting and anticommuting scenarios can then be written as
\begin{widetext}
\begin{align}
    \langle p_{\text{success}}^{[,],\text{QS}} \rangle&=\frac{1}{2}\Big(1+\int_{\mathcal{S}_{[,]}} \! \textup{d}\mu_{\left[,\right]}(U_A,U_B) \ \Re\left[\Tr\left( \mathcal{F}^{(\bar{n},U_A)}\big(U_B \rho_S\big) U_B^{\dagger} \right)\right]\Big) = \frac{1}{2}\Big(1+\sum_{n=0}^{+\infty} X^{[,],\text{QS}}_n(\bar{n})\Big), \label{eq:psucc_com_QS_integral} \\
    \langle p_{\text{success}}^{\{,\}, \text{QS}} \rangle&= \frac{1}{2}\Big(1-\int_{\mathcal{S}_{\{,\}}} \! \textup{d}\mu_{\{,\}}(U_A,U_B) \ \Re\left[\Tr\left( \mathcal{F}^{(\bar{n},U_A)}\big(U_B \rho_S\big) U_B^{\dagger} \right)\right]\Big) = \frac{1}{2}\Big(1-\sum_{n=0}^{+\infty} X^{\{,\},\text{QS}}_n(\bar{n})\Big), \label{eq:psucc_ant_QS_integral} \\
    \langle p_{\text{success}}^{[,],\text{4B}} \rangle&= \frac{1}{2}\Big(1+\int_{\mathcal{S}_{[,]}} \! \textup{d}\mu_{\left[,\right]}(U_A,U_B) \ \Re\left[\Tr\left( \mathcal{G}^{(\bar{n},U_A)}\big(U_B \rho_S\big) U_B^{\dagger} \right)\right]\Big) = \frac{1}{2}\Big(1+\sum_{m=0}^{+\infty}\sum_{n=0}^{+\infty} X^{[,],\text{4B}}_{m,n}(\bar{n})\Big), \label{eq:psucc_com_4B_integral} \\
    \langle p_{\text{success}}^{\{,\}, \text{4B}} \rangle&= \frac{1}{2}\Big(1-\int_{\mathcal{S}_{\{,\}}} \! \textup{d}\mu_{\{,\}}(U_A,U_B) \ \Re\left[\Tr\left( \mathcal{G}^{(\bar{n},U_A)}\big(U_B \rho_S\big) U_B^{\dagger} \right)\right]\Big) = \frac{1}{2}\Big(1-\sum_{m=0}^{+\infty}\sum_{n=0}^{+\infty} X^{\{,\},\text{4B}}_{m,n}(\bar{n})\Big), \label{eq:psucc_ant_4B_integral}
\end{align}
with
\begin{align}
    X^{[,]/\{,\},\text{QS}}_n(\bar{n})&= \int_{\mathcal{S}_{[,]/\{,\}}} \! \textup{d}\mu_{\left[,\right]/\{,\}}(U_A,U_B) \ \Re\left[\Tr\left( A_n^{(\bar{n},U_A)} U_B \rho_S A_n^{(\bar{n},U_A)\dagger} U_B^{\dagger} \right)\right], \label{eq:XnQS} \\
    X^{[,]/\{,\},\text{4B}}_{m,n}(\bar{n})&= \int_{\mathcal{S}_{[,]/\{,\}}} \! \textup{d}\mu_{\left[,\right]/\{,\}}(U_A,U_B) \ \Re\left[\big(c_m^{(\bar{n}/2)}\big)^* c_n^{(\bar{n}/2)} \Tr\left( A_m^{(\bar{n}/2,U_A)} U_B \rho_S A_n^{(\bar{n}/2,U_A)\dagger} U_B^{\dagger} \right)\right], \label{eq:Xmn4B}
\end{align}
where we used the notations $\mathcal{F}^{(\bar{n},U_A)}$, $\mathcal{G}^{(\bar{n},U_A)}$ and $A_n^{(\bar{n},U_A)}$ to indicate explicitly the ideal unitary operation $U_A$ that these correspond to.

To evaluate these expressions further, we can use the following parametrization for the unitaries in the sets $\mathcal{S}_{[,]}$, $\mathcal{S}_{\{,\}}$ and for their measures, in accordance with Eqs.~\eqref{eq:set_com}--\eqref{eq:set_anticom}:
\begin{align}
    \mathcal{S}_{[,]}: & \quad U_A = R_{\vec{\mathbf{u}}}(\theta_A) = R_\phi(\theta_A), \quad U_B = R_{\vec{\mathbf{u}}}(\theta_A) = R_\phi(\theta_B), \quad \int_{\mathcal{S}_{[,]}} \!\textup{d}\mu_{[,]}(U_A,U_B) = \int_{-\pi}^{\pi} \frac{d \phi}{2\pi} \int_{-\pi}^{\pi} \frac{d \theta_A}{2\pi} \int_{-\pi}^{\pi} \frac{d \theta_B}{2\pi}, \label{eq:param_Scom} \\
    \mathcal{S}_{\{,\}}: & \quad U_A = R_{\vec{\mathbf{u}}_A}(\pi) = R_{\phi_A}(\pi), \quad U_B = R_{\vec{\mathbf{u}}_B}(\pi) = R_{\phi_A^\perp,\varphi_B}(\pi), \quad \int_{\mathcal{S}_{\{,\}}} \!\textup{d}\mu_{\{,\}}(U_A,U_B) = \int_{-\pi}^{\pi} \frac{d \phi_A}{2\pi} \int_{-\pi}^{\pi} \frac{d \varphi_B}{2\pi}, \label{eq:param_Sant}
\end{align}

\end{widetext}
where $R_\phi(\theta) = e^{-i \frac{\theta}{2} ( \cos\phi \,\sigma_x+\sin\phi \,\sigma_y )}$ is again a rotation (in the Bloch sphere) of angle $\theta$ around an equatorial axis with azimuthal angle $\phi$, while $R_{\phi_A^\perp,\varphi_B}(\pi) = e^{-i\frac{\pi}{2}(\sin\phi_A\sin\varphi_B \, \sigma_x - \cos\phi_A\sin\varphi_B \, \sigma_y + \cos\varphi_B \, \sigma_z)}$ denotes a rotation of angle $\pi$ around an axis with zenithal angle $\varphi_B$ and azimuthal angle $\phi_A-\pi/2$ (so that it is orthogonal to the equatorial axis with azimuthal angle $\phi_A$). In Eqs.~\eqref{eq:XnQS}--\eqref{eq:Xmn4B}, $A_n^{(\bar{n},U_A)}$ is then given by Eq.~\eqref{eq:Kraus_An}, with the same angles $(\theta,\phi)$ as $U_A = R_\phi(\theta)$ in the parametrizations above.

With this the terms of $X^{[,]/\{,\},QS}_n(\bar{n})$ and $X^{[,]/\{,\},4B}_{m,n}(\bar{n})$ in Eqs.~\eqref{eq:XnQS}--\eqref{eq:Xmn4B} can be evaluated analytically, for any given state $\rho_S$.%
\footnote{Note that, as expected from the rotational symmetries of the two sets of unitaries around the $z$ axis of the Bloch sphere, the results only depend on the $z$ component of the state $\rho_S$, $\Tr[\sigma_z\rho_S]$ (as we see, in particular, in the asymptotic regime below).}
Their explicit forms are however rather tedious to write, so we omit them here, and we now focus on the asymptotic regime.

\subsection{Asymptotic regime}
\label{app:asymptotic_game}

Using $\Tr[{\cal M}(\varrho)\upsilon] = \Tr[\underline{M}\cdot(\varrho^T\otimes\upsilon)]$ and Eqs.~\eqref{eq:choi_Fnbar} and~\eqref{eq:choi_Gnbar} in Eqs.~\eqref{eq:psucc_com_QS_integral}--\eqref{eq:psucc_ant_4B_integral} (or using Eq.~\eqref{eq:Gnbar} directly in Eqs.~\eqref{eq:psucc_com_4B_integral}--\eqref{eq:psucc_ant_4B_integral}), and with the explicit parametrization of the sets $\mathcal{S}_{[,]}$ and $\mathcal{S}_{\{,\}}$ as in Eqs.~\eqref{eq:param_Scom}--\eqref{eq:param_Sant} above, we can evaluate the average success probabilities in the large-$\bar{n}$ limit. We obtain
\begin{align}
   \langle p_{\text{success}}^{\text{QS},[,]} \rangle & = 1-\frac{1}{16 \bar{n}}+O\left( \frac{1}{\bar{n}^2} \right),\\
   \langle p_{\text{success}}^{\text{QS},\{,\}} \rangle & = 1-\frac{2+\pi^2}{16\bar{n}}+O\left( \frac{1}{\bar{n}^2} \right), \\
   \langle p_{\text{success}}^{\text{4B},[,]} \rangle & = 1-\frac{2+\pi^2/3-\Tr[\sigma_z\rho_S]}{16\bar{n}}+O\left( \frac{1}{\bar{n}^2} \right),\\
   \langle p_{\text{success}}^{\text{4B},\{,\}} \rangle & = 1-\frac{4+\pi^2}{16 \bar{n}}+O\left( \frac{1}{\bar{n}^2} \right),
\end{align}
from which we then get
\begin{align}
   &\langle p_{\text{success}}^{\text{QS}} \rangle  = 1-\frac{3+\pi^2}{32\bar{n}}+O\left( \frac{1}{\bar{n}^2} \right), \\
   &\langle p_{\text{success}}^{\text{4B}} \rangle  = 1-\frac{6+4\pi^2/3-\Tr[\sigma_z\rho_S]}{32 \bar{n}}+O\!\left( \frac{1}{\bar{n}^2} \right), \\
   &\langle p_{\text{success}}^{\text{QS}} \rangle - \langle p_{\text{success}}^{\text{4B}} \rangle  = \frac{3+\pi^2/3-\Tr[\sigma_z\rho_S]}{32 \bar{n}}+O\left( \frac{1}{\bar{n}^2} \right).
\end{align}
We thus see that for the task we considered, with the prescribed sets of commuting or anticommuting operations $(U_A,U_B)$, the QS always performs slightly better than the 4B in the asymptotic regime, whatever the initial state $\rho_S$ of the target system. The difference between the two is maximized for $\rho_S = \ketbra{1}{1}$ and minimized for $\rho_S = \ketbra{0}{0}$; for our comparison in the main text (and in Fig.~\ref{avg} in particular) we take the average situation with $\rho_S = \id/2$. 

\section{Circuits with fixed causal order}
\label{app:FCOs}

\subsection{Probabilistic circuits representation}
\label{app:proba_circuit}

The most general circuit that applies two operations (CPTP maps) ${\cal A}$ and ${\cal B}$ in a fixed causal order with ${\cal A}$ preceding ${\cal B}$ (each being applied once and only once), and that produces a binary classical outcome, is depicted on Fig.~\ref{comb} of the main text. It consists in the composition of ${\cal A}$ and ${\cal B}$ with three fixed operations. The first of these operations initalizes the target system as well as a ``memory'' system in some state $\rho \in \mathcal{L}(\mathcal{H}^{A_Ia})$, where  $\mathcal{H}^{A_I}$ is the input Hilbert space of operation ${\cal A}$, $\mathcal{H}^{a}$ is some memory Hilbert space, and where we use the short-hand notation $\mathcal{H}^{XY} = \mathcal{H}^{X} \otimes \mathcal{H}^{Y}$. The second fixed operation $\mathcal{C}$ is a channel (a CPTP map) that connects the ouptut space $\mathcal{H}^{A_O}$ of ${\cal A}$ and the memory space $\mathcal{H}^{a}$ to the input space $\mathcal{H}^{B_I}$ of ${\cal B}$ and some other memory space $\mathcal{H}^{b}$. After operation ${\cal B}$ is applied, the output state of the target and memory systems is finally measured by the third fixed operation, namely a POVM $(E_+,E_-)$.
The probabilities for each outcome $\pm$ are, according to the Born rule:
\begin{align}
    p(\pm|{\cal A},{\cal B}) = \Tr\big[E_\pm \, [({\cal B}\otimes{\cal I}^b)\circ{\cal C}\circ({\cal A}\otimes{\cal I}^a)](\rho)\big],
\end{align}
where $\mathcal{I}^{a/b}$ is the identity channel on the memory space $\mathcal{H}^{a/b}$. It is easily verified that these probabilities can be written in terms of the Choi matrices of the various maps (defined as in Eq.~\eqref{eq:def_Choi}) as in Eq.~\eqref{eq:probas_FCO} of the main text, namely, as
\begin{align}
    p(\pm|{\cal A},{\cal B}) = \Tr\big[W_\pm^T(\underline{A}\otimes\underline{B})\big]
\end{align}
with
\begin{align}
    W_\pm = & \Tr_{ab} \big[ (E_\pm^T\!\otimes\!\id^{A_IA_Oa B_I}) \notag \\[-1mm]
    & \hspace{18mm} (\underline{C}^{T_{ab}}\!\otimes\!\id^{A_IB_O}) (\rho\!\otimes\!\id^{A_OB_IB_Ob}) \big] \label{eq:def_Wpm}
\end{align}
and where $\Tr_{ab}$ denotes the partial trace over the memory systems in ${\cal H}^a$ and ${\cal H}^b$, ${}^{T_{ab}}$ denotes the partial transpose,%
\footnote{The transpose and partial transpose are taken in the ``computational basis'', used to define the Choi representation (see Appendix~\ref{app:asymptotic_behaviours}).}
and $\id^{X}$ is the identity operator in the spaces indicated as a superscript.
More technically speaking: $W_\pm$ is obtained as the so-called ``link product''~\cite{Chiribella08,Chiribella09} of the Choi matrices of the elements $\rho, {\cal C}, E_\pm$ of the FCO circuit.%
\footnote{For the familiar reader: in terms of the link product $*$, $p(\pm|{\cal A},{\cal B}) = E_\pm^T * \underline{B} * \underline{C} * \underline{A} * \rho = (E_\pm^T * \underline{C} * \rho) * (\underline{B} * \underline{A}) = W_\pm * (\underline{A} \otimes \underline{B})$, with $W\pm = E_\pm^T * \underline{C} * \rho$.}

From Eq.~\eqref{eq:def_Wpm}, and using the facts that all operators are positive semidefinite (PSD), that $E_+ + E_- = \id^{B_Ob}$ and that $\Tr_{B_Ib} \underline{C} = \id^{A_Oa}$ (which translates, into the Choi representation, the fact that the channel ${\cal C}$ is trace-preserving), one can easily verify that the pair $(W_+,W_-)$ satisfies
\begin{align}
    & W_\pm \ge 0, \quad W_+ + W_- = W^{A_IA_OB_I} \otimes \id^{B_O}, \notag \\
    & \Tr_{B_I} W^{A_IA_OB_I} = W^{A_I} \otimes \id^{A_O}, \quad \Tr W^{A_I} = 1, \label{eq:cstr_Wpm}
\end{align}
for some PSD matrices $W^{A_IA_OB_I} (= \Tr_{ab} \big[ (\underline{C}^{T_{ab}}\otimes\id^{A_I})(\rho\otimes\id^{A_OB_Ib}) \big])$ and $W^{A_I} (= \Tr_a \rho)$.
As it turns out, the converse is also true: any pair $(W_+,W_-)$---a so-called ``quantum tester''---satisfying Eq.~\eqref{eq:cstr_Wpm} for some PSD matrices $W^{A_IA_OB_I} \in {\cal L}({\cal H}^{A_IA_OB_I})$ and $W^{A_I} \in {\cal L}({\cal H}^{A_I})$ can also be obtained from a quantum circuit with fixed causal order of the form of Fig.~\ref{comb}, for some appropriate choice of $\rho, {\cal C}, (E_+,E_-)$~\cite{Chiribella08,Chiribella09}. Hence, optimizing over all possible FCO circuits amounts to optimizing over pairs of operators $(W_+,W_-)$ satisfying the constraints above---or similar constraints for the order where ${\cal B}$ comes before ${\cal A}$.

\subsection{Success probabilities at the commuting-vs-anticommuting discrimination task}
\label{app:subsec:psucc_FCO}

In the task under consideration, the outcomes $\pm$ of the POVM $(E_+, E_-)$, or equivalently of the quantum tester $(W_+, W_-)$, correspond to the guess that the ideal operations $U_A$ and $U_B$ under consideration commute or anticommute, which leads to the form of Eq.~\eqref{eq:Psucc_FCO} for the success probability $p_{\text{success}}^{\text{FCO}}$. Optimizing it over all circuits with fixed causal order---i.e., over all testers $(W_+,W_-)$ satisfying the relevant constraints---is then a semidefinite programming (SDP) problem~\cite{araujo_witnessing_2015}, which (for some fixed operators $G_\pm$) can be solved efficiently.

The operators $G_+$ and $G_-$ from Eqs.~\eqref{eq:def_Gp}--\eqref{eq:def_Gm} are obtained from the Choi representations of the maps $\mathcal{A}$ and $\mathcal{B}$, and from the definitions of the sets $\mathcal{S}_{[,]}$ and $\mathcal{S}_{\{,\}}$ given in Eqs.~\eqref{eq:set_com} and~\eqref{eq:set_anticom}, which can be parametrized as in Eqs.~\eqref{eq:param_Scom}--\eqref{eq:param_Sant}.
Recall that in the scenario we consider, the operations $\mathcal{B}$ are always taken to be unitary, of the form $\mathcal{B}: \rho\mapsto U_B\rho U_B^\dagger$. On the other hand, in the finite-energy case the operations $\mathcal{A}$ are obtained from the Kraus operators $A_n^{(\bar{n})}$ of Eq.~\eqref{eq:Kraus_An}, according to $\mathcal{A}: \rho\mapsto \sum_n A_n^{(\bar{n})}\rho A_n^{(\bar{n})\,\dagger}$. These are meant to approximate the unitary operations $\mathcal{A}: \rho\mapsto U_A\rho U_A^\dagger$, which are reached only in the infinite-energy limit.

\medskip

Let us derive the explicit forms of $G_+$ and $G_-$ in the infinite-energy limit, precisely. Writing (according to Eqs.~\eqref{eq:def_Gp}--\eqref{eq:def_Gm} and~\eqref{eq:param_Scom}--\eqref{eq:param_Sant}, and in terms of the Choi matrices $\underline{R}_\phi(\theta)$ and $\underline{R}_{\phi_A^\perp,\varphi_B}(\pi)$ of the CPTP maps corresponding to the rotations $R_\phi(\theta)$ and $R_{\phi_A^\perp,\varphi_B}(\pi)$)
\begin{align}
    G_+ = & \int_{-\pi}^{\pi} \frac{d \phi}{2\pi} \int_{-\pi}^{\pi} \frac{d \theta_A}{2\pi} \int_{-\pi}^{\pi} \frac{d \theta_B}{2\pi} \underline{R}_\phi(\theta_A) \otimes \underline{R}_\phi(\theta_B), \\
    G_- = & \int_{-\pi}^{\pi} \frac{d \phi_A}{2\pi} \int_{-\pi}^{\pi} \frac{d \varphi_B}{2\pi} \underline{R}_{\phi_A}(\pi) \otimes \underline{R}_{\phi_A^\perp,\varphi_B}(\pi),
\end{align}
after some calculations one finds (written in the space $\mathcal{L}(\mathcal{H}^{A_I})\otimes\mathcal{L}(\mathcal{H}^{A_O})\otimes\mathcal{L}(\mathcal{H}^{B_I})\otimes\mathcal{L}(\mathcal{H}^{B_O})$, with implicit tensor products)
\begin{align}
    G_+ = & \frac14 \Big[ \big(\id\id+{\textstyle\frac12}\sigma_x\sigma_x-{\textstyle\frac12}\sigma_y\sigma_y\big)\big(\id\id+{\textstyle\frac12}\sigma_x\sigma_x-{\textstyle\frac12}\sigma_y\sigma_y\big) \notag \\
    & \qquad + {\textstyle\frac{1}{8}} \big(\sigma_x\sigma_x+\sigma_y\sigma_y\big)\big(\sigma_x\sigma_x+\sigma_y\sigma_y\big) \notag \\
    & \qquad + {\textstyle\frac{1}{8}} \big(\sigma_x\sigma_y-\sigma_y\sigma_x\big)\big(\sigma_x\sigma_y-\sigma_y\sigma_x\big) \Big] , \label{eq:Gp_explicit} \\
    G_- = & \frac14 \Big[ \big(\id\id-\sigma_z\sigma_z\big)\big(\id\id-{\textstyle\frac12}\sigma_x\sigma_x+{\textstyle\frac12}\sigma_y\sigma_y\big) \notag \\
    & \qquad - {\textstyle\frac{1}{4}} \big(\sigma_x\sigma_x+\sigma_y\sigma_y\big)\big(\sigma_x\sigma_x+\sigma_y\sigma_y\big) \notag \\
    & \qquad - {\textstyle\frac{1}{4}} \big(\sigma_x\sigma_y-\sigma_y\sigma_x\big)\big(\sigma_x\sigma_y-\sigma_y\sigma_x\big) \Big] . \label{eq:Gm_explicit}
\end{align}

With these operators $G_\pm$, optimizing $p_{\text{success}}^{\text{FCO}}$ from Eq.~\eqref{eq:Psucc_FCO} under the constraints of Eq.~\eqref{eq:cstr_Wpm}, i.e.\ for FCO circuits with ${\cal A}$ before ${\cal B}$, we found an optimal success probablity $p^{\text{FCO}}_{\text{success}}\simeq0.9489$.

However, for the analogous constraints corresponding to FCO circuits with ${\cal B}$ before ${\cal A}$, we found $p^{\text{FCO}}_{\text{success}}=1$. Indeed there exists such a circuit that allows one to discriminate perfectly commuting pairs $(U_A,U_B)$ drawn from $\mathcal{S}_{[,]}$, from anticommuting pairs drawn from $\mathcal{S}_{\{,\}}$. This circuit can be reconstructed from the results of the SDP optimization; it is described in the next section.

\begin{figure}
\centering   
\includegraphics[width=80mm]{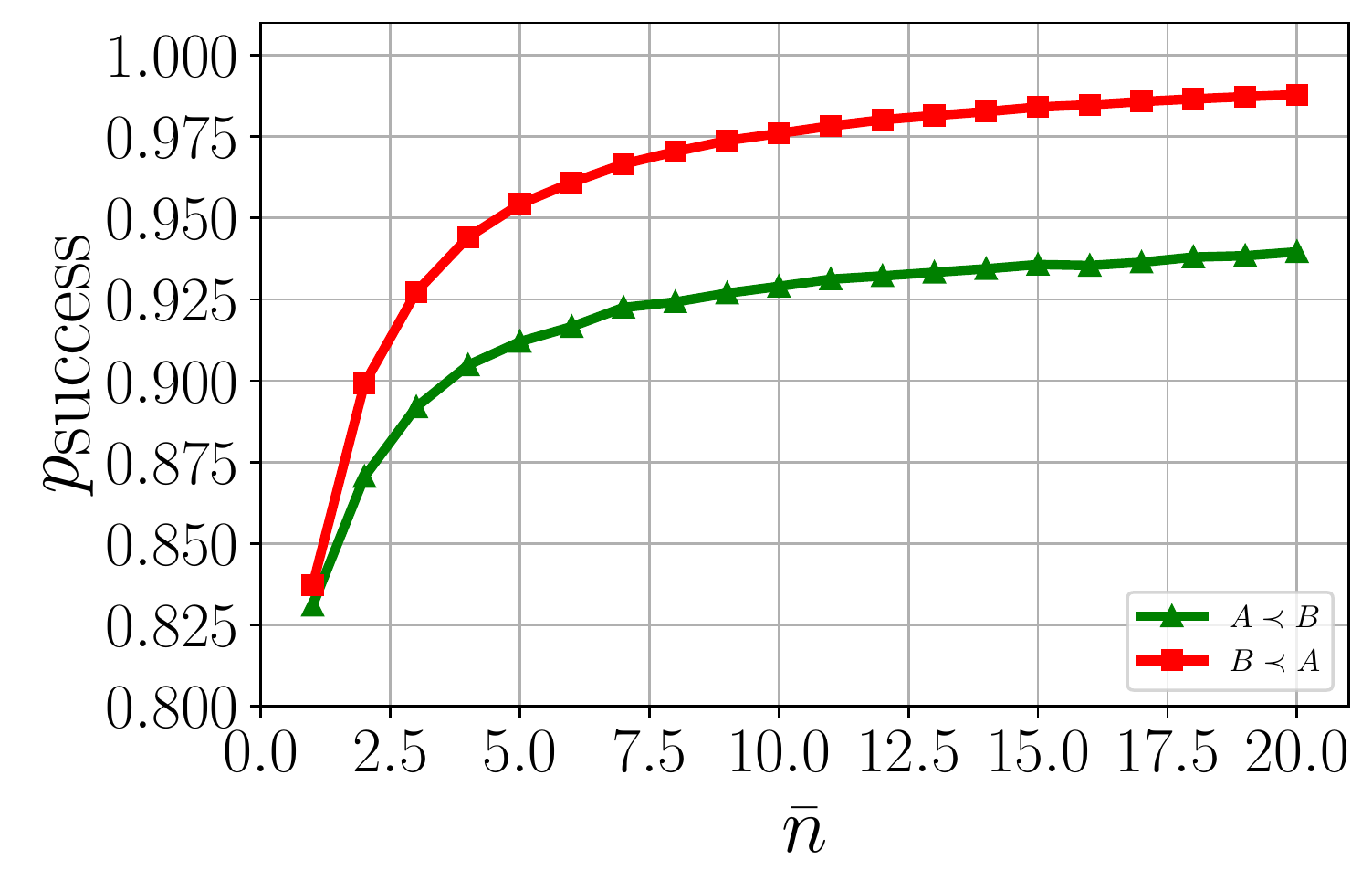}
\caption{Success probabilities at our discrimination task for FCO circuits with ${\cal A}$ before ${\cal B}$ ($A\prec B$) or ${\cal B}$ before ${\cal A}$ ($B \prec A$), for $1 \leq \bar{n} \leq 20$.}
\label{fig:AB_BA_finite_nbar}
\end{figure}

The success probabilities for FCO circuits in the finite-energy regime are shown in Fig.~\ref{fig:AB_BA_finite_nbar}. For both fixed orders between ${\cal A}$ and ${\cal B}$, we see that these success probabilities decrease when $\bar{n}$ decreases, as the operation ${\cal A}$ is more and more noisy. As in the ideal unitary case, the best FCO circuits with ${\cal B}$ before ${\cal A}$ are found to outperform the best circuits with ${\cal A}$ before ${\cal B}$.

\subsection{Optimal FCO circuit for unitary operations $U_A$ and $U_B$ from $\mathcal{S}_{[,]}$ or $\mathcal{S}_{\{,\}}$}
\label{combBA}

As just claimed, one can find a FCO circuit with ${\cal B}$ before ${\cal A}$ which correctly guesses whether $U_A$ and $U_B$ are drawn from $\mathcal{S}_{[,]}$ or $\mathcal{S}_{\{,\}}$ (when both ${\cal A}$ and ${\cal B}$ implement $U_A$ and $U_B$ perfectly, i.e., in the infinite-energy limit). This circuit is of the form depicted on Fig.~\ref{comb} of the main text, with the roles of ${\cal A}$ and ${\cal B}$ being exchanged.

Specifically, the input state $\rho$ can be taken to be a maximally entangled state
\begin{align}
    \rho = \ketbra{\Phi^+}{\Phi^+}^{B_Ib}
\end{align}
with $\ket{\Phi^+}^{B_Ib}=\frac{1}{\sqrt{2}}(\ket{00}^{B_Ib}+\ket{11}^{B_Ib})$, where we introduced a 2-dimensional memory space ${\cal H}^b$, while the channel ${\cal C}$ can be taken to be an isometric channel ${\cal C}: \rho \mapsto C \rho C^\dagger$ with
\begin{align}
C = & \ket{\Phi^+}^{A_Ia_1}\ket{0}^{a_2}\bra{00}^{B_Ob} + \ket{\Phi^+}^{A_Ia_1}\ket{1}^{a_2}\bra{11}^{B_Ob} \notag \\
& + \ket{01}^{A_Ia_1}\ket{2}^{a_2}\bra{01}^{B_Ob} + \ket{10}^{A_Ia_1}\ket{2}^{a_2}\bra{10}^{B_Ob},
\end{align}
where we introduced two more memory spaces: a 2-dimensional space ${\cal H}^{a_1}$ and a 3-dimensional space ${\cal H}^{a_2}$.

Let us indeed check that these choices allow one to solve the task perfectly. For any unitary operations $U_A$ and $U_B$, the output state of the circuit before the POVM is $\ket{\psi}^{A_Oa_1a_2} = (U_A\otimes \id^{a_1a_2})C(U_B\otimes\id^b)\ket{\Phi^+}$. Considering that either $U_A = R_\phi(\theta_A), U_B = R_\phi(\theta_B)$ if these are drawn from $\mathcal{S}_{[,]}$, or $U_A = R_{\phi_A}(\pi), U_B = R_{\phi_A^\perp,\varphi_B}(\pi)$ if these are drawn from $\mathcal{S}_{\{,\}}$ (see Eqs.~\eqref{eq:param_Scom}--\eqref{eq:param_Sant}), the corresponding output states are easily calculated to be
\begin{align}
    & \ket{\psi_{[,]}} = \cos\!{\textstyle\frac{\theta_B}{2}} \Big( \!\cos\!{\textstyle\frac{\theta_A}{2}} \!\ket{\Phi^+}^{A_Oa_1} \!-\! i \sin\!{\textstyle\frac{\theta_A}{2}} \!\ket{\Psi_{\phi}}^{A_Oa_1} \!\Big)\!\ket{+}^{a_2} \notag \\
    & \quad \qquad - \sin\!{\textstyle\frac{\theta_B}{2}} \Big( \!\sin\!{\textstyle\frac{\theta_A}{2}} \!\ket{\Phi^+}^{A_Oa_1} \!+\! i \cos\!{\textstyle\frac{\theta_A}{2}} \!\ket{\Psi_{\phi}}^{A_Oa_1} \!\Big)\!\ket{2}^{a_2}\!\!, \notag \\[1mm]
    & \ket{\psi_{\{,\}}} = -\cos\varphi_B \ket{\Psi_{\phi_A}}^{A_Oa_1} \ket{-}^{a_2} \notag \\
    & \qquad \qquad + i \sin\varphi_B \ket{\Phi^-}^{A_Oa_1}\ket{2}^{a_2}
\end{align}
with $\ket{\Phi^\pm}^{A_Oa_1} = \frac{1}{\sqrt{2}} ( \ket{00}^{A_Oa_1} \pm \ket{11}^{A_Oa_1})$, $\ket{\Psi_{\phi}}^{A_Oa_1} = \frac{1}{\sqrt{2}} ( e^{-i\phi} \ket{01}^{A_Oa_1} + e^{i\phi} \ket{10}^{A_Oa_1})$ and $\ket{\pm}^{a_2} = \frac{1}{\sqrt{2}} (\ket{0}^{a_2} \pm \ket{1}^{a_2})$. From these expressions we can clearly see (using in particular that $\ket{\Phi^-}$ is orthogonal to both $\ket{\Phi^+}$ and $\ket{\Psi_{\phi}}$) that $\ket{\psi_{[,]}}$ and $\ket{\psi_{\{,\}}}$ are orthogonal (whatever the values of $\phi,\theta_A,\theta_B,\phi_A,\varphi_B$), so that one can find a POVM that discriminates the 2 states---and hence, the commuting and anticommuting cases---perfectly.

\subsection{Isotropic FCO circuits}
\label{app:isoFCO}

Despite the previous finding, it is known that no FCO circuit can perfectly discriminate between any general pairs of either commuting or anticommuting unitaries~\cite{chiribella12,araujo_witnessing_2015}. The existence of a FCO circuit that discriminates perfectly between pairs in $\mathcal{S}_{[,]}$ or $\mathcal{S}_{\{,\}}$ is due to the fact that these sets are restricted to certain orientations of the unitaries (e.g., the rotation axes of all $U_A$'s are in the equatorial plane of the Bloch sphere).

\begin{figure}
\centering   
\includegraphics[width=\columnwidth]{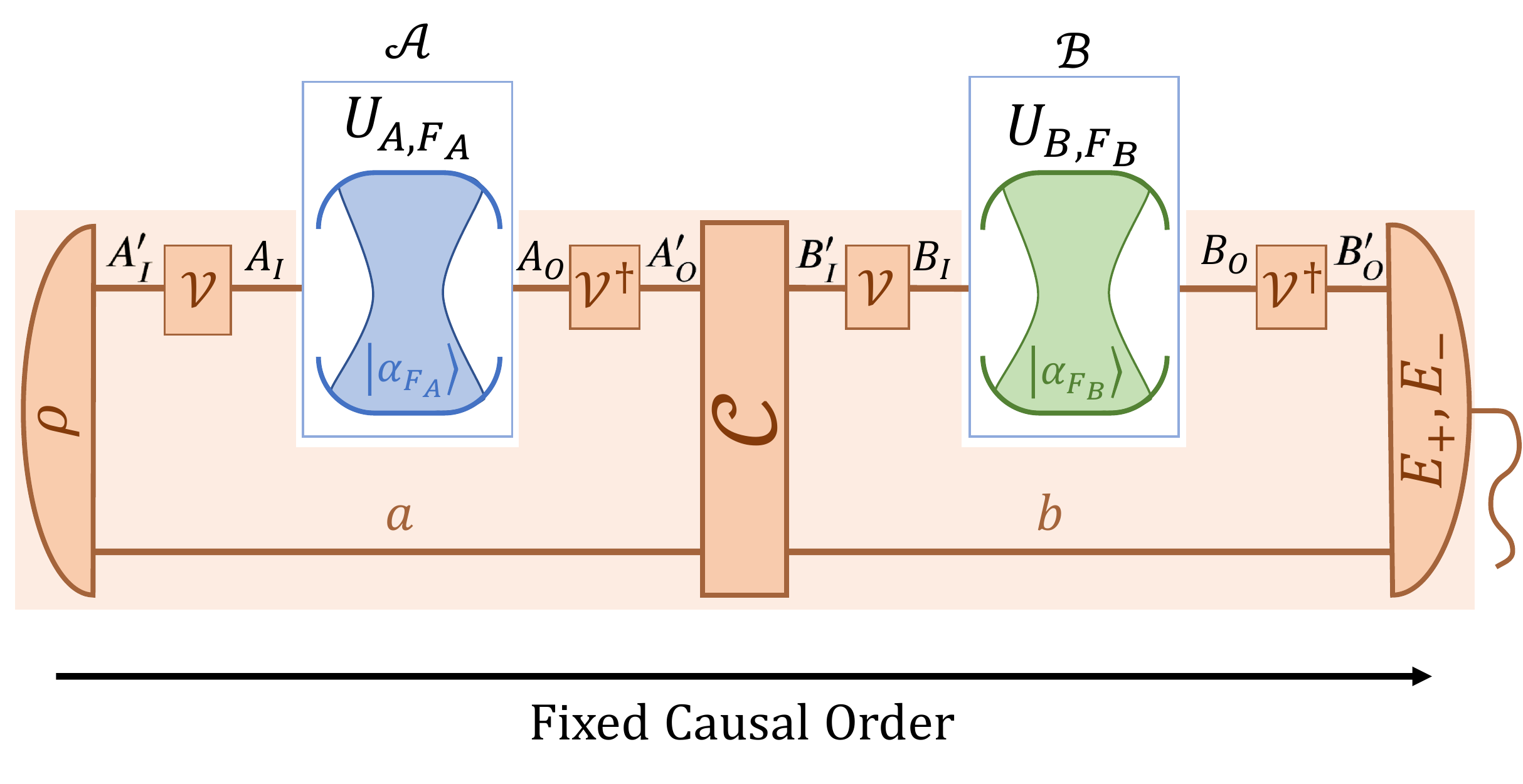}
\caption{Isotropic FCO circuits are required, for any fixed operations ${\cal A}$ and ${\cal B}$, to provide the same measurement statistics whatever the unitary operations ${\cal V}$ and ${\cal V}^\dagger$ inserted at each input and output port.}
\label{fig:isotropic_FCO}
\end{figure}

As discussed in the main text, it is also insightful to see how ``isotropic'' FCO circuits, which cannot take advantage of any specific orientation of the unitaries, perform at the discrimination task. By such circuits, we mean circuits of the form of Fig.~\ref{comb} (or with ${\cal B}$ before ${\cal A}$) which are required, for any fixed operations (any CP maps) ${\cal A}$ and ${\cal B}$, to act in the same way on $({\cal A},{\cal B})$ as on any operations $({\cal V}^\dagger\circ{\cal A}\circ{\cal V},{\cal V}^\dagger\circ{\cal B}\circ{\cal V})$, for any unitary channels ${\cal V}:\rho\mapsto V\rho V^\dagger$ and ${\cal V}^\dagger:\rho\mapsto V^\dagger\rho V$ (where $V$ is some unitary operator): see Fig.~\ref{fig:isotropic_FCO}.

Isotropic FCO circuits can most generally be obtained as in Fig.~\ref{fig:isotropic_FCO}, by starting from any FCO circuit and averaging over the unitaries $V$ sampled according to the Haar measure $d\mu_\text{Haar}(V)$. Technically speaking, in the case of circuits (or testers) with binary outcomes as considered here, these are of the form
\begin{align}
    W_\pm^\text{iso} = & \Tr_{A_I'A_O'B_I'B_O'}\Big[\big(W_\pm^{A_I'A_O'B_I'B_O'}\otimes\id^{A_IA_OB_IB_O}\big)^T \notag \\[-1mm]
    & \hspace{40mm} \Tilde{H}^{A_I'A_IA_OA_O'B_I'B_IB_OB_O'}\Big] \label{eq:Wiso_v0}
\end{align}
for some tester $(W_\pm^{A_I'A_O'B_I'B_O'})$ satisfying Eq.~\eqref{eq:cstr_Wpm} (in the primed spaces, introduced as in Fig.~\ref{fig:isotropic_FCO}) and with the Haar-randomized operator
\begin{align}
    & \Tilde{H}^{A_I'A_IA_OA_O'B_I'B_IB_OB_O'} \notag \\
    & = \int d\mu_\text{Haar}(V) \ \underline{V}^{A_I'A_I}\!\otimes\!\underline{V^\dagger}^{A_OA_O'}\!\otimes\!\underline{V}^{B_I'B_I}\!\otimes\!\underline{V^\dagger}^{B_OB_O'},
\end{align}
where $\underline{V}$ and $\underline{V^\dagger}$ are the Choi matrices of the unitary maps ${\cal V}$ and ${\cal V}^\dagger$ introduced above.%
\footnote{Note that in general $\underline{V^\dagger} \neq (\underline{V})^\dagger = \underline{V}$.}

The operator $\Tilde{H}^{A_I'A_IA_OA_O'B_I'B_IB_OB_O'}$ can be calculated explicitly, but is too long and too tedious to write here. It is however possible to further simplify the characterization above: one indeed finds that matrices of the form of Eq.~\eqref{eq:Wiso_v0} are just restricted to be in the (only 14-dimensional) subspace
\begin{widetext}
\begin{align}
    {\cal L}_\text{iso} = \text{span} \big( & \id^{A_IA_OB_IB_O}, \kketbra{\id}{\id}^{A_IA_O}\otimes\id^{B_IB_O}, \id^{A_IA_O}\otimes\kketbra{\id}{\id}^{B_IB_O}, \kketbra{\id}{\id}^{A_IA_O}\otimes\kketbra{\id}{\id}^{B_IB_O}, \notag \\
    & \kketbra{\id}{\id}^{A_IB_O}\otimes\id^{B_IA_O}, \id^{A_IB_O}\otimes\kketbra{\id}{\id}^{B_IA_O}, \kketbra{\id}{\id}^{A_IB_O}\otimes\kketbra{\id}{\id}^{B_IA_O}, \notag \\
    & \kketbra{\sigma_y}{\sigma_y}^{A_IB_I}\otimes\id^{A_OB_O}, \id^{A_IB_I}\otimes\kketbra{\sigma_y}{\sigma_y}^{A_OB_O}, \kketbra{\sigma_y}{\sigma_y}^{A_IB_I}\otimes\kketbra{\sigma_y}{\sigma_y}^{A_OB_O}, \notag \\
    & (\sigma_x\sigma_y\sigma_z+\sigma_x\sigma_z\sigma_y+\sigma_y\sigma_x\sigma_z-\sigma_z\sigma_y\sigma_x-\sigma_z\sigma_x\sigma_y-\sigma_y\sigma_z\sigma_x)^{A_IA_OB_I}\id^{B_O}, \notag \\
    & (\sigma_x\sigma_y\sigma_z+\sigma_x\sigma_z\sigma_y+\sigma_y\sigma_x\sigma_z-\sigma_z\sigma_y\sigma_x-\sigma_z\sigma_x\sigma_y-\sigma_y\sigma_z\sigma_x)^{A_OB_IB_O}\id^{A_I}, \notag \\
    & (\sigma_x\sigma_y\sigma_z+\sigma_x\sigma_z\sigma_y+\sigma_y\sigma_x\sigma_z-\sigma_z\sigma_y\sigma_x-\sigma_z\sigma_x\sigma_y-\sigma_y\sigma_z\sigma_x)^{B_IB_OA_I}\id^{A_O}, \notag \\
    & (\sigma_x\sigma_y\sigma_z+\sigma_x\sigma_z\sigma_y+\sigma_y\sigma_x\sigma_z-\sigma_z\sigma_y\sigma_x-\sigma_z\sigma_x\sigma_y-\sigma_y\sigma_z\sigma_x)^{B_OA_IA_O}\id^{B_I} \big)
\end{align}
\end{widetext}
with $\kket{\id} = \ket{00} + \ket{11}$, $\kket{\sigma_y} = (\id\otimes\sigma_y)\kket{\id} = i ( \ket{01} - \ket{10} )$ and with implicit tensor products on the last four lines. 

All in all, we thus find that isotropic FCO testers are simply required to satisfy Eq.~\eqref{eq:cstr_Wpm} (or the analogous conditions for ${\cal B}$ before ${\cal A}$), with the additional constraint that $W_\pm^\text{iso}\in{\cal L}_\text{iso}$.
Optimizing Eq.~\eqref{eq:Psucc_FCO} under these constraints gives, in the infinite-energy limit, an optimal probability of success $p^{\text{FCO}}_{\text{success}}\simeq0.9288$ for isotropic FCO circuits, for both orders where ${\cal A}$ comes before ${\cal B}$ and where ${\cal B}$ comes before ${\cal A}$. The results for the finite-energy regime are shown on Fig.~\ref{avg} of the main text. As it turns out, we find here, for $1\leq \bar{n} \leq 20$, that FCO circuits with ${\cal B}$ before ${\cal A}$ outperform slightly those with ${\cal A}$ before ${\cal B}$ (with differences in the success probabilities of the order of $10^{-2}$).

\medskip

Note, finally, that testing the performance of isotropic FCO circuits at discriminating between pairs of unitaries $(U_A,U_B)$ in $\mathcal{S}_{[,]}$ or $\mathcal{S}_{\{,\}}$ is equivalent to testing the performance of general FCO circuits at discriminating between pairs of unitaries of the form $(V^\dagger U_AV,V^\dagger U_BV)$ with $(U_A,U_B)$ in $\mathcal{S}_{[,]}$ or $\mathcal{S}_{\{,\}}$ and with $V$ a random unitary drawn according to the Haar measure (as both cases correspond to the same physical situation of Fig.~\ref{fig:isotropic_FCO}). To analyse the latter situation, one can simply replace the sets $\mathcal{S}_{[,]}$, $\mathcal{S}_{\{,\}}$ considered so far by the thus obtained sets $\mathcal{S}_{[,]}'$, $\mathcal{S}_{\{,\}}'$ of such pairs, that include the Haar-randomization. Note that $(V^\dagger U_AV,V^\dagger U_BV)$ have the same commuting or anticommuting property as $(U_A,U_B)$, so that the interpretation of the task in terms of a commuting versus anticommuting discrimination problem is preserved; note also that both the QS and 4B would give the same probabilities of success for $\mathcal{S}_{[,]}'$, $\mathcal{S}_{\{,\}}'$ as for $\mathcal{S}_{[,]}$, $\mathcal{S}_{\{,\}}$ when the target system is initialized in the state $\rho_S = \id/2$. However, by inserting the Haar-random $V$'s we lose the physical motivation coming from the Jaynes-Cummings model, which led us to restrict $U_A$ to rotations around an equatorial axis.

Replacing $\mathcal{S}_{[,]}$ and $\mathcal{S}_{\{,\}}$ by $\mathcal{S}_{[,]}'$ and $\mathcal{S}_{\{,\}}'$ in Eqs.~\eqref{eq:def_Gp}--\eqref{eq:def_Gm}, the $G_\pm$ operators calculated previously, in the infinite-energy limit (cf.\ Eqs.~\eqref{eq:Gp_explicit}--\eqref{eq:Gm_explicit}), become
\begin{align}
    G_+' = & \frac14 \Big[ \id + {\textstyle\frac13} \id^{A_IA_O}\mathfrak{S}_1^{B_IB_O} + {\textstyle\frac13} \mathfrak{S}_1^{A_IA_O}\id^{B_IB_O} \notag \\
    & \qquad + {\textstyle\frac{1}{15}} \mathfrak{S}_1^{A_IA_O}\mathfrak{S}_1^{B_IB_O} + {\textstyle\frac{1}{15}} \mathfrak{S}_2^{A_IA_OB_IB_O} \Big] , \\
    G_-' = & \frac14 \Big[ \id - {\textstyle\frac13} \id^{A_IA_O}\mathfrak{S}_1^{B_IB_O} - {\textstyle\frac13} \mathfrak{S}_1^{A_IA_O}\id^{B_IB_O} \notag \\
    & \qquad + {\textstyle\frac{1}{5}} \mathfrak{S}_1^{A_IA_O}\mathfrak{S}_1^{B_IB_O} - {\textstyle\frac{2}{15}} \mathfrak{S}_2^{A_IA_OB_IB_O} \Big]
\end{align}
with $\mathfrak{S}_1 = \sum_{i=x,y,z} \sigma_i\sigma_i^T$, $\mathfrak{S}_2 = \sum_{i,j=x,y,z} \sigma_i\sigma_j^T\sigma_i\sigma_j^T + \sigma_i\sigma_j^T\sigma_j\sigma_i^T$.
(It can be verified that $G_\pm' \in {\cal L}_\text{iso}$, as expected.)
Optimizing Eq.~\eqref{eq:Psucc_FCO} for these operators, over all pairs $(W_+,W_-)$ satisfying Eq.~\eqref{eq:cstr_Wpm}, we indeed find again $p^{\text{FCO}}_{\text{success}}\simeq0.9288$ (for both orders between ${\cal A}$ and ${\cal B}$), as in the case where we restricted to isotropic FCO circuits above.

\bibliographystyle{apsrev4-2} 
\bibliography{bibliography}

\begin{thebibliography}{52}%
\makeatletter
\providecommand \@ifxundefined [1]{%
 \@ifx{#1\undefined}
}%
\providecommand \@ifnum [1]{%
 \ifnum #1\expandafter \@firstoftwo
 \else \expandafter \@secondoftwo
 \fi
}%
\providecommand \@ifx [1]{%
 \ifx #1\expandafter \@firstoftwo
 \else \expandafter \@secondoftwo
 \fi
}%
\providecommand \natexlab [1]{#1}%
\providecommand \enquote  [1]{``#1''}%
\providecommand \bibnamefont  [1]{#1}%
\providecommand \bibfnamefont [1]{#1}%
\providecommand \citenamefont [1]{#1}%
\providecommand \href@noop [0]{\@secondoftwo}%
\providecommand \href [0]{\begingroup \@sanitize@url \@href}%
\providecommand \@href[1]{\@@startlink{#1}\@@href}%
\providecommand \@@href[1]{\endgroup#1\@@endlink}%
\providecommand \@sanitize@url [0]{\catcode `\\12\catcode `\$12\catcode
  `\&12\catcode `\#12\catcode `\^12\catcode `\_12\catcode `\%12\relax}%
\providecommand \@@startlink[1]{}%
\providecommand \@@endlink[0]{}%
\providecommand \url  [0]{\begingroup\@sanitize@url \@url }%
\providecommand \@url [1]{\endgroup\@href {#1}{\urlprefix }}%
\providecommand \urlprefix  [0]{URL }%
\providecommand \Eprint [0]{\href }%
\providecommand \doibase [0]{https://doi.org/}%
\providecommand \selectlanguage [0]{\@gobble}%
\providecommand \bibinfo  [0]{\@secondoftwo}%
\providecommand \bibfield  [0]{\@secondoftwo}%
\providecommand \translation [1]{[#1]}%
\providecommand \BibitemOpen [0]{}%
\providecommand \bibitemStop [0]{}%
\providecommand \bibitemNoStop [0]{.\EOS\space}%
\providecommand \EOS [0]{\spacefactor3000\relax}%
\providecommand \BibitemShut  [1]{\csname bibitem#1\endcsname}%
\let\auto@bib@innerbib\@empty
%</preamble>
\bibitem [{\citenamefont {Oreshkov}\ \emph {et~al.}(2012)\citenamefont
  {Oreshkov}, \citenamefont {Costa},\ and\ \citenamefont
  {Brukner}}]{oreshkov12}%
  \BibitemOpen
  \bibfield  {author} {\bibinfo {author} {\bibfnamefont {O.}~\bibnamefont
  {Oreshkov}}, \bibinfo {author} {\bibfnamefont {F.}~\bibnamefont {Costa}},\
  and\ \bibinfo {author} {\bibfnamefont {{\v{C}}.}~\bibnamefont {Brukner}},\
  }\href {https://doi.org/10.1038/ncomms2076} {\bibfield  {journal} {\bibinfo
  {journal} {Nat. Commun.}\ }\textbf {\bibinfo {volume} {3}},\ \bibinfo {pages}
  {1092} (\bibinfo {year} {2012})}\BibitemShut {NoStop}%
\bibitem [{\citenamefont {Chiribella}\ \emph {et~al.}(2013)\citenamefont
  {Chiribella}, \citenamefont {D'Ariano}, \citenamefont {Perinotti},\ and\
  \citenamefont {Valiron}}]{chiribella_quantum_2013}%
  \BibitemOpen
  \bibfield  {author} {\bibinfo {author} {\bibfnamefont {G.}~\bibnamefont
  {Chiribella}}, \bibinfo {author} {\bibfnamefont {G.~M.}\ \bibnamefont
  {D'Ariano}}, \bibinfo {author} {\bibfnamefont {P.}~\bibnamefont
  {Perinotti}},\ and\ \bibinfo {author} {\bibfnamefont {B.}~\bibnamefont
  {Valiron}},\ }\href {https://doi.org/10.1103/PhysRevA.88.022318} {\bibfield
  {journal} {\bibinfo  {journal} {Phys. Rev. A}\ }\textbf {\bibinfo {volume}
  {88}},\ \bibinfo {pages} {022318} (\bibinfo {year} {2013})}\BibitemShut
  {NoStop}%
\bibitem [{\citenamefont {Wechs}\ \emph {et~al.}(2021)\citenamefont {Wechs},
  \citenamefont {Dourdent}, \citenamefont {Abbott},\ and\ \citenamefont
  {Branciard}}]{wechs21}%
  \BibitemOpen
  \bibfield  {author} {\bibinfo {author} {\bibfnamefont {J.}~\bibnamefont
  {Wechs}}, \bibinfo {author} {\bibfnamefont {H.}~\bibnamefont {Dourdent}},
  \bibinfo {author} {\bibfnamefont {A.~A.}\ \bibnamefont {Abbott}},\ and\
  \bibinfo {author} {\bibfnamefont {C.}~\bibnamefont {Branciard}},\ }\href
  {https://doi.org/10.1103/PRXQuantum.2.030335} {\bibfield  {journal} {\bibinfo
   {journal} {PRX Quantum}\ }\textbf {\bibinfo {volume} {2}},\ \bibinfo {pages}
  {030335} (\bibinfo {year} {2021})}\BibitemShut {NoStop}%
\bibitem [{\citenamefont {Chiribella}(2012)}]{chiribella12}%
  \BibitemOpen
  \bibfield  {author} {\bibinfo {author} {\bibfnamefont {G.}~\bibnamefont
  {Chiribella}},\ }\href {https://doi.org/10.1103/PhysRevA.86.040301}
  {\bibfield  {journal} {\bibinfo  {journal} {Phys. Rev. A}\ }\textbf {\bibinfo
  {volume} {86}},\ \bibinfo {pages} {040301(R)} (\bibinfo {year}
  {2012})}\BibitemShut {NoStop}%
\bibitem [{\citenamefont {Ara{\'u}jo}\ \emph {et~al.}(2014)\citenamefont
  {Ara{\'u}jo}, \citenamefont {Costa},\ and\ \citenamefont
  {Brukner}}]{araujo_computational_2014}%
  \BibitemOpen
  \bibfield  {author} {\bibinfo {author} {\bibfnamefont {M.}~\bibnamefont
  {Ara{\'u}jo}}, \bibinfo {author} {\bibfnamefont {F.}~\bibnamefont {Costa}},\
  and\ \bibinfo {author} {\bibfnamefont {{\v C}.}~\bibnamefont {Brukner}},\
  }\href {https://doi.org/10.1103/PhysRevLett.113.250402} {\bibfield  {journal}
  {\bibinfo  {journal} {Phys. Rev. Lett.}\ }\textbf {\bibinfo {volume} {113}},\
  \bibinfo {pages} {250402} (\bibinfo {year} {2014})}\BibitemShut {NoStop}%
\bibitem [{\citenamefont {Facchini}\ and\ \citenamefont
  {Perdrix}(2015)}]{facchini15}%
  \BibitemOpen
  \bibfield  {author} {\bibinfo {author} {\bibfnamefont {S.}~\bibnamefont
  {Facchini}}\ and\ \bibinfo {author} {\bibfnamefont {S.}~\bibnamefont
  {Perdrix}},\ }in\ \href {https://doi.org/10.1007/978-3-319-17142-5_28} {\emph
  {\bibinfo {booktitle} {Theory and Applications of Models of Computation}}},\
  \bibinfo {editor} {edited by\ \bibinfo {editor} {\bibfnamefont
  {R.}~\bibnamefont {Jain}}, \bibinfo {editor} {\bibfnamefont {S.}~\bibnamefont
  {Jain}},\ and\ \bibinfo {editor} {\bibfnamefont {F.}~\bibnamefont
  {Stephan}}}\ (\bibinfo  {publisher} {Springer International Publishing},\
  \bibinfo {address} {Cham},\ \bibinfo {year} {2015})\ pp.\ \bibinfo {pages}
  {324--331}\BibitemShut {NoStop}%
\bibitem [{\citenamefont {Feix}\ \emph {et~al.}(2015)\citenamefont {Feix},
  \citenamefont {Ara\'ujo},\ and\ \citenamefont {Brukner}}]{feix15}%
  \BibitemOpen
  \bibfield  {author} {\bibinfo {author} {\bibfnamefont {A.}~\bibnamefont
  {Feix}}, \bibinfo {author} {\bibfnamefont {M.}~\bibnamefont {Ara\'ujo}},\
  and\ \bibinfo {author} {\bibfnamefont {{\v C}.}~\bibnamefont {Brukner}},\
  }\href {https://doi.org/10.1103/PhysRevA.92.052326} {\bibfield  {journal}
  {\bibinfo  {journal} {Phys. Rev. A}\ }\textbf {\bibinfo {volume} {92}},\
  \bibinfo {pages} {052326} (\bibinfo {year} {2015})}\BibitemShut {NoStop}%
\bibitem [{\citenamefont {Gu\'erin}\ \emph {et~al.}(2016)\citenamefont
  {Gu\'erin}, \citenamefont {Feix}, \citenamefont {Ara\'ujo},\ and\
  \citenamefont {Brukner}}]{guerin16}%
  \BibitemOpen
  \bibfield  {author} {\bibinfo {author} {\bibfnamefont {P.~A.}\ \bibnamefont
  {Gu\'erin}}, \bibinfo {author} {\bibfnamefont {A.}~\bibnamefont {Feix}},
  \bibinfo {author} {\bibfnamefont {M.}~\bibnamefont {Ara\'ujo}},\ and\
  \bibinfo {author} {\bibfnamefont {{\v C}.}~\bibnamefont {Brukner}},\ }\href
  {https://doi.org/10.1103/PhysRevLett.117.100502} {\bibfield  {journal}
  {\bibinfo  {journal} {Phys. Rev. Lett.}\ }\textbf {\bibinfo {volume} {117}},\
  \bibinfo {pages} {100502} (\bibinfo {year} {2016})}\BibitemShut {NoStop}%
\bibitem [{\citenamefont {Ebler}\ \emph {et~al.}(2018)\citenamefont {Ebler},
  \citenamefont {Salek},\ and\ \citenamefont {Chiribella}}]{Ebler2018}%
  \BibitemOpen
  \bibfield  {author} {\bibinfo {author} {\bibfnamefont {D.}~\bibnamefont
  {Ebler}}, \bibinfo {author} {\bibfnamefont {S.}~\bibnamefont {Salek}},\ and\
  \bibinfo {author} {\bibfnamefont {G.}~\bibnamefont {Chiribella}},\ }\href
  {https://doi.org/10.1103/PhysRevLett.120.120502} {\bibfield  {journal}
  {\bibinfo  {journal} {Phys. Rev. Lett.}\ }\textbf {\bibinfo {volume} {120}},\
  \bibinfo {pages} {120502} (\bibinfo {year} {2018})}\BibitemShut {NoStop}%
\bibitem [{\citenamefont {Salek}\ \emph {et~al.}(2018)\citenamefont {Salek},
  \citenamefont {Ebler},\ and\ \citenamefont {Chiribella}}]{salek2018quantum}%
  \BibitemOpen
  \bibfield  {author} {\bibinfo {author} {\bibfnamefont {S.}~\bibnamefont
  {Salek}}, \bibinfo {author} {\bibfnamefont {D.}~\bibnamefont {Ebler}},\ and\
  \bibinfo {author} {\bibfnamefont {G.}~\bibnamefont {Chiribella}},\ }\Eprint
  {https://arxiv.org/abs/1809.06655} {arXiv:1809.06655 [quant-ph]}  (\bibinfo
  {year} {2018})\BibitemShut {NoStop}%
\bibitem [{\citenamefont {Taddei}\ \emph {et~al.}(2021)\citenamefont {Taddei},
  \citenamefont {Cari\~ne}, \citenamefont {Mart\'{\i}nez}, \citenamefont
  {Garc\'{\i}a}, \citenamefont {Guerrero}, \citenamefont {Abbott},
  \citenamefont {Ara\'ujo}, \citenamefont {Branciard}, \citenamefont {G\'omez},
  \citenamefont {Walborn}, \citenamefont {Aolita},\ and\ \citenamefont
  {Lima}}]{taddei21}%
  \BibitemOpen
  \bibfield  {author} {\bibinfo {author} {\bibfnamefont {M.~M.}\ \bibnamefont
  {Taddei}}, \bibinfo {author} {\bibfnamefont {J.}~\bibnamefont {Cari\~ne}},
  \bibinfo {author} {\bibfnamefont {D.}~\bibnamefont {Mart\'{\i}nez}}, \bibinfo
  {author} {\bibfnamefont {T.}~\bibnamefont {Garc\'{\i}a}}, \bibinfo {author}
  {\bibfnamefont {N.}~\bibnamefont {Guerrero}}, \bibinfo {author}
  {\bibfnamefont {A.~A.}\ \bibnamefont {Abbott}}, \bibinfo {author}
  {\bibfnamefont {M.}~\bibnamefont {Ara\'ujo}}, \bibinfo {author}
  {\bibfnamefont {C.}~\bibnamefont {Branciard}}, \bibinfo {author}
  {\bibfnamefont {E.~S.}\ \bibnamefont {G\'omez}}, \bibinfo {author}
  {\bibfnamefont {S.~P.}\ \bibnamefont {Walborn}}, \bibinfo {author}
  {\bibfnamefont {L.}~\bibnamefont {Aolita}},\ and\ \bibinfo {author}
  {\bibfnamefont {G.}~\bibnamefont {Lima}},\ }\href
  {https://doi.org/10.1103/PRXQuantum.2.010320} {\bibfield  {journal} {\bibinfo
   {journal} {PRX Quantum}\ }\textbf {\bibinfo {volume} {2}},\ \bibinfo {pages}
  {010320} (\bibinfo {year} {2021})}\BibitemShut {NoStop}%
\bibitem [{\citenamefont {Chiribella}\ \emph
  {et~al.}(2021{\natexlab{a}})\citenamefont {Chiribella}, \citenamefont
  {Banik}, \citenamefont {Bhattacharya}, \citenamefont {Guha}, \citenamefont
  {Alimuddin}, \citenamefont {Roy}, \citenamefont {Saha}, \citenamefont
  {Agrawal},\ and\ \citenamefont {Kar}}]{Chiribella_2021}%
  \BibitemOpen
  \bibfield  {author} {\bibinfo {author} {\bibfnamefont {G.}~\bibnamefont
  {Chiribella}}, \bibinfo {author} {\bibfnamefont {M.}~\bibnamefont {Banik}},
  \bibinfo {author} {\bibfnamefont {S.~S.}\ \bibnamefont {Bhattacharya}},
  \bibinfo {author} {\bibfnamefont {T.}~\bibnamefont {Guha}}, \bibinfo {author}
  {\bibfnamefont {M.}~\bibnamefont {Alimuddin}}, \bibinfo {author}
  {\bibfnamefont {A.}~\bibnamefont {Roy}}, \bibinfo {author} {\bibfnamefont
  {S.}~\bibnamefont {Saha}}, \bibinfo {author} {\bibfnamefont {S.}~\bibnamefont
  {Agrawal}},\ and\ \bibinfo {author} {\bibfnamefont {G.}~\bibnamefont {Kar}},\
  }\href {https://doi.org/10.1088/1367-2630/abe7a0} {\bibfield  {journal}
  {\bibinfo  {journal} {New J. Phys.}\ }\textbf {\bibinfo {volume} {23}},\
  \bibinfo {pages} {033039} (\bibinfo {year} {2021}{\natexlab{a}})}\BibitemShut
  {NoStop}%
\bibitem [{\citenamefont {Chiribella}\ \emph
  {et~al.}(2021{\natexlab{b}})\citenamefont {Chiribella}, \citenamefont
  {Wilson},\ and\ \citenamefont {Chau}}]{Chiribella_2021b}%
  \BibitemOpen
  \bibfield  {author} {\bibinfo {author} {\bibfnamefont {G.}~\bibnamefont
  {Chiribella}}, \bibinfo {author} {\bibfnamefont {M.}~\bibnamefont {Wilson}},\
  and\ \bibinfo {author} {\bibfnamefont {H.~F.}\ \bibnamefont {Chau}},\ }\href
  {https://doi.org/10.1103/PhysRevLett.127.190502} {\bibfield  {journal}
  {\bibinfo  {journal} {Phys. Rev. Lett.}\ }\textbf {\bibinfo {volume} {127}},\
  \bibinfo {pages} {190502} (\bibinfo {year} {2021}{\natexlab{b}})}\BibitemShut
  {NoStop}%
\bibitem [{\citenamefont {Chiribella}\ \emph
  {et~al.}(2008{\natexlab{a}})\citenamefont {Chiribella}, \citenamefont
  {D'Ariano},\ and\ \citenamefont {Perinotti}}]{chiribella08a}%
  \BibitemOpen
  \bibfield  {author} {\bibinfo {author} {\bibfnamefont {G.}~\bibnamefont
  {Chiribella}}, \bibinfo {author} {\bibfnamefont {G.~M.}\ \bibnamefont
  {D'Ariano}},\ and\ \bibinfo {author} {\bibfnamefont {P.}~\bibnamefont
  {Perinotti}},\ }\href {https://doi.org/10.1209/0295-5075/83/30004} {\bibfield
   {journal} {\bibinfo  {journal} {EPL}\ }\textbf {\bibinfo {volume} {83}},\
  \bibinfo {pages} {30004} (\bibinfo {year} {2008}{\natexlab{a}})}\BibitemShut
  {NoStop}%
\bibitem [{\citenamefont {Friis}\ \emph {et~al.}(2014)\citenamefont {Friis},
  \citenamefont {Dunjko}, \citenamefont {D\"ur},\ and\ \citenamefont
  {Briegel}}]{Friis2014}%
  \BibitemOpen
  \bibfield  {author} {\bibinfo {author} {\bibfnamefont {N.}~\bibnamefont
  {Friis}}, \bibinfo {author} {\bibfnamefont {V.}~\bibnamefont {Dunjko}},
  \bibinfo {author} {\bibfnamefont {W.}~\bibnamefont {D\"ur}},\ and\ \bibinfo
  {author} {\bibfnamefont {H.~J.}\ \bibnamefont {Briegel}},\ }\href
  {https://doi.org/10.1103/PhysRevA.89.030303} {\bibfield  {journal} {\bibinfo
  {journal} {Phys. Rev. A}\ }\textbf {\bibinfo {volume} {89}},\ \bibinfo
  {pages} {030303} (\bibinfo {year} {2014})}\BibitemShut {NoStop}%
\bibitem [{\citenamefont {Procopio}\ \emph {et~al.}(2015)\citenamefont
  {Procopio}, \citenamefont {Moqanaki}, \citenamefont {Ara{\'u}jo},
  \citenamefont {Costa}, \citenamefont {Alonso~Calafell}, \citenamefont {Dowd},
  \citenamefont {Hamel}, \citenamefont {Rozema}, \citenamefont {Brukner},\ and\
  \citenamefont {Walther}}]{procopio_experimental_2015}%
  \BibitemOpen
  \bibfield  {author} {\bibinfo {author} {\bibfnamefont {L.~M.}\ \bibnamefont
  {Procopio}}, \bibinfo {author} {\bibfnamefont {A.}~\bibnamefont {Moqanaki}},
  \bibinfo {author} {\bibfnamefont {M.}~\bibnamefont {Ara{\'u}jo}}, \bibinfo
  {author} {\bibfnamefont {F.}~\bibnamefont {Costa}}, \bibinfo {author}
  {\bibfnamefont {I.}~\bibnamefont {Alonso~Calafell}}, \bibinfo {author}
  {\bibfnamefont {E.~G.}\ \bibnamefont {Dowd}}, \bibinfo {author}
  {\bibfnamefont {D.~R.}\ \bibnamefont {Hamel}}, \bibinfo {author}
  {\bibfnamefont {L.~A.}\ \bibnamefont {Rozema}}, \bibinfo {author}
  {\bibfnamefont {{\v C}.}~\bibnamefont {Brukner}},\ and\ \bibinfo {author}
  {\bibfnamefont {P.}~\bibnamefont {Walther}},\ }\href
  {https://doi.org/10.1038/ncomms8913} {\bibfield  {journal} {\bibinfo
  {journal} {Nat. Commun.}\ }\textbf {\bibinfo {volume} {6}},\ \bibinfo {pages}
  {7913} (\bibinfo {year} {2015})}\BibitemShut {NoStop}%
\bibitem [{\citenamefont {Rubino}\ \emph {et~al.}(2017)\citenamefont {Rubino},
  \citenamefont {Rozema}, \citenamefont {Feix}, \citenamefont {Ara{\'u}jo},
  \citenamefont {Zeuner}, \citenamefont {Procopio}, \citenamefont {Brukner},\
  and\ \citenamefont {Walther}}]{rubino17}%
  \BibitemOpen
  \bibfield  {author} {\bibinfo {author} {\bibfnamefont {G.}~\bibnamefont
  {Rubino}}, \bibinfo {author} {\bibfnamefont {L.~A.}\ \bibnamefont {Rozema}},
  \bibinfo {author} {\bibfnamefont {A.}~\bibnamefont {Feix}}, \bibinfo {author}
  {\bibfnamefont {M.}~\bibnamefont {Ara{\'u}jo}}, \bibinfo {author}
  {\bibfnamefont {J.~M.}\ \bibnamefont {Zeuner}}, \bibinfo {author}
  {\bibfnamefont {L.~M.}\ \bibnamefont {Procopio}}, \bibinfo {author}
  {\bibfnamefont {{\v C}.}~\bibnamefont {Brukner}},\ and\ \bibinfo {author}
  {\bibfnamefont {P.}~\bibnamefont {Walther}},\ }\href
  {https://doi.org/10.1126/sciadv.1602589} {\bibfield  {journal} {\bibinfo
  {journal} {Sci. Adv.}\ }\textbf {\bibinfo {volume} {3}},\ \bibinfo {pages}
  {e1602589} (\bibinfo {year} {2017})}\BibitemShut {NoStop}%
\bibitem [{\citenamefont {Goswami}\ \emph {et~al.}(2018)\citenamefont
  {Goswami}, \citenamefont {Giarmatzi}, \citenamefont {Kewming}, \citenamefont
  {Costa}, \citenamefont {Branciard}, \citenamefont {Romero},\ and\
  \citenamefont {White}}]{goswami18}%
  \BibitemOpen
  \bibfield  {author} {\bibinfo {author} {\bibfnamefont {K.}~\bibnamefont
  {Goswami}}, \bibinfo {author} {\bibfnamefont {C.}~\bibnamefont {Giarmatzi}},
  \bibinfo {author} {\bibfnamefont {M.}~\bibnamefont {Kewming}}, \bibinfo
  {author} {\bibfnamefont {F.}~\bibnamefont {Costa}}, \bibinfo {author}
  {\bibfnamefont {C.}~\bibnamefont {Branciard}}, \bibinfo {author}
  {\bibfnamefont {J.}~\bibnamefont {Romero}},\ and\ \bibinfo {author}
  {\bibfnamefont {A.~G.}\ \bibnamefont {White}},\ }\href
  {https://doi.org/10.1103/PhysRevLett.121.090503} {\bibfield  {journal}
  {\bibinfo  {journal} {Phys. Rev. Lett.}\ }\textbf {\bibinfo {volume} {121}},\
  \bibinfo {pages} {090503} (\bibinfo {year} {2018})}\BibitemShut {NoStop}%
\bibitem [{\citenamefont {Wei}\ \emph {et~al.}(2019)\citenamefont {Wei},
  \citenamefont {Tischler}, \citenamefont {Zhao}, \citenamefont {Li},
  \citenamefont {Arrazola}, \citenamefont {Liu}, \citenamefont {Zhang},
  \citenamefont {Li}, \citenamefont {You}, \citenamefont {Wang}, \citenamefont
  {Chen}, \citenamefont {Sanders}, \citenamefont {Zhang}, \citenamefont
  {Pryde}, \citenamefont {Xu},\ and\ \citenamefont {Pan}}]{Wei2019}%
  \BibitemOpen
  \bibfield  {author} {\bibinfo {author} {\bibfnamefont {K.}~\bibnamefont
  {Wei}}, \bibinfo {author} {\bibfnamefont {N.}~\bibnamefont {Tischler}},
  \bibinfo {author} {\bibfnamefont {S.-R.}\ \bibnamefont {Zhao}}, \bibinfo
  {author} {\bibfnamefont {Y.-H.}\ \bibnamefont {Li}}, \bibinfo {author}
  {\bibfnamefont {J.~M.}\ \bibnamefont {Arrazola}}, \bibinfo {author}
  {\bibfnamefont {Y.}~\bibnamefont {Liu}}, \bibinfo {author} {\bibfnamefont
  {W.}~\bibnamefont {Zhang}}, \bibinfo {author} {\bibfnamefont
  {H.}~\bibnamefont {Li}}, \bibinfo {author} {\bibfnamefont {L.}~\bibnamefont
  {You}}, \bibinfo {author} {\bibfnamefont {Z.}~\bibnamefont {Wang}}, \bibinfo
  {author} {\bibfnamefont {Y.-A.}\ \bibnamefont {Chen}}, \bibinfo {author}
  {\bibfnamefont {B.~C.}\ \bibnamefont {Sanders}}, \bibinfo {author}
  {\bibfnamefont {Q.}~\bibnamefont {Zhang}}, \bibinfo {author} {\bibfnamefont
  {G.~J.}\ \bibnamefont {Pryde}}, \bibinfo {author} {\bibfnamefont
  {F.}~\bibnamefont {Xu}},\ and\ \bibinfo {author} {\bibfnamefont {J.-W.}\
  \bibnamefont {Pan}},\ }\href {https://doi.org/10.1103/PhysRevLett.122.120504}
  {\bibfield  {journal} {\bibinfo  {journal} {Phys. Rev. Lett.}\ }\textbf
  {\bibinfo {volume} {122}},\ \bibinfo {pages} {120504} (\bibinfo {year}
  {2019})}\BibitemShut {NoStop}%
\bibitem [{\citenamefont {Goswami}\ \emph {et~al.}(2020)\citenamefont
  {Goswami}, \citenamefont {Cao}, \citenamefont {Paz-Silva}, \citenamefont
  {Romero},\ and\ \citenamefont {White}}]{Goswami2020}%
  \BibitemOpen
  \bibfield  {author} {\bibinfo {author} {\bibfnamefont {K.}~\bibnamefont
  {Goswami}}, \bibinfo {author} {\bibfnamefont {Y.}~\bibnamefont {Cao}},
  \bibinfo {author} {\bibfnamefont {G.~A.}\ \bibnamefont {Paz-Silva}}, \bibinfo
  {author} {\bibfnamefont {J.}~\bibnamefont {Romero}},\ and\ \bibinfo {author}
  {\bibfnamefont {A.~G.}\ \bibnamefont {White}},\ }\href
  {https://doi.org/10.1103/PhysRevResearch.2.033292} {\bibfield  {journal}
  {\bibinfo  {journal} {Phys. Rev. Research}\ }\textbf {\bibinfo {volume}
  {2}},\ \bibinfo {pages} {033292} (\bibinfo {year} {2020})}\BibitemShut
  {NoStop}%
\bibitem [{\citenamefont {Guo}\ \emph {et~al.}(2020)\citenamefont {Guo},
  \citenamefont {Hu}, \citenamefont {Hou}, \citenamefont {Cao}, \citenamefont
  {Cui}, \citenamefont {Liu}, \citenamefont {Huang}, \citenamefont {Li},
  \citenamefont {Guo},\ and\ \citenamefont {Chiribella}}]{guo20}%
  \BibitemOpen
  \bibfield  {author} {\bibinfo {author} {\bibfnamefont {Y.}~\bibnamefont
  {Guo}}, \bibinfo {author} {\bibfnamefont {X.-M.}\ \bibnamefont {Hu}},
  \bibinfo {author} {\bibfnamefont {Z.-B.}\ \bibnamefont {Hou}}, \bibinfo
  {author} {\bibfnamefont {H.}~\bibnamefont {Cao}}, \bibinfo {author}
  {\bibfnamefont {J.-M.}\ \bibnamefont {Cui}}, \bibinfo {author} {\bibfnamefont
  {B.-H.}\ \bibnamefont {Liu}}, \bibinfo {author} {\bibfnamefont {Y.-F.}\
  \bibnamefont {Huang}}, \bibinfo {author} {\bibfnamefont {C.-F.}\ \bibnamefont
  {Li}}, \bibinfo {author} {\bibfnamefont {G.-C.}\ \bibnamefont {Guo}},\ and\
  \bibinfo {author} {\bibfnamefont {G.}~\bibnamefont {Chiribella}},\ }\href
  {https://doi.org/10.1103/PhysRevLett.124.030502} {\bibfield  {journal}
  {\bibinfo  {journal} {Phys. Rev. Lett.}\ }\textbf {\bibinfo {volume} {124}},\
  \bibinfo {pages} {030502} (\bibinfo {year} {2020})}\BibitemShut {NoStop}%
\bibitem [{\citenamefont {Rubino}\ \emph {et~al.}(2021)\citenamefont {Rubino},
  \citenamefont {Rozema}, \citenamefont {Ebler}, \citenamefont
  {Kristj\'ansson}, \citenamefont {Salek}, \citenamefont {Allard~Gu\'erin},
  \citenamefont {Abbott}, \citenamefont {Branciard}, \citenamefont {Brukner},
  \citenamefont {Chiribella},\ and\ \citenamefont
  {Walther}}]{rubino_experimental_2021}%
  \BibitemOpen
  \bibfield  {author} {\bibinfo {author} {\bibfnamefont {G.}~\bibnamefont
  {Rubino}}, \bibinfo {author} {\bibfnamefont {L.~A.}\ \bibnamefont {Rozema}},
  \bibinfo {author} {\bibfnamefont {D.}~\bibnamefont {Ebler}}, \bibinfo
  {author} {\bibfnamefont {H.}~\bibnamefont {Kristj\'ansson}}, \bibinfo
  {author} {\bibfnamefont {S.}~\bibnamefont {Salek}}, \bibinfo {author}
  {\bibfnamefont {P.}~\bibnamefont {Allard~Gu\'erin}}, \bibinfo {author}
  {\bibfnamefont {A.~A.}\ \bibnamefont {Abbott}}, \bibinfo {author}
  {\bibfnamefont {C.}~\bibnamefont {Branciard}}, \bibinfo {author}
  {\bibfnamefont {{\v C}.}~\bibnamefont {Brukner}}, \bibinfo {author}
  {\bibfnamefont {G.}~\bibnamefont {Chiribella}},\ and\ \bibinfo {author}
  {\bibfnamefont {P.}~\bibnamefont {Walther}},\ }\href
  {https://doi.org/10.1103/PhysRevResearch.3.013093} {\bibfield  {journal}
  {\bibinfo  {journal} {Phys. Rev. Research}\ }\textbf {\bibinfo {volume}
  {3}},\ \bibinfo {pages} {013093} (\bibinfo {year} {2021})}\BibitemShut
  {NoStop}%
\bibitem [{\citenamefont {Rubino}\ \emph {et~al.}(2022)\citenamefont {Rubino},
  \citenamefont {Rozema}, \citenamefont {Massa}, \citenamefont {Ara{\'{u}}jo},
  \citenamefont {Zych}, \citenamefont {Brukner},\ and\ \citenamefont
  {Walther}}]{Rubino2022experimental}%
  \BibitemOpen
  \bibfield  {author} {\bibinfo {author} {\bibfnamefont {G.}~\bibnamefont
  {Rubino}}, \bibinfo {author} {\bibfnamefont {L.~A.}\ \bibnamefont {Rozema}},
  \bibinfo {author} {\bibfnamefont {F.}~\bibnamefont {Massa}}, \bibinfo
  {author} {\bibfnamefont {M.}~\bibnamefont {Ara{\'{u}}jo}}, \bibinfo {author}
  {\bibfnamefont {M.}~\bibnamefont {Zych}}, \bibinfo {author} {\bibfnamefont
  {{\v{C}}.}~\bibnamefont {Brukner}},\ and\ \bibinfo {author} {\bibfnamefont
  {P.}~\bibnamefont {Walther}},\ }\href
  {https://doi.org/10.22331/q-2022-01-11-621} {\bibfield  {journal} {\bibinfo
  {journal} {{Quantum}}\ }\textbf {\bibinfo {volume} {6}},\ \bibinfo {pages}
  {621} (\bibinfo {year} {2022})}\BibitemShut {NoStop}%
\bibitem [{\citenamefont {Cao}\ \emph {et~al.}(2022)\citenamefont {Cao},
  \citenamefont {Bavaresco}, \citenamefont {Wang}, \citenamefont {Rozema},
  \citenamefont {Zhang}, \citenamefont {Huang}, \citenamefont {Liu},
  \citenamefont {Li}, \citenamefont {Guo},\ and\ \citenamefont
  {Walther}}]{cao22}%
  \BibitemOpen
  \bibfield  {author} {\bibinfo {author} {\bibfnamefont {H.}~\bibnamefont
  {Cao}}, \bibinfo {author} {\bibfnamefont {J.}~\bibnamefont {Bavaresco}},
  \bibinfo {author} {\bibfnamefont {N.-N.}\ \bibnamefont {Wang}}, \bibinfo
  {author} {\bibfnamefont {L.~A.}\ \bibnamefont {Rozema}}, \bibinfo {author}
  {\bibfnamefont {C.}~\bibnamefont {Zhang}}, \bibinfo {author} {\bibfnamefont
  {Y.-F.}\ \bibnamefont {Huang}}, \bibinfo {author} {\bibfnamefont {B.-H.}\
  \bibnamefont {Liu}}, \bibinfo {author} {\bibfnamefont {C.-F.}\ \bibnamefont
  {Li}}, \bibinfo {author} {\bibfnamefont {G.-C.}\ \bibnamefont {Guo}},\ and\
  \bibinfo {author} {\bibfnamefont {P.}~\bibnamefont {Walther}},\ }\Eprint
  {https://arxiv.org/abs/2202.05346} {arXiv:2202.05346 [quant-ph]}  (\bibinfo
  {year} {2022})\BibitemShut {NoStop}%
\bibitem [{\citenamefont {MacLean}\ \emph {et~al.}(2017)\citenamefont
  {MacLean}, \citenamefont {Ried}, \citenamefont {Spekkens},\ and\
  \citenamefont {Resch}}]{maclean17}%
  \BibitemOpen
  \bibfield  {author} {\bibinfo {author} {\bibfnamefont {J.-P.~W.}\
  \bibnamefont {MacLean}}, \bibinfo {author} {\bibfnamefont {K.}~\bibnamefont
  {Ried}}, \bibinfo {author} {\bibfnamefont {R.~W.}\ \bibnamefont {Spekkens}},\
  and\ \bibinfo {author} {\bibfnamefont {K.~J.}\ \bibnamefont {Resch}},\ }\href
  {https://doi.org/10.1038/ncomms15149} {\bibfield  {journal} {\bibinfo
  {journal} {Nat. Commun.}\ }\textbf {\bibinfo {volume} {8}},\ \bibinfo {pages}
  {15149} (\bibinfo {year} {2017})}\BibitemShut {NoStop}%
\bibitem [{\citenamefont {Oreshkov}(2019)}]{oreshkov19}%
  \BibitemOpen
  \bibfield  {author} {\bibinfo {author} {\bibfnamefont {O.}~\bibnamefont
  {Oreshkov}},\ }\href {https://doi.org/10.22331/q-2019-12-02-206} {\bibfield
  {journal} {\bibinfo  {journal} {{Quantum}}\ }\textbf {\bibinfo {volume}
  {3}},\ \bibinfo {pages} {206} (\bibinfo {year} {2019})}\BibitemShut {NoStop}%
\bibitem [{\citenamefont {Paunkovi\'{c}}\ and\ \citenamefont
  {Vojinovi\'{c}}(2020)}]{paunkovic20}%
  \BibitemOpen
  \bibfield  {author} {\bibinfo {author} {\bibfnamefont {N.}~\bibnamefont
  {Paunkovi\'{c}}}\ and\ \bibinfo {author} {\bibfnamefont {M.}~\bibnamefont
  {Vojinovi\'{c}}},\ }\href {https://doi.org/10.22331/q-2020-05-28-275}
  {\bibfield  {journal} {\bibinfo  {journal} {Quantum}\ }\textbf {\bibinfo
  {volume} {4}},\ \bibinfo {pages} {275} (\bibinfo {year} {2020})}\BibitemShut
  {NoStop}%
\bibitem [{\citenamefont {Kristj\'ansson}\ \emph {et~al.}(2021)\citenamefont
  {Kristj\'ansson}, \citenamefont {Mao},\ and\ \citenamefont
  {Chiribella}}]{kristjansson20}%
  \BibitemOpen
  \bibfield  {author} {\bibinfo {author} {\bibfnamefont {H.}~\bibnamefont
  {Kristj\'ansson}}, \bibinfo {author} {\bibfnamefont {W.}~\bibnamefont
  {Mao}},\ and\ \bibinfo {author} {\bibfnamefont {G.}~\bibnamefont
  {Chiribella}},\ }\href {https://doi.org/10.1103/PhysRevResearch.3.043147}
  {\bibfield  {journal} {\bibinfo  {journal} {Phys. Rev. Research}\ }\textbf
  {\bibinfo {volume} {3}},\ \bibinfo {pages} {043147} (\bibinfo {year}
  {2021})}\BibitemShut {NoStop}%
\bibitem [{\citenamefont {Vilasini}\ and\ \citenamefont
  {Renner}(2022)}]{vilasini22}%
  \BibitemOpen
  \bibfield  {author} {\bibinfo {author} {\bibfnamefont {V.}~\bibnamefont
  {Vilasini}}\ and\ \bibinfo {author} {\bibfnamefont {R.}~\bibnamefont
  {Renner}},\ }\Eprint {https://arxiv.org/abs/2203.11245} {arXiv:2203.11245
  [quant-ph]}  (\bibinfo {year} {2022})\BibitemShut {NoStop}%
\bibitem [{\citenamefont {Ormrod}\ \emph {et~al.}(2022)\citenamefont {Ormrod},
  \citenamefont {Vanrietvelde},\ and\ \citenamefont {Barrett}}]{ormrod22}%
  \BibitemOpen
  \bibfield  {author} {\bibinfo {author} {\bibfnamefont {N.}~\bibnamefont
  {Ormrod}}, \bibinfo {author} {\bibfnamefont {A.}~\bibnamefont
  {Vanrietvelde}},\ and\ \bibinfo {author} {\bibfnamefont {J.}~\bibnamefont
  {Barrett}},\ }\Eprint {https://arxiv.org/abs/2204.10273} {arXiv:2204.10273
  [quant-ph]}  (\bibinfo {year} {2022})\BibitemShut {NoStop}%
\bibitem [{\citenamefont {Felce}\ and\ \citenamefont
  {Vedral}(2020)}]{felce_quantum_2020}%
  \BibitemOpen
  \bibfield  {author} {\bibinfo {author} {\bibfnamefont {D.}~\bibnamefont
  {Felce}}\ and\ \bibinfo {author} {\bibfnamefont {V.}~\bibnamefont {Vedral}},\
  }\href {https://doi.org/10.1103/PhysRevLett.125.070603} {\bibfield  {journal}
  {\bibinfo  {journal} {Phys. Rev. Lett.}\ }\textbf {\bibinfo {volume} {125}},\
  \bibinfo {pages} {070603} (\bibinfo {year} {2020})}\BibitemShut {NoStop}%
\bibitem [{\citenamefont {Guha}\ \emph {et~al.}(2020)\citenamefont {Guha},
  \citenamefont {Alimuddin},\ and\ \citenamefont {Parashar}}]{Guha2020}%
  \BibitemOpen
  \bibfield  {author} {\bibinfo {author} {\bibfnamefont {T.}~\bibnamefont
  {Guha}}, \bibinfo {author} {\bibfnamefont {M.}~\bibnamefont {Alimuddin}},\
  and\ \bibinfo {author} {\bibfnamefont {P.}~\bibnamefont {Parashar}},\ }\href
  {https://doi.org/10.1103/PhysRevA.102.032215} {\bibfield  {journal} {\bibinfo
   {journal} {Phys. Rev. A}\ }\textbf {\bibinfo {volume} {102}},\ \bibinfo
  {pages} {032215} (\bibinfo {year} {2020})}\BibitemShut {NoStop}%
\bibitem [{\citenamefont {Simonov}\ \emph {et~al.}(2022)\citenamefont
  {Simonov}, \citenamefont {Francica}, \citenamefont {Guarnieri},\ and\
  \citenamefont {Paternostro}}]{Simonov2022}%
  \BibitemOpen
  \bibfield  {author} {\bibinfo {author} {\bibfnamefont {K.}~\bibnamefont
  {Simonov}}, \bibinfo {author} {\bibfnamefont {G.}~\bibnamefont {Francica}},
  \bibinfo {author} {\bibfnamefont {G.}~\bibnamefont {Guarnieri}},\ and\
  \bibinfo {author} {\bibfnamefont {M.}~\bibnamefont {Paternostro}},\ }\href
  {https://doi.org/10.1103/PhysRevA.105.032217} {\bibfield  {journal} {\bibinfo
   {journal} {Phys. Rev. A}\ }\textbf {\bibinfo {volume} {105}},\ \bibinfo
  {pages} {032217} (\bibinfo {year} {2022})}\BibitemShut {NoStop}%
\bibitem [{\citenamefont {Chen}\ and\ \citenamefont
  {Hasegawa}(2021)}]{Chen2021}%
  \BibitemOpen
  \bibfield  {author} {\bibinfo {author} {\bibfnamefont {Y.}~\bibnamefont
  {Chen}}\ and\ \bibinfo {author} {\bibfnamefont {Y.}~\bibnamefont
  {Hasegawa}},\ }\Eprint {https://arxiv.org/abs/2105.12466} {arXiv:2105.12466
  [quant-ph]}  (\bibinfo {year} {2021})\BibitemShut {NoStop}%
\bibitem [{\citenamefont {Zhao}\ and\ \citenamefont {Xu}(2022)}]{Zhao2022}%
  \BibitemOpen
  \bibfield  {author} {\bibinfo {author} {\bibfnamefont {J.}~\bibnamefont
  {Zhao}}\ and\ \bibinfo {author} {\bibfnamefont {Y.}~\bibnamefont {Xu}},\
  }\href {https://doi.org/10.1088/1572-9494/ac490d} {\bibfield  {journal}
  {\bibinfo  {journal} {Commun. Theor. Phys.}\ }\textbf {\bibinfo {volume}
  {74}},\ \bibinfo {pages} {025601} (\bibinfo {year} {2022})}\BibitemShut
  {NoStop}%
\bibitem [{\citenamefont {Nie}\ \emph {et~al.}(2022)\citenamefont {Nie},
  \citenamefont {Feng}, \citenamefont {Longden},\ and\ \citenamefont
  {Vedral}}]{Nie2022}%
  \BibitemOpen
  \bibfield  {author} {\bibinfo {author} {\bibfnamefont {H.}~\bibnamefont
  {Nie}}, \bibinfo {author} {\bibfnamefont {T.}~\bibnamefont {Feng}}, \bibinfo
  {author} {\bibfnamefont {S.}~\bibnamefont {Longden}},\ and\ \bibinfo {author}
  {\bibfnamefont {V.}~\bibnamefont {Vedral}},\ }\Eprint
  {https://arxiv.org/abs/2201.06954} {arXiv:2201.06954 [quant-ph]}  (\bibinfo
  {year} {2022})\BibitemShut {NoStop}%
\bibitem [{\citenamefont {Nie}\ \emph {et~al.}(2020)\citenamefont {Nie},
  \citenamefont {Zhu}, \citenamefont {Xi}, \citenamefont {Long}, \citenamefont
  {Lin}, \citenamefont {Tian}, \citenamefont {Qiu}, \citenamefont {Yang},
  \citenamefont {Dong}, \citenamefont {Li}, \citenamefont {Xin},\ and\
  \citenamefont {Lu}}]{Nie2020}%
  \BibitemOpen
  \bibfield  {author} {\bibinfo {author} {\bibfnamefont {X.}~\bibnamefont
  {Nie}}, \bibinfo {author} {\bibfnamefont {X.}~\bibnamefont {Zhu}}, \bibinfo
  {author} {\bibfnamefont {C.}~\bibnamefont {Xi}}, \bibinfo {author}
  {\bibfnamefont {X.}~\bibnamefont {Long}}, \bibinfo {author} {\bibfnamefont
  {Z.}~\bibnamefont {Lin}}, \bibinfo {author} {\bibfnamefont {Y.}~\bibnamefont
  {Tian}}, \bibinfo {author} {\bibfnamefont {C.}~\bibnamefont {Qiu}}, \bibinfo
  {author} {\bibfnamefont {X.}~\bibnamefont {Yang}}, \bibinfo {author}
  {\bibfnamefont {Y.}~\bibnamefont {Dong}}, \bibinfo {author} {\bibfnamefont
  {J.}~\bibnamefont {Li}}, \bibinfo {author} {\bibfnamefont {T.}~\bibnamefont
  {Xin}},\ and\ \bibinfo {author} {\bibfnamefont {D.}~\bibnamefont {Lu}},\
  }\Eprint {https://arxiv.org/abs/2011.12580} {arXiv:2011.12580 [quant-ph]}
  (\bibinfo {year} {2020})\BibitemShut {NoStop}%
\bibitem [{\citenamefont {Cao}\ \emph {et~al.}(2021)\citenamefont {Cao},
  \citenamefont {ning Wang}, \citenamefont {Jia}, \citenamefont {Zhang},
  \citenamefont {Guo}, \citenamefont {Liu}, \citenamefont {Huang},
  \citenamefont {Li},\ and\ \citenamefont {Guo}}]{cao21}%
  \BibitemOpen
  \bibfield  {author} {\bibinfo {author} {\bibfnamefont {H.}~\bibnamefont
  {Cao}}, \bibinfo {author} {\bibfnamefont {N.}~\bibnamefont {ning Wang}},
  \bibinfo {author} {\bibfnamefont {Z.-A.}\ \bibnamefont {Jia}}, \bibinfo
  {author} {\bibfnamefont {C.}~\bibnamefont {Zhang}}, \bibinfo {author}
  {\bibfnamefont {Y.}~\bibnamefont {Guo}}, \bibinfo {author} {\bibfnamefont
  {B.-H.}\ \bibnamefont {Liu}}, \bibinfo {author} {\bibfnamefont {Y.-F.}\
  \bibnamefont {Huang}}, \bibinfo {author} {\bibfnamefont {C.-F.}\ \bibnamefont
  {Li}},\ and\ \bibinfo {author} {\bibfnamefont {G.-C.}\ \bibnamefont {Guo}},\
  }\Eprint {https://arxiv.org/abs/2101.07979} {arXiv:2101.07979 [quant-ph]}
  (\bibinfo {year} {2021})\BibitemShut {NoStop}%
\bibitem [{\citenamefont {Felce}\ \emph {et~al.}(2021)\citenamefont {Felce},
  \citenamefont {Vedral},\ and\ \citenamefont {Tennie}}]{Felce2021}%
  \BibitemOpen
  \bibfield  {author} {\bibinfo {author} {\bibfnamefont {D.}~\bibnamefont
  {Felce}}, \bibinfo {author} {\bibfnamefont {V.}~\bibnamefont {Vedral}},\ and\
  \bibinfo {author} {\bibfnamefont {F.}~\bibnamefont {Tennie}},\ }\Eprint
  {https://arxiv.org/abs/2107.12413} {arXiv:2107.12413 [quant-ph]}  (\bibinfo
  {year} {2021})\BibitemShut {NoStop}%
\bibitem [{\citenamefont {Haroche}\ and\ \citenamefont
  {Raimond}(2006)}]{haroche2006exploring}%
  \BibitemOpen
  \bibfield  {author} {\bibinfo {author} {\bibfnamefont {S.}~\bibnamefont
  {Haroche}}\ and\ \bibinfo {author} {\bibfnamefont {J.-M.}\ \bibnamefont
  {Raimond}},\ }\href
  {https://doi.org/10.1093/acprof:oso/9780198509141.001.0001} {\emph {\bibinfo
  {title} {Exploring the Quantum: Atoms, Cavities, and Photons}}}\ (\bibinfo
  {publisher} {Oxford University Press},\ \bibinfo {address} {Oxford},\
  \bibinfo {year} {2006})\BibitemShut {NoStop}%
\bibitem [{\citenamefont {Barnes}\ and\ \citenamefont
  {Warren}(1999)}]{Barnes99}%
  \BibitemOpen
  \bibfield  {author} {\bibinfo {author} {\bibfnamefont {J.~P.}\ \bibnamefont
  {Barnes}}\ and\ \bibinfo {author} {\bibfnamefont {W.~S.}\ \bibnamefont
  {Warren}},\ }\href {https://doi.org/10.1103/PhysRevA.60.4363} {\bibfield
  {journal} {\bibinfo  {journal} {Phys. Rev. A}\ }\textbf {\bibinfo {volume}
  {60}},\ \bibinfo {pages} {4363} (\bibinfo {year} {1999})}\BibitemShut
  {NoStop}%
\bibitem [{\citenamefont
  {Gea-Banacloche}(2002{\natexlab{a}})}]{Gea-Banacloche02}%
  \BibitemOpen
  \bibfield  {author} {\bibinfo {author} {\bibfnamefont {J.}~\bibnamefont
  {Gea-Banacloche}},\ }\href {https://doi.org/10.1103/PhysRevA.65.022308}
  {\bibfield  {journal} {\bibinfo  {journal} {Phys. Rev. A}\ }\textbf {\bibinfo
  {volume} {65}},\ \bibinfo {pages} {022308} (\bibinfo {year}
  {2002}{\natexlab{a}})}\BibitemShut {NoStop}%
\bibitem [{\citenamefont
  {Gea-Banacloche}(2002{\natexlab{b}})}]{Gea-Banacloche02_2}%
  \BibitemOpen
  \bibfield  {author} {\bibinfo {author} {\bibfnamefont {J.}~\bibnamefont
  {Gea-Banacloche}},\ }\href {https://doi.org/10.1103/PhysRevLett.89.217901}
  {\bibfield  {journal} {\bibinfo  {journal} {Phys. Rev. Lett.}\ }\textbf
  {\bibinfo {volume} {89}},\ \bibinfo {pages} {217901} (\bibinfo {year}
  {2002}{\natexlab{b}})}\BibitemShut {NoStop}%
\bibitem [{\citenamefont {Ozawa}(2002)}]{Ozawa02}%
  \BibitemOpen
  \bibfield  {author} {\bibinfo {author} {\bibfnamefont {M.}~\bibnamefont
  {Ozawa}},\ }\href {https://doi.org/10.1103/PhysRevLett.89.057902} {\bibfield
  {journal} {\bibinfo  {journal} {Phys. Rev. Lett.}\ }\textbf {\bibinfo
  {volume} {89}},\ \bibinfo {pages} {057902} (\bibinfo {year}
  {2002})}\BibitemShut {NoStop}%
\bibitem [{\citenamefont {Grynberg}\ \emph {et~al.}(2010)\citenamefont
  {Grynberg}, \citenamefont {Aspect},\ and\ \citenamefont
  {Fabre}}]{AspectandFabrebook}%
  \BibitemOpen
  \bibfield  {author} {\bibinfo {author} {\bibfnamefont {G.}~\bibnamefont
  {Grynberg}}, \bibinfo {author} {\bibfnamefont {A.}~\bibnamefont {Aspect}},\
  and\ \bibinfo {author} {\bibfnamefont {C.}~\bibnamefont {Fabre}},\ }\href
  {https://doi.org/10.1017/CBO9780511778261} {\emph {\bibinfo {title}
  {{Introduction} {to} {Quantum} {Optics:} {From the Semi-classical Approach to
  Quantized Light}}}}\ (\bibinfo  {publisher} {Cambridge University Press},\
  \bibinfo {address} {Cambridge},\ \bibinfo {year} {2010})\BibitemShut
  {NoStop}%
\bibitem [{\citenamefont {Chiribella}\ \emph
  {et~al.}(2008{\natexlab{b}})\citenamefont {Chiribella}, \citenamefont
  {D'Ariano},\ and\ \citenamefont {Perinotti}}]{Chiribella08}%
  \BibitemOpen
  \bibfield  {author} {\bibinfo {author} {\bibfnamefont {G.}~\bibnamefont
  {Chiribella}}, \bibinfo {author} {\bibfnamefont {G.~M.}\ \bibnamefont
  {D'Ariano}},\ and\ \bibinfo {author} {\bibfnamefont {P.}~\bibnamefont
  {Perinotti}},\ }\href {https://doi.org/10.1103/PhysRevLett.101.060401}
  {\bibfield  {journal} {\bibinfo  {journal} {Phys. Rev. Lett.}\ }\textbf
  {\bibinfo {volume} {101}},\ \bibinfo {pages} {060401} (\bibinfo {year}
  {2008}{\natexlab{b}})}\BibitemShut {NoStop}%
\bibitem [{\citenamefont {Ara{\'{u}}jo}\ \emph {et~al.}(2015)\citenamefont
  {Ara{\'{u}}jo}, \citenamefont {Branciard}, \citenamefont {Costa},
  \citenamefont {Feix}, \citenamefont {Giarmatzi},\ and\ \citenamefont
  {Brukner}}]{araujo_witnessing_2015}%
  \BibitemOpen
  \bibfield  {author} {\bibinfo {author} {\bibfnamefont {M.}~\bibnamefont
  {Ara{\'{u}}jo}}, \bibinfo {author} {\bibfnamefont {C.}~\bibnamefont
  {Branciard}}, \bibinfo {author} {\bibfnamefont {F.}~\bibnamefont {Costa}},
  \bibinfo {author} {\bibfnamefont {A.}~\bibnamefont {Feix}}, \bibinfo {author}
  {\bibfnamefont {C.}~\bibnamefont {Giarmatzi}},\ and\ \bibinfo {author}
  {\bibfnamefont {{\v{C}}.}~\bibnamefont {Brukner}},\ }\href
  {https://doi.org/10.1088/1367-2630/17/10/102001} {\bibfield  {journal}
  {\bibinfo  {journal} {New J. Phys.}\ }\textbf {\bibinfo {volume} {17}},\
  \bibinfo {pages} {102001} (\bibinfo {year} {2015})}\BibitemShut {NoStop}%
\bibitem [{\citenamefont {Chiribella}\ \emph {et~al.}(2009)\citenamefont
  {Chiribella}, \citenamefont {D'Ariano},\ and\ \citenamefont
  {Perinotti}}]{Chiribella09}%
  \BibitemOpen
  \bibfield  {author} {\bibinfo {author} {\bibfnamefont {G.}~\bibnamefont
  {Chiribella}}, \bibinfo {author} {\bibfnamefont {G.~M.}\ \bibnamefont
  {D'Ariano}},\ and\ \bibinfo {author} {\bibfnamefont {P.}~\bibnamefont
  {Perinotti}},\ }\href {https://doi.org/10.1103/PhysRevA.80.022339} {\bibfield
   {journal} {\bibinfo  {journal} {Phys. Rev. A}\ }\textbf {\bibinfo {volume}
  {80}},\ \bibinfo {pages} {022339} (\bibinfo {year} {2009})}\BibitemShut
  {NoStop}%
\bibitem [{\citenamefont {Cuomo}\ \emph {et~al.}(2021)\citenamefont {Cuomo},
  \citenamefont {Caleffi},\ and\ \citenamefont {Cacciapuoti}}]{cuomo}%
  \BibitemOpen
  \bibfield  {author} {\bibinfo {author} {\bibfnamefont {D.}~\bibnamefont
  {Cuomo}}, \bibinfo {author} {\bibfnamefont {M.}~\bibnamefont {Caleffi}},\
  and\ \bibinfo {author} {\bibfnamefont {A.~S.}\ \bibnamefont {Cacciapuoti}},\
  }in\ \href {https://doi.org/10.1109/SPAWC51858.2021.9593171} {\emph {\bibinfo
  {booktitle} {2021 {IEEE} 22nd {International} {Workshop} on {Signal}
  {Processing} {Advances} in {Wireless} {Communications} ({SPAWC})}}}\
  (\bibinfo {year} {2021})\BibitemShut {NoStop}%
\bibitem [{\citenamefont {Bowdrey}\ \emph {et~al.}(2002)\citenamefont
  {Bowdrey}, \citenamefont {Oi}, \citenamefont {Short}, \citenamefont
  {Banaszek},\ and\ \citenamefont {Jones}}]{Bowdrey2002}%
  \BibitemOpen
  \bibfield  {author} {\bibinfo {author} {\bibfnamefont {M.~D.}\ \bibnamefont
  {Bowdrey}}, \bibinfo {author} {\bibfnamefont {D.~K.}\ \bibnamefont {Oi}},
  \bibinfo {author} {\bibfnamefont {A.~J.}\ \bibnamefont {Short}}, \bibinfo
  {author} {\bibfnamefont {K.}~\bibnamefont {Banaszek}},\ and\ \bibinfo
  {author} {\bibfnamefont {J.~A.}\ \bibnamefont {Jones}},\ }\href
  {https://doi.org/https://doi.org/10.1016/S0375-9601(02)00069-5} {\bibfield
  {journal} {\bibinfo  {journal} {Phys. Lett. A}\ }\textbf {\bibinfo {volume}
  {294}},\ \bibinfo {pages} {258} (\bibinfo {year} {2002})}\BibitemShut
  {NoStop}%
\bibitem [{\citenamefont {Choi}(1975)}]{choi75}%
  \BibitemOpen
  \bibfield  {author} {\bibinfo {author} {\bibfnamefont {M.-D.}\ \bibnamefont
  {Choi}},\ }\href {https://doi.org/10.1016/0024-3795(75)90075-0} {\bibfield
  {journal} {\bibinfo  {journal} {Linear Algebra Appl.}\ }\textbf {\bibinfo
  {volume} {10}},\ \bibinfo {pages} {285} (\bibinfo {year} {1975})}\BibitemShut
  {NoStop}%
\bibitem [{\citenamefont {Haight}(1967)}]{Haight1967handbook}%
  \BibitemOpen
  \bibfield  {author} {\bibinfo {author} {\bibfnamefont {F.~A.}\ \bibnamefont
  {Haight}},\ }\href@noop {} {\emph {\bibinfo {title} {Handbook of the
  {P}oisson distribution}}}\ (\bibinfo  {publisher} {John Wiley \& Sons},\
  \bibinfo {address} {New York},\ \bibinfo {year} {1967})\BibitemShut {NoStop}%
\end{thebibliography}%

\end{document}